\documentclass[aps,prm,amsmath,amssymb,reprint,superscriptaddress,nofootinbib]{revtex4-2}
\usepackage{charter,graphicx,verbatim,threeparttable,float,amssymb,gensymb }
\usepackage{amsfonts}
\usepackage{upgreek}
\usepackage{booktabs}
\begin{document}

\newcommand{\oss}{\textit{AM}$_6$\textit{X}$_6$}
\newcommand{\nboss}{\textit{Ln}Nb$_6$Sn$_6$}
\newcommand{\luoss}{LuNb$_6$Sn$_6$}
\newcommand{\hfgesn}{HfFe$_6$Ge$_6$}

\preprint{APS/123-QED}

\title{
Stability frontiers in the \textit{AM}$_6$\textit{X}$_6$ kagome metals: The \textit{Ln}Nb$_6$Sn$_6$ (\textit{Ln}:Ce--Lu,Y) family and density-wave transition in LuNb$_6$Sn$_6$}

\author{Brenden R. Ortiz}
\email{ortizbr@ornl.gov}
\affiliation{Materials Science and Technology Division, Oak Ridge National Laboratory, Oak Ridge, TN 37831, USA}

\author{William R. Meier} 
\email{javamocham@gmail.com}
\affiliation{Materials Science and Engineering Department, University of Tennessee Knoxville, Knoxville, TN 37996, USA}

\author{Ganesh Pokharel} 
\affiliation{Perry College of Sciences, University of West Georgia, Carrollton, GA 30118, USA}

\author{Juan Chamorro} 
\affiliation{Materials Science and Engineering Department, Carnegie Mellon University, Pittsburgh, PA 15213, USA}

\author{Fazhi Yang} 
\affiliation{Materials Science and Technology Division, Oak Ridge National Laboratory, Oak Ridge, TN 37831, USA}

\author{Shirin Mozaffari} 
\affiliation{Materials Science and Engineering Department, University of Tennessee Knoxville, Knoxville, TN 37996, USA}

\author{Alex Thaler} 
\affiliation{Materials Science and Engineering Department, University of Tennessee Knoxville, Knoxville, TN 37996, USA}

\author{Steven J. Gomez Alvarado} 
\affiliation{Materials Department, University of California, Santa Barbara, California 93106, USA}

\author{Heda Zhang} 
\affiliation{Materials Science and Technology Division, Oak Ridge National Laboratory, Oak Ridge, TN 37831, USA}

\author{David S. Parker}
\affiliation{Materials Science and Technology Division, Oak Ridge National Laboratory, Oak Ridge, TN 37831, USA}

\author{German D. Samolyuk}
\affiliation{Materials Science and Technology Division, Oak Ridge National Laboratory, Oak Ridge, TN 37831, USA}

\author{Joseph A. M. Paddison} 
\affiliation{Neutron Scattering Division, Oak Ridge National Laboratory, Oak Ridge, TN 37831, USA}

\author{Jiaqiang Yan} 
\affiliation{Materials Science and Technology Division, Oak Ridge National Laboratory, Oak Ridge, TN 37831, USA}

\author{Feng Ye} 
\affiliation{Neutron Scattering Division, Oak Ridge National Laboratory, Oak Ridge, TN 37831, USA}

\author{Suchismita Sarker} 
\affiliation{Cornell High Energy Synchrotron Source, Cornell University, Ithaca, New York 14853, USA}

\author{Stephen D. Wilson} 
\affiliation{Materials Department, University of California, Santa Barbara, California 93106, USA}

\author{Hu Miao} 
\affiliation{Materials Science and Technology Division, Oak Ridge National Laboratory, Oak Ridge, TN 37831, USA}

\author{David Mandrus} 
\affiliation{Materials Science and Technology Division, Oak Ridge National Laboratory, Oak Ridge, TN 37831, USA}
\affiliation{Materials Science and Engineering Department, University of Tennessee Knoxville, Knoxville, TN 37996, USA}

\author{Michael A. McGuire} 
\affiliation{Materials Science and Technology Division, Oak Ridge National Laboratory, Oak Ridge, TN 37831, USA}

\date{\today}% It is always \today, today,

\begin{abstract}

The kagome motif is a versatile platform for condensed matter physics, hosting rich interactions between magnetic, electronic, and structural degrees of freedom. 
In recent years, the discovery of a charge density wave (CDW) in the AV$_3$Sb$_5$ superconductors and structurally-derived bond density waves (BDW) in FeGe and ScV$_6$Sn$_6$ have stoked the search for new kagome platforms broadly exhibiting density wave (DW) transitions. 
In this work, we evaluate the known \textit{AM}$_6$\textit{X}$_6$ chemistries and construct a stability diagram that summarizes the structural relationships between the $>$125 member family. 
Subsequently, we introduce our discovery of the broader \textit{Ln}Nb$_6$Sn$_6$ (\textit{Ln}:Ce--Nd,Sm,Gd--Tm,Lu,Y) family of kagome metals and an analogous DW transition in LuNb$_6$Sn$_6$. 
Our X-ray scattering measurements clearly indicate a (1/3, 1/3, 1/3) ordering wave vector ($\sqrt{3}\times\sqrt{3}\times3$ superlattice) and diffuse scattering on half-integer $L$-planes. 
Our analysis of the structural data supports the ``rattling mode'' DW model proposed for ScV$_6$Sn$_6$ and paints a detailed picture of the steric interactions between the rare-earth filler element and the host Nb--Sn kagome scaffolding. 
We also provide a broad survey of the magnetic properties within the HfFe$_6$Ge$_6$-type \textit{Ln}Nb$_6$Sn$_6$ members, revealing a number of complex antiferromagnetic and metamagnetic transitions throughout the family. 
This work integrates our new \nboss~series of compounds into the broader \oss~family, providing new material platforms and forging a new route forward at the frontier of kagome metal research.
\footnote{Notice: This manuscript has been authored by UT-Battelle, LLC, under contract DE-AC05-00OR22725 with the US Department of Energy (DOE). The US government retains and the publisher, by accepting the article for publication, acknowledges that the US government retains a nonexclusive, paid-up, irrevocable, worldwide license to publish or reproduce the published form of this manuscript, or allow others to do so, for US government purposes. DOE will provide public access to these results of federally sponsored research in accordance with the DOE Public Access Plan (https://www.energy.gov/doe-public-access-plan).} 

\end{abstract}
\maketitle
\section{Introduction}
\label{sec:Intro}

Despite the geometrical simplicity of the kagome motif, the diverse array of chemistries and structures available to the solid-state community has produced an explosion of complex and nuanced materials. Kagome insulators were first popularized for the potential to realize a quantum spin liquid by decorating the structurally frustrated lattice with quantum spins.\cite{han2012fractionalized,depenbrock2012nature,Norman2016HerbertsmithiteReview,Shores2005Atacamite,Okamoto2009_Vesignieite,Han2014_Barlowite,Pasco2018_Barlowite+Herbertsmithite} However, the consequences of geometric frustration are not limited to magnetic insulators, and there exists a purely electronic analog to geometric frustration in kagome metals.

The kagome tiling leads to hopping-based interference effects\cite{kiesel2012sublattice,wang2013competing,Beugeling2012_TopologicalPhases2DLattices,Guo2009_TopoInsuOnKagome} that promote strong electronic interactions. Tight binding models of the prototypical kagome motif produce electronic structures with particle-hole asymmetric saddle points, Dirac crossings, and flat-band features.\cite{wang2013competing,kiesel2013unconventional,Wen2010_TopoInsulInKagome,Park2021_InstabilitesOfKagomeMetals} Theoretically, chemical tuning can align the Fermi level with the aforementioned features, increasing the probability of correlated electronic instabilities. For example, electron filling towards the saddle points at filling fractions of $f=5/12$ and $f=3/12$ has been suggested as the impetus for a wide range of correlated effects including density-wave order,\cite{Christensen2021_135CDWTheory,Denner2021_ChargeOrderAV3Sb5,Ferrari2022_CDW-KagomeHubbard} orbital magnetism,\cite{Feng2021_FluxPhaseKagomeAV3Sb5,Feng2021_LowEFluxPhaseInKagome} topological insulator phases,\cite{Wen2010_TopoInsulInKagome} and superconductivity.\cite{Kiesel2013_FermiSurfaceInstabilites-KagomeHubbard}

The discovery of the \textit{A}V$_3$Sb$_5$ (\textit{A}: K, Rb, Cs) family of kagome superconductors, which exhibit the unusual combination of a charge density wave (CDW) and superconducting ground state, exemplified the latent potential of the kagome metals.\cite{ortiz2019new,ortizCsV3Sb5,ortiz2020KV3Sb5,RbV3Sb5SC} However, despite attempts to expand the \textit{A}\textit{M}$_3$\textit{X}$_5$ family, the suite of known compounds has remained limited (\textit{A}V$_3$Sb$_5$ (\textit{A}: K, Rb, Cs)\cite{ortiz2019new}, \textit{A}Ti$_3$Bi$_5$ (\textit{A}: Rb, Cs)\cite{werhahn2022kagome}, CsCr$_3$Sb$_5$\cite{liu2024superconductivity}). As such, in parallel with continued research into the \textit{AM}$_3$\textit{X}$_5$ family, the community continues to search for structural families with more flexibility.

When evaluating based on chemical diversity alone, the CoSn-derived kagome metals (and the \oss~derivatives) are excellent candidate materials, with over 125 compounds currently known. Furthermore, both FeGe (CoSn-prototype)\cite{teng2022discovery} and ScV$_6$Sn$_6$ (HfFe$_6$Ge$_6$-prototype, e.g. filled CoSn)\cite{arachchige2022ScV6Sn6} were initially reported as possessing CDW transitions. Recent research has revealed that these materials are likely driven by complex, structurally-derived modulations more akin to bond-density-waves.\cite{Hu2024_PhononPromotedCDW-ScV6Sn6,Meier2023tiny,Pokharel2023_FrustratedCO+CooperativeDistortionsScV6Sn6,Cao2023_CompetingChargeOrderScV6Sn6,Lee2024_NatureCDWScV6Sn6,Korshunov2023_SofteningPhononScV6Sn6,Hu2023_ScV6Sn6-Theory-FlatPhonons+UnconventionalCDW,Liu2024_DrivingMechanismScV6Sn6,Yu2024_MagAndCorrelationsScV6Sn6,Wang2023_EnhancedSpinPolarizationViaDimerizationFeGe,Wen2024_UnconventionalCDW-FeGe,Chen2024_LongRangeGeDimerizationFeGe}
Nevertheless, the diversity of chemical choice for all three sites in the \oss~structure has continued to attract the interest of both the chemistry and physics communities. 

In this work we present a broad outlook at the \oss~family of kagome metals, evaluating chemical and structural trends to produce a stability diagram of known \oss~materials. We subsequently expand upon the known chemistries by presenting the discovery and single-crystal synthesis of the \textit{Ln}Nb$_6$Sn$_6$ (\textit{Ln}:Ce--Tm,Lu,Y) family. We also present the discovery of a DW-like, bond modulation in the new kagome metal LuNb$_6$Sn$_6$, and further demonstrate a consistent interpretation \textit{via} the ``rattling'' interpretation developed for ScV$_6$Sn$_6$. Our X-ray scattering measurements clearly indicate the presence of a (1/3, 1/3, 1/3) ordering wave vector and diffuse scattering on half-integer $L$-planes. A broad survey of the magnetic properties within the ordered (HfFe$_6$Ge$_6$-type) \textit{Ln}Nb$_6$Sn$_6$ members further reveals a number of complex antiferromagnetic and metamagnetic transitions throughout the family. Our work provides a unique stability map of the \oss~ family, and further integrates the 4-\textit{d}-based \nboss~ kagome metals as a new platform to explore the coupling between structural chemistry, electronic instabilities, and magnetism.

\section{Experimental Methods}
\label{sec:Methods}

\subsection{Single Crystal Synthesis}
\label{sec:Methods_Synthesis}

Single crystal growth of \textit{Ln}Nb$_6$Sn$_6$ (\textit{Ln}:Ce--Tm,Y) single crystals was performed using the self-flux technique. Elemental reagents were combined in a ratio of 8:2:90 \textit{Ln}:Nb:Sn. The exact ratio of \textit{Ln}:Nb:Sn is flexible and growths have succeeded with compositions as rich as 12\% rare-earth and as as poor as 4\%. We utilized Ames Lab rare-earth metals (Ce-Lu, Y), Nb powder (Alfa, 99.9\%), and Sn shot (Alfa, 99.9\%). Reagents were placed into 5~mL Al$_2$O$_3$ (Canfield) crucibles fitted with a catch crucible/porous frit and sealed within fused quartz ampoules with approximately 0.6--0.7~atm of argon cover gas.\cite{canfield2016use} Samples were heated to 1150\degree C and thermalized for 18~h. For growths utilizing Ce--Tm (and Y), samples are cooled to 780\degree C at a rate of 2\degree C/hr before centrifugation at 780\degree C. Growths targeting LuNb$_6$Sn$_6$ are cooled to 900\degree C at a rate of 0.5-1\degree C/hr and subsequently centrifuged at 900\degree C. 

Single crystals are small, well-faceted, gray-metallic, hexagonal plates and blocks. The samples are stable in air, water, and common solvents. Crystals resist attack by concentrated HCl and dilute HNO$_3$, with the exception of LuNb$_6$Sn$_6$. 

\subsection{Scattering and ARPES}
\label{sec:Methods_Scattering+ARPES}

Single crystals were mounted on kapton loops with Paratone oil for single crystal x-ray diffraction (SCXRD).
Diffraction data at 100~K was collected on a Bruker D8 Advance Quest diffractometer with a graphite monochromator. Supplementary diffraction data at 45~K for LuNb$_6$Sn$_6$ were collected with a Rigaku XtaLab PRO equipped with a Rigaku HyPix6000HE detector and an Oxford N-HeliX cryocooler. Both instruments used Mo K$\alpha$ radiation ($\lambda$ = 0.71073~\AA). Data integration, reduction, and structure solution was performed using the Bruker APEX4 software package, Rigaku Oxford Diffraction CrysAlisPro\cite{crysalispro2014agilent}, JANA,\cite{Petricek2023_JANA2020} or Shelx.\cite{sheldrick2008short} Diffuse X-ray scattering measurements were performed at the Cornell High Energy X-ray Synchrotron Source (CHESS), beamline IDB4-QM2 ($\lambda$ = 0.27021\AA). The experiment was conducted in transmission geometry using a 6-megapixel photon-counting pixel-array detector with a silicon sensor layer and a flowing He cryostream for temperature control. Temperature-dependent powder diffraction was performed on a PANalytical Xpert Pro diffactometer ($\lambda$ = 1.5406~\AA) equipped with a Oxford PheniX closed-cycle helium cryostat. Diffuse data was analyzed and visualized using the NeXpy/NXRefine software package.\cite{NexPy,GomezNexPy} The Topas V6 software package\cite{coelho2018topas} was used to analyze the polycrystalline data. 

The ARPES experiments were performed at beamline 21-ID-1 of NSLS-II at BNL. The LuNb$_6$Sn$_6$ samples were cleaved \textit{in-situ} under vacuum ($<$3$\times$10$^{-11}$~Torr). Measurements were taken using an incident energy of 79~eV and a energy resolution of 15~meV. 

\subsection{Bulk Characterization}
\label{sec:Methods_BulkCharacterization}

Magnetization measurements (300--1.8~K) on crystals of \textit{Ln}Nb$_6$Sn$_6$ (\textit{Ln}:Ce--Lu,Y) were performed in a 7~T Quantum Design Magnetic Property Measurement System (MPMS3) SQUID magnetometer in vibrating-sample magnetometry (VSM) mode. Additional measurements below 1.8~K utilized the Quantum Design iHe-3 insert for the MPMS3 (1.8--0.4~K). For consistency, the same sample was utilized for both measurements wherever possible. The data sets were matched at 1.8~K, with deference given to the MPMS3 (lower background) dataset. Small errors (5--10\% of the absolute magnetization) can often be observed when transitioning to the $^3$He regime, predominately attributed to the difficulty of aligning the small crystals in the $^3$He setup. Both field-cooled (FC) and zero-field-cooled (ZFC) measurements were performed, though ZFC curves are only shown where hysteresis is noted. Measurements were made for both \textbf{\textit{H}}\,$\parallel$\,\textbf{\textit{c}} and \textbf{\textit{H}}\,$\perp $\,\textbf{\textit{c}}. When possible, orientations with \textbf{\textit{H}}\,$\perp$\,\textbf{\textit{c}} are also oriented such that \textbf{\textit{H}}\,$\parallel$\,\textbf{\textit{a}}. Temperature-dependent measurements are typically performed at an applied field of 500~Oe, except for the non-magnetic YNb$_6$Sn$_6$ and LuNb$_6$Sn$_6$ ($H=$~10~kOe). 

Heat capacity measurements (300--1.8~K) on crystals of \textit{Ln}Nb$_6$Sn$_6$ (\textit{Ln}:Gd--Lu,Y) were performed in a Quantum Design 9~T Dynacool Physical Property Measurement System (PPMS), and a Quantum Design 14~T PPMS equipped with a $^3$He (9--0.4~K) insert. Additional measurements were performed for \textit{Ln}Nb$_6$Sn$_6$ (\textit{Ln}:Ho--Lu,Y) utilizing a Quantum Design dilution refrigerator insert (4--0.1~K) for the 9~T Dynacool PPMS. The same samples were used for both measurements wherever possible. Similar to magnetization measurements, curves were matched in the crossover regime around 2~K. A systematic thermometry offset of approximately 0.2~K is observed in the $^3$He data.

Resistivity measurements (300--1.8~K) were performed using a Quantum Design 9~T Dynacool Physical Property Measurement System (PPMS). Resistivity bars were constructed from single crystals of \textit{Ln}Nb$_6$Sn$_6$ (\textit{Ln}:Gd--Lu,Y) \textit{via} polishing. Naturally faceted crystals were mounted on a Struers AccuStop sample holder using Crystalbond 509 and polished into rectangular prisms with approximate dimensions of $1\times0.3\times0.1$~mm. Crystalbond was subsequently removed using acetone. Where possible, samples were polished such that $I\,\parallel$\,\textit{\textbf{a}}. Electrical contact was achieved using silver paint (DuPont cp4929N-100) and platinum wire (Alfa, 0.05~mm Premion 99.995\%) in a four-wire configuration.

\subsection{Electronic Structure Calculations}
\label{sec:Methods_ElectronicStructureCalcs}

First-principles calculations were performed within the density functional theory\cite{kohn1965self} approximation using the linearized augmented plane-wave (LAPW) method \cite{andersen1975linear,singh2006planewaves,sjostedt2000alternative} as implemented in the WIEN2K code.\cite{blaha2020wien2k} The LAPW “muffin-tin” spheres of radii 2.5~Bohr for all three components, with $RK_\text{max}$ = 9.0 were used. The experimental lattice parameters were used for all calculations, and atomic positions were relaxed until atomic forces fell below 1~mRy/Bohr. The exchange-correlation energy was calculated within generalized gradient approximation \cite{kohn1996density} with the parametrization by Perdew, Burke, and Ernzerhof (PBE).\cite{perdew1996generalized} Spin-orbit coupling was included using the second-variation approach.\cite{koelling1977technique} Brillouin zone (BZ) summation in electronic system self-consistency procedure was carried out over 13$\times$13$\times$6 $k$-points mesh, while Fermi surface was built using 22$\times$22$\times$11 mesh. To build the Fermi surface the FermiSurfer \cite{kawamura2019fermisurfer}  and XCrySDen \cite{kokalj1999xcrysden} graphical packages were used.

\section{Results and Discussion}
\label{sec:ResultsDiscussion}
\subsection{\textit{AM}$_6$\textit{X}$_6$ Structural Trends}
\label{sec:ResulthsDiscussion_StructureTrends}

\begin{figure*}
\includegraphics[width=\linewidth]{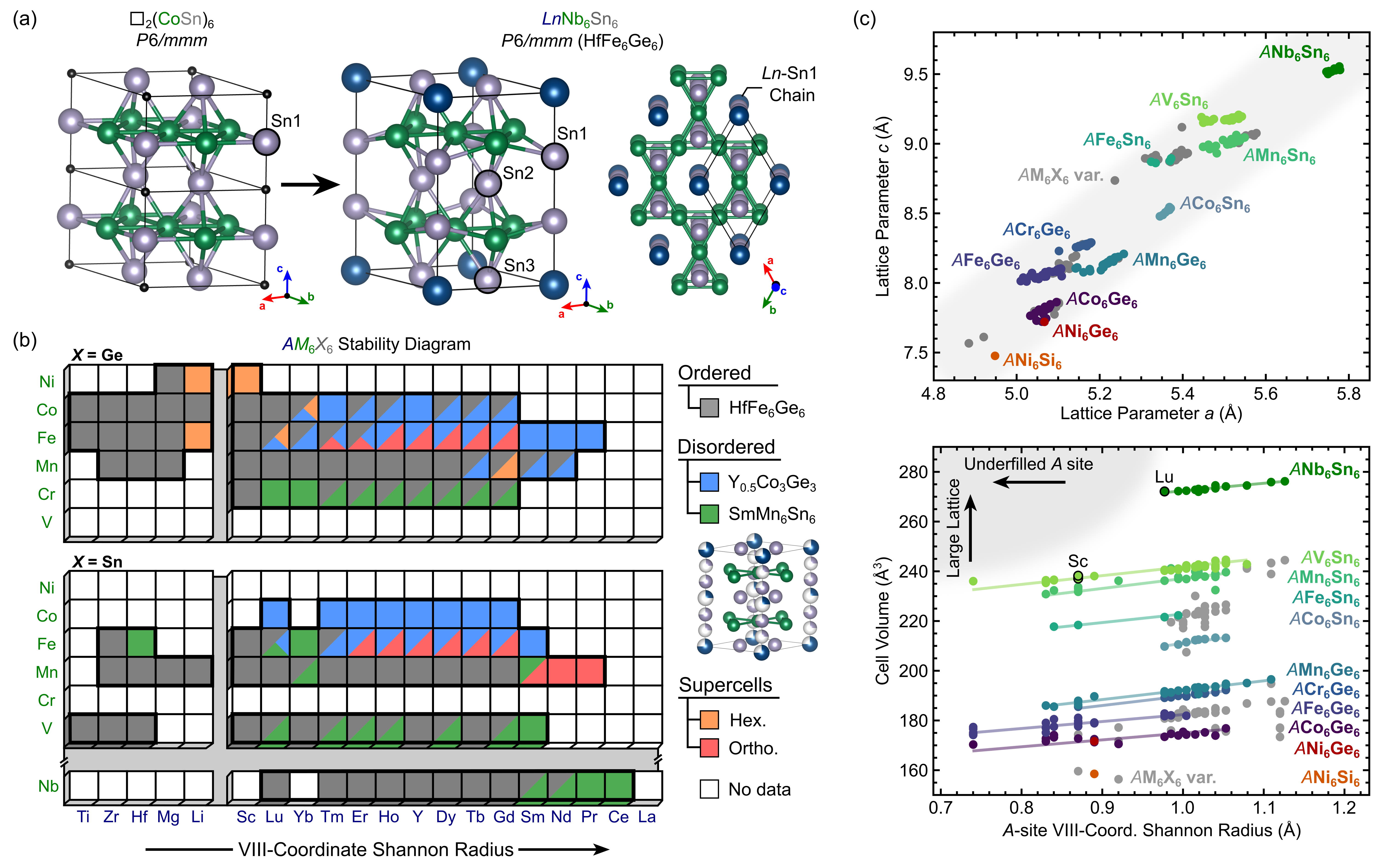}
\caption{(a) The CoSn family is one of the simplest kagome prototypes. It's unusual structure contains large interstitial voids (black spheres) that can be filled with electropositive cations to form the chemically diverse \oss~materials. (b) Here we show a structural stability diagram of all currently known \oss~materials.\cite{gschneidner2005handbook,olenich1981crystal,mazet1999neutron,venturini1992crystallographic,skolozdra1991magnetic,buchholz1981intermetallische,mazet2008first,romaka2024structure,brabers1994magnetic,yang2024crystal,romaka2022interaction,yang2024crystal,schobinger1997ferrimagnetism,konyk2019er,schobinger1997atomic,konyk2021phase,ihou2006magnetic,mazet2001macroscopic,chabot1983lufe6ge6,raghavan2002fe,oleksyn1997crystal,bodak1992er,el1991crystal,cadogan2009study,eproprcr6ce6,koretskaya1986new,mruz1984new,el1994magnetic,konyk2021phase,hu2019measurement,kun2023phase,buchholz1978verbindungen,szytula2004crystal,weiland2020refine,dzyanyj1995crystal,he2024quantum,arachchige2022ScV6Sn6,romaka2011peculiarities,romaka2019lu,Meier2023tiny,guo2023triangular,zhang2022electronic,huang2023anisotropic,zeng2024magnetic,pokharel2022highly,rosenberg2022uniaxial,pokharel2021electronic,lee2022anisotropic,mazet1999study,el1991refinement,mazet1999study,xia2006ybmn6sn6,weitzer1993structural,clatterbuck1999magnetic,venturini1993magnetic,mazet2006magnetic,mazet2000local,el1991crystal,mazet2002study,mazet2000evidence,schobinger1998fe,wang1994structural,raghavan2001fe,stȩpien2000ternary,skolozdra1984crystal,Skolozdra1997StannidesRE+TM,skolozdra2000ternary,zhuang2008phase,oshchapovsky2010tbnb6sn6,yue2012syntheses,savidan2010experimental,malaman1997magnetic}  The \hfgesn~(gray) is the most commonly studied \oss, though there exist disordered (blue, green) compounds and various supercells/stacking variants (orange,red). The subject of this work, the \nboss, stand apart as the only 4$d$-based kagome metals. (c,top) Of the known \oss~series, \nboss~possesses the largest unit cell and heaviest lattice. By extension, the rare-earth site has the most room to rattle and larger atoms can be accommodated. (c,bottom) Previous work predicts DW instabilities when the combination of a large host lattice and small filler atom is realized -- both ScV$_6$Sn$_6$ and the new LuNb$_6$Sn$_6$ display DW-like instabilities and are marked.}
\label{fig:1}
\end{figure*}

Of the binary kagome compounds, the CoSn prototype is one of the most fundamental families. 
The simple unit cell and isolated kagome planes leads to an electronic structure that clearly exhibits the hallmark flat-bands, Dirac points, and saddle points predicted by theory.\cite{jovanovic2022simple} 
Unfortunately, the family is limited in scope, and only CoSn, FeGe, NiIn, FeSn, PtTl, and RhPb are known.\cite{Meier2020_FlatBandsCoSnType}
It has been proposed that the limited stability of the CoSn structure may relate to the low density structure, which is filled with large interstitial voids.\cite{venturini2006filling,simak1997stability} 
Figure~\ref{fig:1}(a) shows two stacked CoSn unit cells, highlighting the interstitial voids with small black spheres. 
This unusual feature of CoSn allows integration of ``filler'' atoms, producing \oss-type structures. The most frequently encountered \oss~structure is the \hfgesn~prototype, formed when the (0,0,0) position in the $1\times1\times2$ CoSn supercell is filled with electropositive cations. As one may expect, the connection between the filler size and host lattice sterics is complex, producing vast structural and stacking diversity in the \oss~family. Over 100 unique compounds are known.\cite{gschneidner2005handbook,olenich1981crystal,mazet1999neutron,venturini1992crystallographic,skolozdra1991magnetic,buchholz1981intermetallische,mazet2008first,romaka2024structure,brabers1994magnetic,yang2024crystal,romaka2022interaction,yang2024crystal,schobinger1997ferrimagnetism,konyk2019er,schobinger1997atomic,konyk2021phase,ihou2006magnetic,mazet2001macroscopic,chabot1983lufe6ge6,raghavan2002fe,oleksyn1997crystal,bodak1992er,el1991crystal,cadogan2009study,eproprcr6ce6,koretskaya1986new,mruz1984new,el1994magnetic,konyk2021phase,hu2019measurement,kun2023phase,buchholz1978verbindungen,szytula2004crystal,weiland2020refine,dzyanyj1995crystal,he2024quantum,arachchige2022ScV6Sn6,romaka2011peculiarities,romaka2019lu,Meier2023tiny,guo2023triangular,zhang2022electronic,huang2023anisotropic,zeng2024magnetic,pokharel2022highly,rosenberg2022uniaxial,pokharel2021electronic,lee2022anisotropic,mazet1999study,el1991refinement,mazet1999study2,xia2006ybmn6sn6,weitzer1993structural,clatterbuck1999magnetic,venturini1993magnetic,mazet2006magnetic,mazet2000local,el1991crystal,mazet2002study,mazet2000evidence,schobinger1998fe,wang1994structural,raghavan2001fe,stȩpien2000ternary,skolozdra1984crystal,Skolozdra1997StannidesRE+TM,skolozdra2000ternary,zhuang2008phase,oshchapovsky2010tbnb6sn6,yue2012syntheses,savidan2010experimental} with a plethora of stacking variations, intergrowths, and incommensurate structures.\cite{venturini2006filling,fredrickson2008origins}

The most recognizable \oss~prototype is the \hfgesn~structure. Fig.~\ref{fig:1}(a) shows the relationship between the filled CoSn supercell and the \hfgesn~prototype, represented here by our \textit{Ln}Nb$_6$Sn$_6$ family. Note that the volumetric expansion from the filler atoms pushes the in-plane (Sn1) atoms out of the (0,0,1/4) and (0,0,3/4) kagome plane. 
Where the original Sn1--Sn1 distance in CoSn was nearly 4.3\,\AA, the new Sn1--Sn1 bond is $\approx$3.2\,\AA. 
This is comparable with other bonding distances in the \oss~cell ($d_\text{Nb-Nb}\approx$2.9\,\AA, $d_\text{Nb-Sn}\approx$3.0\,\AA). 

Furthermore, the length of the Sn1--Sn1 bond is highly sensitive to the choice of the filler atom. Between the largest (e.g. NdNb$_6$Sn$_6$) and smallest (e.g. LuNb$_6$Sn$_6$) members of the family, the Sn1--Sn1 bond length compresses nearly 4\% from 3.29\,\AA~ (Lu) to less than 3.16\,\AA~with large atoms (Nd). Compression of the Sn1--Sn1 bond appears to be the structure's preferred means to relieve strain from the incorporation of large atoms. For comparison, from NdNb$_6$Sn$_6$ to LuNb$_6$Sn$_6$, the $c$-axis lattice parameter only expands 0.6\%, the volume expands $<$2\%, and the $z$-position of the kagome lattice shifts only 0.4\%. Thus, it appears that the Nb-(Sn2,3) sublattices act as a largely rigid scaffolding interlaced with flexible \textit{Ln}1--Sn1--Sn1--\textit{Ln}1 chains.

These flexible, less constrained $Ln$1--Sn1--Sn1--$Ln1$ chains are key to our prior work detailing a ``rattling'' bond modulation in ScV$_6$Sn$_6$,\cite{Meier2023tiny} previously identified as a potential CDW instability.\cite{arachchige2022ScV6Sn6} 
The ``rattling'' is more constrained than the analogous terminology applied in the Skutterudites and filled clathrates, but draws from similar steric principles.\cite{Goto2004_Tunneling+RattlingInClathrate,Ciesielski2023_RattlingThermalConductivityClathrate,Dong2000_ChemTrendsRattlingClathrates,Sales1999_DisplacementsThermalConductivityClathrateLike}  
Thus, we have an interesting confluence of chemical pressures inside the \oss~compounds. 
The filler atom stabilizes compositions by filling anomalously large interstitial voids, but it clearly affects the bond distances and crystal structure in a highly non-isotropic manner. 
It naturally follows that some filler atoms will be too large, while others will be too small. 
The boundaries between the structural stability and internal strain is the impetus for our discussion -- a \oss~stability field.

Figure \ref{fig:1}(b) presents a stability diagram of all known \oss~phases where \textit{X} is Ge or Sn. 
Given the limited number of silicon-based compounds (MgNi$_6$Si$_6$, ScNi$_6$Si$_6$, LiNi$_6$Si$_6$)\cite{buchholz1981intermetallische,morozkin2016ni}, we have omitted the corresponding silicide diagram. 
The \textit{M}-site has been organized according to atomic number, and the \textit{A}-site has been organized based on the VIII-coordinate Shannon radius. 
This diagram is an agglomeration of all publicly available structural data on the \oss~compounds.\cite{gschneidner2005handbook,olenich1981crystal,mazet1999neutron,venturini1992crystallographic,skolozdra1991magnetic,buchholz1981intermetallische,mazet2008first,romaka2024structure,brabers1994magnetic,yang2024crystal,romaka2022interaction,yang2024crystal,schobinger1997ferrimagnetism,konyk2019er,schobinger1997atomic,konyk2021phase,ihou2006magnetic,mazet2001macroscopic,chabot1983lufe6ge6,raghavan2002fe,oleksyn1997crystal,bodak1992er,el1991crystal,cadogan2009study,eproprcr6ce6,koretskaya1986new,mruz1984new,el1994magnetic,konyk2021phase,hu2019measurement,kun2023phase,buchholz1978verbindungen,szytula2004crystal,weiland2020refine,dzyanyj1995crystal,he2024quantum,arachchige2022ScV6Sn6,romaka2011peculiarities,romaka2019lu,Meier2023tiny,guo2023triangular,zhang2022electronic,huang2023anisotropic,zeng2024magnetic,pokharel2022highly,rosenberg2022uniaxial,pokharel2021electronic,lee2022anisotropic,mazet1999study,el1991refinement,mazet1999study,xia2006ybmn6sn6,weitzer1993structural,clatterbuck1999magnetic,venturini1993magnetic,mazet2006magnetic,mazet2000local,el1991crystal,mazet2002study,mazet2000evidence,schobinger1998fe,wang1994structural,raghavan2001fe,stȩpien2000ternary,skolozdra1984crystal,Skolozdra1997StannidesRE+TM,skolozdra2000ternary,zhuang2008phase,oshchapovsky2010tbnb6sn6,yue2012syntheses,savidan2010experimental,malaman1997magnetic} 
In our diagram, there are five major categories of structures: 1) HfFe$_6$Ge$_6$ (gray), 2) disordered SmMn$_6$Sn$_6$ (green), 3) disordered Y$_{0.5}$Co$_3$Ge$_3$ (blue), 4) ``other'' hexagonal cells (orange), and 5) ``other'' orthorhombic cells (red). 
We have preserved the classification given by the author of each work.

The HfFe$_6$Ge$_6$ cell is what is typically regarded as the ``pristine,'' ordered, \oss-type compound. However, we can clearly see that many compounds exhibit varying degrees of disorder. 
Both disordered SmMn$_6$Sn$_6$ and Y$_{0.5}$Co$_3$Ge$_3$ are characterized by partial suboccupancy of the \textit{A}-site and partial occupancy on the (normally vacant) (0,0,0.5) site. 
In the case of Y$_{0.5}$Co$_3$Ge$_3$, the  (0,0,0.5) and (0,0,0) site are both nearly half occupied, causing the unit cell to reduce to a smaller, single kagome-layer compound. 
We suspect a similar phenomenon occurs in our PrNb$_6$Sn$_6$ and CeNb$_6$Sn$_6$, which exhibit clear signatures of disorder in diffraction and subtle changes in crystal habit. 
For the purposes of this work we classify them as non-\hfgesn~structures, though we hope the ordered structure can be stabilized with more refined synthesis methods.

We note that the Y$_{0.5}$Co$_3$Ge$_3$ and SmMn$_6$Sn$_6$-type structures seem to appear on opposite sides of the diagram. 
Empirically, Y$_{0.5}$Co$_3$Ge$_3$ seems to favor the smaller Co- and Fe- host lattices, whereas SmMn$_6$Sn$_6$-type occurs more frequently in larger V- and Cr-based lattices. 
The other observation of note is the striking and complex variations in stacking seen in the iron-containing compounds, with HfFe$_6$Ge$_6$, ErFe$_6$Sn$_6$, HoFe$_6$Sn$_6$, YFe$_6$Sn$_6$, DyFe$_6$Sn$_6$, and TbFe$_6$Sn$_6$ all representing unique commensurate superstructures arising from variations in the stacking/filling of the host lattice.\cite{el1991crystal, fredrickson2008origins} 
We have only examined those compounds with \oss-type stoichiometry, though many other structures and stackings (e.g. \textit{AM}$_3$X$_4$ kagome metals)\cite{ortiz2023ybv,ortiz2023evolution,ortiz2024intricate,ovchinnikov2018synthesis,ovchinnikov2019bismuth,motoyama2018magnetic,chen2023134,guo2024tunable} can be derived from the CoSn and \hfgesn~prototypes.

On our diagram, the \nboss~family stands apart as the only 4$d$-element series of \oss~compounds. Hints of the family's existence were noted with both YNb$_6$Sn$_6$ and TbNb$_6$Sn$_6$ reported as side products in exploratory reactions, and served as a launching point for our synthetic endeavors.\cite{yue2012syntheses,oshchapovsky2010tbnb6sn6} The \nboss~system appears to tolerate larger (e.g. Sm, Nd) and reject smaller (e.g. Sc) \textit{A}-site filler atoms, with the notable absence of YbNb$_6$Sn$_6$. The inclusion of Nb will produce the largest \textit{M}$_6$\textit{X}$_6$ scaffolding and the largest interstitial voids. To assist in visualiation, Figure~\ref{fig:1}(c,top) presents a plot of the \textit{a} and \textit{c} lattice parameters for all \oss~compounds. Notable series of \hfgesn-type \oss~have been colored to highlight chemical families, with the various structural distortions and disordered compounds left gray. The exception is \textit{A}Co$_6$Sn$_6$, which is highlighted even though no \hfgesn~representatives exist for the stannides. The general trends are visually striking, with the \nboss~family sitting  isolated from the rest of the data.

Meier et al.\cite{Meier2023tiny} previously proposed that the structural instability in ScV$_6$Sn$_6$ arises from underfilling of the interstitial voids.\cite{Meier2023tiny} Figure~\ref{fig:1}(c,bottom) provides a graphical interpretation of this hypothesis by plotting the \textit{A}-site Shannon radius (filler size) and the unit cell volume (proxy for host lattice voids). Based on the model, we expect the ``rattling'' DW instability to be favored for the combination of a small \textit{A}-atom and a large host lattice (shaded gray region in upper left). We have marked the two known DW compounds, ScV$_6$Sn$_6$ and the our newly discovered LuNb$_6$Sn$_6$, with black-bordered points. Coincidentally, while writing this manuscript, Feng. et al. computationally cataloged the instabilities in the broader \oss~compounds, suggesting that the ``hypothetical'' Nb-based \cite{Feng2024_MT6Z6-PhononInstabilities} compounds would be unstable to Sn1-Sn1 bond modulation mode.

Both LuNb$_6$Sn$_6$ and ScV$_6$Sn$_6$ are among the smallest \textit{A}-site end members of the V- and Nb-based \oss~series, respectively, which agrees with expectations. One point of curiosity on Figure \ref{fig:1}(c, bottom), however, would be the three V-based compounds with even \textit{smaller} Shannon radii than ScV$_6$Sn$_6$. These compounds correspond to the recently reported TiV$_6$Sn$_6$, ZrV$_6$Sn$_6$ and HfV$_6$Sn$_6$.\cite{he2024quantum} TiV$_6$Sn$_6$ likely exhibits some Ti$_\text{V}$ disorder, but the absence of a structural (DW-like) instability in the Hf- and Zr-based compounds is curious. However, unlike the rare-earth compounds, it is unlikely that Hf and Zr can be treated within the ionic (Zr$^{4+}$, Hf$^{4+}$, 8-coordinate) limit. Further, recent computational assessments have indicated that Ti, Zr, and Hf-based compounds are not expected to show bond modulations, suggesting the importance of appropriate band-filling and $d$-orbital states.\cite{Feng2024_MT6Z6-PhononInstabilities} Fermi level alignment with the prototypical kagome bands is persistent theme in kagome metal research -- and thus we turn to examine the characteristic electronic structure of the \nboss~family.

\subsection{Electronic structure of \textit{Ln}Nb$_6$Sn$_6$}
\label{sec:ResultsDiscussion_ElectronicStructure}

\begin{figure}
\includegraphics[width=3.33in]{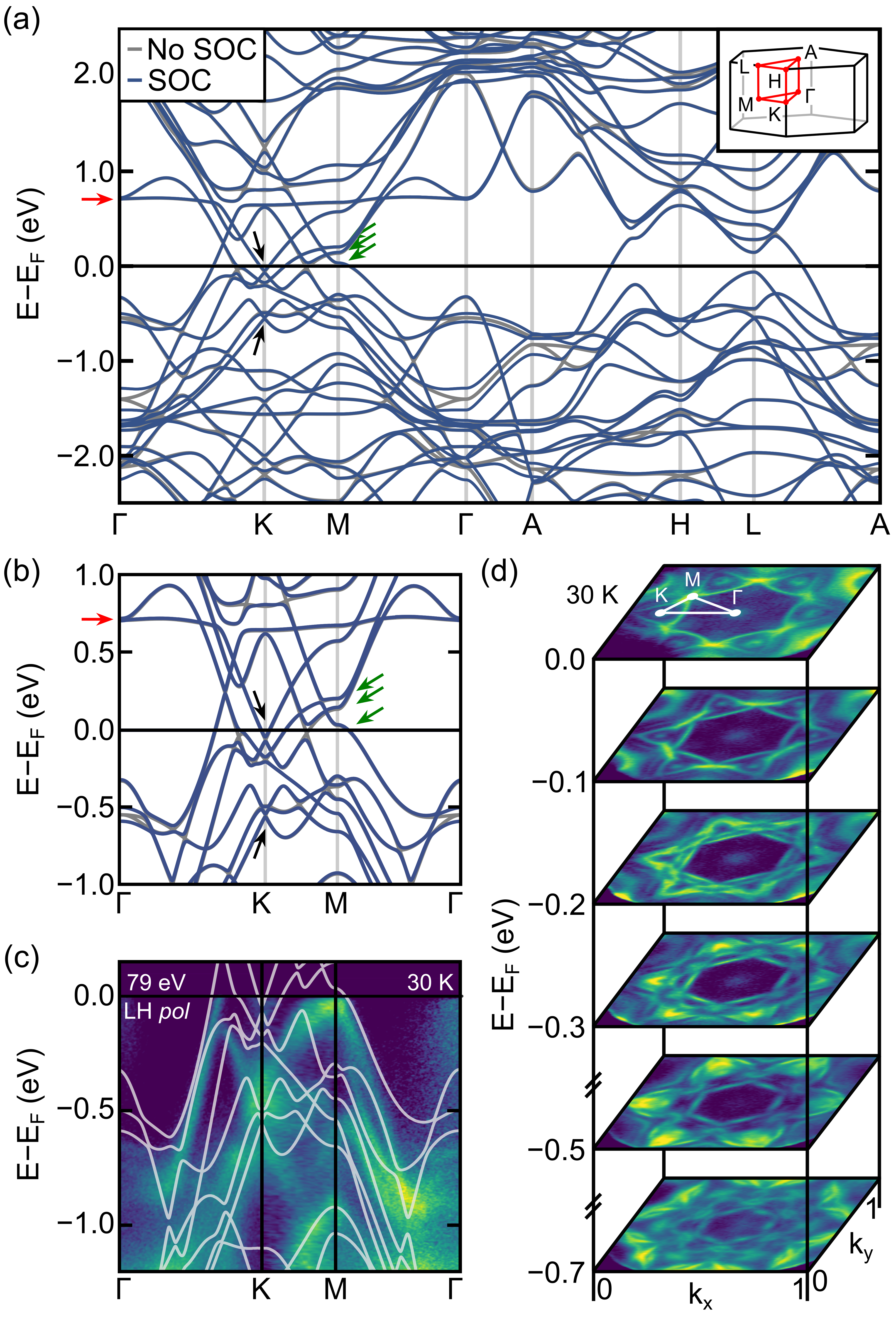}
\caption{Here we investigate the electronic structure of the nonmagnetic LuNb$_6$Sn$_6$ as a proxy for the \nboss~family of kagome metals. (a,b) The electronic structure of the normal state LuNb$_6$Sn$_6$ structure exhibits many of the classic characteristics of the prototypical kagome band structure including Dirac points (black), flat bands (red), and VHS (green). Inclusion of spin-orbit coupling (SOC) gaps many of the Dirac points, however, particularly near $E_\text{F}$. (c) experimental ARPES $\Gamma$-$K$-$M$-$\Gamma$ data overlaid with DFT (+SOC). The $\Gamma$-$K$-$M$-$\Gamma$ and ARPES Fermi surface cuts (d) both possess a mixture of bulk bands and surface states, consistent with observations in ScV$_6$Sn$_6$.}
\label{fig:dft}
\end{figure}

The tight-binding model of the kagome motif provides an underlying motivation for research into kagome metals,\cite{kiesel2012sublattice,wang2013competing,Beugeling2012_TopologicalPhases2DLattices,Guo2009_TopoInsuOnKagome} though the actual manifestation of ``kagome'' bands in real materials is often substantially more complex. Figure~\ref{fig:dft} presents the electronic structure of non-magnetic LuNb$_6$Sn$_6$, which serves as a convenient proxy for the electronic structure of the wider \nboss~family. Here we are intentionally neglecting the influence of magnetism for the other rare-earth compounds in an effort to simplify the discussion and provide a general perspective on the electronic structure in the Nb-Sn compounds. We anticipate that more in-depth computational works will follow, particularly with regards to some of the more complex magnetic members.

Figure~\ref{fig:dft} demonstrates the density functional theory (DFT) calculation for LuNb$_6$Sn$_6$ within $\sim$2~eV of the Fermi level. We have shown data with (blue) and without (gray) spin-orbit coupling. Figure~\ref{fig:dft}(b) provides a magnified view of the electronic structure within 1\,eV of the Fermi level for the $\Gamma$-$K$-$M$-$\Gamma$ path. From both figures, there are several key features that are easily identified. There is a flat-band (red arrow) that extends throughout much of the Brilluoin zone at approximately +0.6\,eV. While the distant flat band is unlikely to have an appreciable effect on the physical properties, both a Van Hove singularity (VHS) at $M$ and a Dirac-like feature at $K$ are within 0.1\,eV of the calculated Fermi level. Several other VHS's (green arrows) and Dirac-like features (black arrows) have been highlighted on Figure \ref{fig:dft}(a,b). We note that spin-orbit coupling (relevant with heavier Nb) gaps many of the Dirac cones, including the one near the Fermi level at $K$.

To verify the electronic structure near the Fermi level, and to check the alignment of the computationally derived Fermi level, we performed a series of ARPES measurements ($T=30$~K) on cleaved single crystals of LuNb$_6$Sn$_6$. Figure~\ref{fig:dft}(c) is a reconstructed image of the ARPES intensity along the $\Gamma$-$K$-$M$-$\Gamma$ high symmetry path. While the ARPES measurements were performed at $k_z = \pi$, we observe minimal changes in $k_z$-dependent scans and are comfortable utilizing the data for a qualitative comparison with DFT. A overlay of the DFT band structure has been provided as a faint white trace. We see generally good agreement between the DFT and theory, particularly with regards to the high DOS near the Van Hove singularity at $K$, and the faint signal from the Dirac-like feature at $M$. Figure~\ref{fig:dft}(d) highlights a series of Fermi surface contours at different isoenergy cuts from which the band structure in Fig.~\ref{fig:dft}(c) is derived. 

Prior knowledge from ScV$_6$Sn$_6$ helps to identify substantial signal from surface-states in both the reconstructed band dispersions and the Fermi surface maps. The potential for topologically non-trivial surface states is an intriguing prospect. Within ScV$_6$Sn$_6$, there have been many recent ARPES, STM, and computational efforts directed at investigating surface states.\cite{Cheng2024_ScV6Sn6-ARPES-BulkSurface,Hu2024_PhononPromotedCDW-ScV6Sn6} Given their chemical similarity, we suspect LuNb$_6$Sn$_6$ to exhibit equally complex features. Though our ARPES results are below the DW transition temperature (68~K), we did not observe strong changes in the ARPES intensities above and below the transition. The ScV$_6$Sn$_6$ system also shows relatively subtle changes through the DW transition. However, more detailed studies have shown a rich landscape of subtle changes,\cite{Hu2023_ScV6Sn6-Theory-FlatPhonons+UnconventionalCDW,Tan2023_AbundantInstabilitesScV6Sn6,Tuniz2023_CDW-DynamicsScV6Sn6,Cheng2024_STM+ARPES-ScV6Sn6,Kang2023_LifshitzScV6Sn6} and while the evolution of the surface states and DW in ARPES is intriguing, these efforts are beyond the scope of this manuscript. A more in-depth investigation into the ARPES and scattering results are anticipated in our follow-up work. 

\subsection{Density Wave Instability in LuNb$_6$Sn$_6$}
\label{sec:ResultsDiscussion_CDWInstability}

\begin{figure*}
\includegraphics[width=\linewidth]{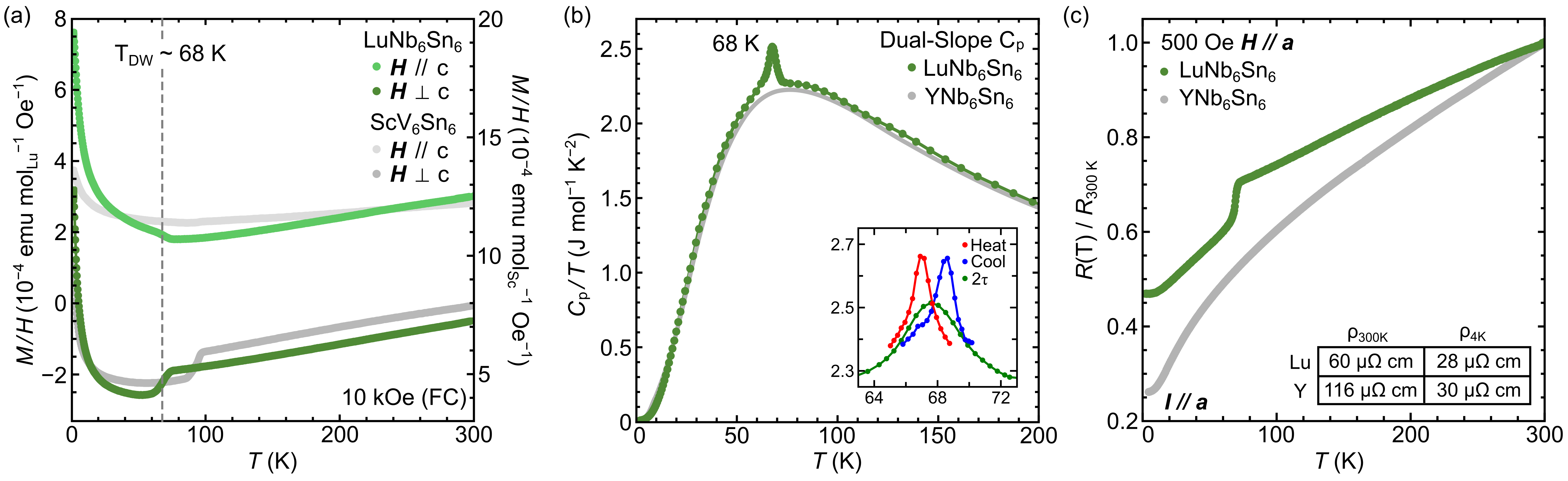}
\caption{Bulk characterization on \luoss~single crystals highlight the emergence of a density-wave transition at 68\,K. (a) Magnetic susceptibility measurements are consistent with nonmagnetic Lu and Nb sublattices and indicate a sharp drop in the susceptibility at the DW transition. Both \textbf{\textit{H}}\,$\parallel$\,\textbf{\textit{c}} and \textbf{\textit{H}}\,$\perp$\,\textbf{\textit{c}} orientations are shown(green), with reference traces for ScV$_6$Sn$_6$ provided as well (gray). (b) Heat capacity measurements reveal a first-order transition with clear splitting between heating and cooling curves. Both the standard 2$\tau$ dual-slope (green) and large pulse single-slope (red/blue) measurements are shown. (c) Electronic transport in single crystals of \luoss~ with $I\,\parallel$\,\textbf{\textit{a}} and \textbf{\textit{H}}\,$\parallel$\,\textbf{\textit{c}} exhibit a strong drop in resistivity at the anomaly. Broadly, we note that crystals of \luoss~show low RRR (2-3) and relatively flat temperature dependence. A trace of analogous transport in YNb$_6$Sn$_6$ is shown for comparison.}
\label{fig:cdw}
\end{figure*}

The DW instability in the structurally analogous ScV$_6$Sn$_6$\cite{arachchige2022ScV6Sn6} manifests with thermodynamic signatures remarkably similar to those observed in the CDW AV$_3$Sb$_5$ systems.\cite{ortizCsV3Sb5,ortiz2019new,ortiz2020KV3Sb5,RbV3Sb5SC} This led to initial claims of an analogous CDW state in ScV$_6$Sn$_6$. Combined with the similarities in their electronic structures and the near alignment with the VHS ($M$-point) and Dirac point ($K$-point), this was a reasonable hypothesis. However, we now understand the transition in ScV$_6$Sn$_6$ as more analogous to a bond-modulated DW, and can use this knowledge to help interpret the transition in \luoss. 

Figure~\ref{fig:cdw}(a) shows the magnetic susceptibility for single crystals of \luoss~(green) measured using a 10\,kOe magnetic field. A sample of ScV$_6$Sn$_6$ was measured as well (gray) and is included as a reference. A sharp decrease in the susceptibility occurs around 68\,K, coinciding with the onset of the DW transition. There is a relatively strong shift in the magnetization between the \textbf{\textit{H}}\,$\parallel$\,\textbf{\textit{c}} and \textbf{\textit{H}}\,$\perp$\,\textbf{\textit{c}} orientations, which is likely tied to anisotropy of Landau diamagnetism. A similar, albeit weaker effect can be seen in ScV$_6$Sn$_6$. 

The step in magnetic susceptibility directly corresponds an anomaly in the specific heat (Fig.~\ref{fig:cdw}(b)). We have identified that the transition is first-order, with strong differences between data analyzed using the standard $2\tau$ (dual-slope, 2\% rise) and data analyzed with the large pulse (single-slope, 20--30\% rise) methodology. The inset shows the clear splitting between heating and cooling curves. The 2$\tau$ heat capacity for YNb$_6$Sn$_6$, corrected for a small difference in the molar mass of YNb$_6$Sn$_6$ and \luoss~is shown for comparison.

\begin{figure*}
\includegraphics[width=1\textwidth]{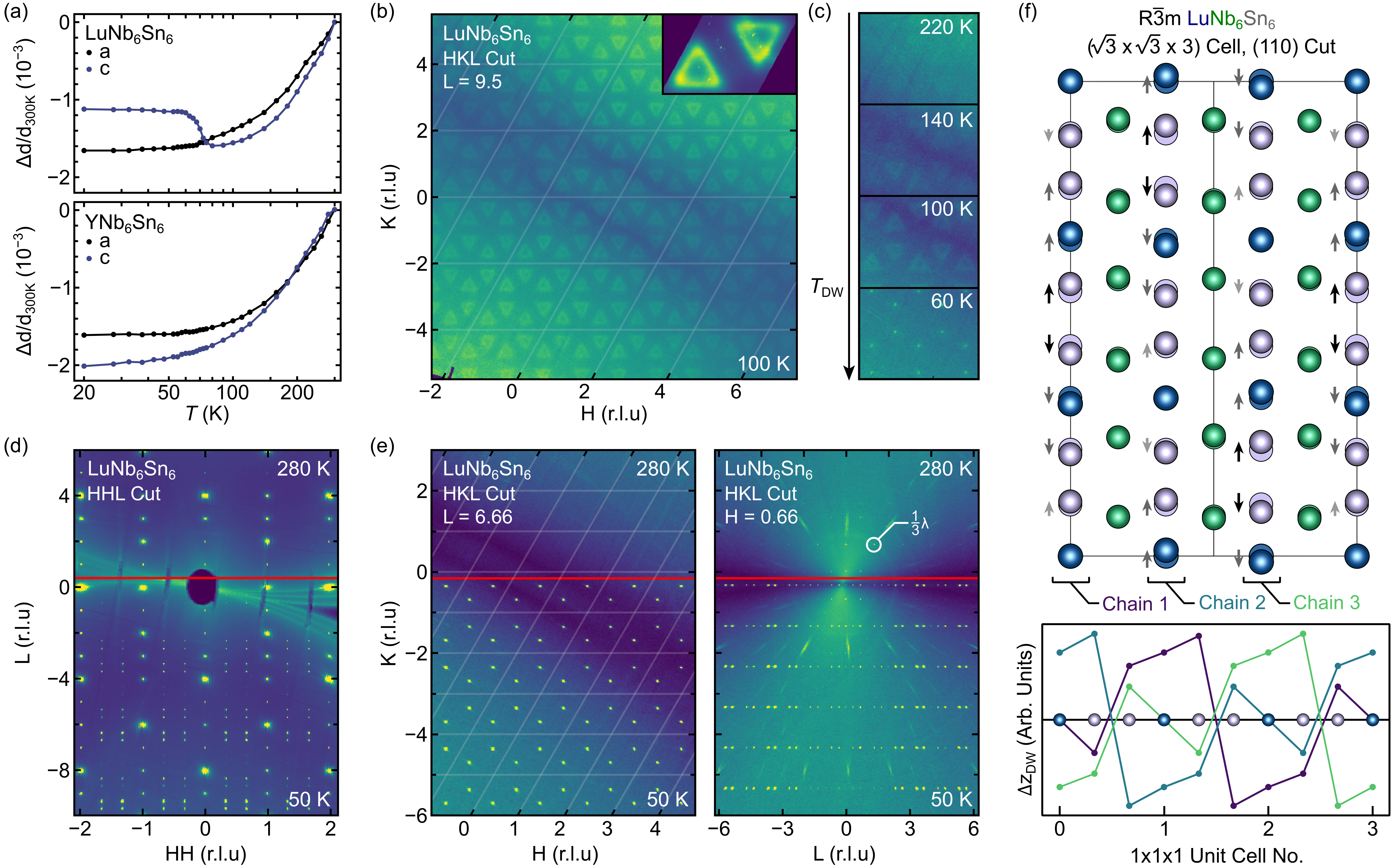}
\caption{(a) Temperature-dependent powder x-ray diffraction data highlights the effect on \textit{c} and \textit{a} through the DW transition. Note that all data has been indexed using the original $1\times1\times1$ unit cell. Remarkably, the \textit{a}-axis contraction is largely unaffected and resembles that of YNb$_6$Sn$_6$. (b) Upon cooling, we observe clear diffuse scattering on half integer $L$ planes. The diffuse scattering is striking, revealing a pattern of  ``hollow triangles'' with slightly more intensity at the corners. (c) While the diffuse scattering can be observed above 200~K, it is strongest near the phase transition. Below 68~K, the diffuse intensity coalesces into the long-range ordered superlattice. The superlattice is clearly visible on reconstructions of the HHL plane (d) and on cuts through the raw data (e). The superlattice is consistent with a (1/3,1/3,1/3) wavevector. Structural refinement of the \luoss~structure below 68~K produces a $\sqrt{3}\times\sqrt{3}\times3$ supercell with staggered displacements along the three unique chains. (f) Shows a (110) slice of the \luoss~superlattice with atomic displacements exaggerated and marked. A alternate visualization (below) is shown to highlight the phased offset of the displacements along different chains.}
\label{fig:scatter}
\end{figure*}

Figure \ref{fig:cdw}(c) shows the electrical resistivity of single crystal \luoss~polished into a rectangular bar with approximate dimensions of $1\times0.3\times0.1$~mm. A precipitous drop in the resistivity is observed at 68\,K, corresponding to the susceptibility drop and the heat capacity anomaly. Current was applied along the [100] direction. A weak 500\,Oe field was applied along the [001] direction to quench trace superconductivity arising from elemental Sn. The field has no other appreciable effects on the transport. As before, an analogous result on YNb$_6$Sn$_6$ is shown for comparison. The data has been scaled according to $R/R_{\text{300~K}}$, but quantitative resistivity values have been provided in the figure as well. Both samples exhibit similar resistivity values at 4\,K ($\sim$30\,\textmu$\Omega$\,cm), but with different RRR values. 

We've established the presence of a suspected DW-like, first-order phase transition near 68~K. We now turn to examine the temperature-dependent scattering data for evidence of an emergent superlattice and diffuse scattering. Figure~\ref{fig:scatter} presents a suite of X-ray scattering results for both polycrystalline and single crystal samples of LuNb$_6$Sn$_6$. All results are indexed in the standard $1\times1\times1$ hexagonal unit cell. Beginning with Fig.~\ref{fig:scatter}(a), we examine the temperature dependence of the lattice parameters for both \luoss~and YNb$_6$Sn$_6$. The transition is clear, predominately observed in the abrupt change of the $c$-axis lattice parameter. Remarkably, the $a$-axis is nearly unperturbed, contracting at a rate consistent with the featureless YNb$_6$Sn$_6$ reference. 

While cooling the sample, yet well above the transition temperature (e.g. 220--100\,K), we observed clear evidence of diffuse X-ray scattering by \luoss. Figure \ref{fig:scatter}(b) presents a slice of the diffuse scattering on the $HK$($L$=9.5) plane (CHESS-QM2 data). A remarkable tiling of hollow triangles appear across the entire dynamic range. The inset shows a higher resolution scan (BNL 21-ID-1) over a smaller range, clearly showing the hollow triangles and highlighting some finer features like the increased scattering intensity on triangle corners. The temperature dependence of the diffuse scattering (Fig.~\ref{fig:scatter}(c)) increases in intensity immediately above the phase transition before coalescing into superlattice peaks below 68\,K. 
Intriguingly, we observed that the diffuse pattern arises most strongly on planes where $L=3n+0.5$ (e.g. 6.5, 9.5), nearly vanishes on $L=3n+1.5$ (e.g. 7.5, 10.5), and is only weakly visible on $L=3n+2.5$ (e.g. 8.5, 11.5). 

Though the hollow triangles are a feature seemingly unique to LuNb$_6$Sn$_6$, to first-order, the diffuse scattering at half integer $L$ is reminiscent of our prior results in ScV$_6$Sn$_6$ \cite{Korshunov2023_SofteningPhononScV6Sn6,Pokharel2023_FrustratedCO+CooperativeDistortionsScV6Sn6,Alvarado2024_FrustratedIsingChargeCorrScV6Sn6}. We previously revealed that the diffuse scattering in ScV$_6$Sn$_6$ could be reproduced using a minimal model of two-dimensional Ising-like displacements of Sn atoms that are frustrated via repulsive strain fields across the kagome network.\cite{Alvarado2024_FrustratedIsingChargeCorrScV6Sn6} A rough application of our previous methods with a similar number of interchain interactions did not readily reproduce the hollow triangles, suggesting that a more complex analysis may be needed. However, an in-depth analysis of the diffuse data is beyond the scope of this foundational work.

Figure \ref{fig:scatter}(d,e) show three different slices through reciprocal space to visualize the emergent superlattice that forms at low temperatures. All graphics show the same sample at high temperature (280~K) and deep within the DW phase (50~K) to help omit artifacts from the diffraction experiment (e.g. background, $\lambda$/3 monochromator contamination). Figure \ref{fig:scatter}(d) is a reconstructed slice of the $HHL$ plane,\cite{GomezNexPy} clearly indicating superlattice peaks with a wavevector of (1/3,1/3,1/3). The bright streaks extending from the zone center are artifacts from the $HHL$ reconstruction, and are to be ignored. Figure~\ref{fig:scatter}(e) shows two slices through the raw data $HK(L=6.66)$ and $(H=0.66)KL$ to highlight the (1/3,1/3,1/3) wavevector. The superlattice peaks are relatively weak, nearly three orders of magnitude weaker than typical integer Bragg reflections. Note that the streaking in Fig.~\ref{fig:scatter}(e) arises from the bleed-over of the much stronger integer $HKL$ reflections into the $H$=0.66 cut. 

The refined structural modulation in \luoss~ is essentially identical to that observed in ScV$_6$Sn$_6$.\cite{arachchige2022ScV6Sn6} We refined the data both using a supercell and a supersymmetry approach. The supercell method indexes the sample using the rotated $\sqrt{3}\times\sqrt{3}\times3$ supercell, which produces the structural solution in Figure \ref{fig:scatter}(f). For graphical clarity, the displacement of individual atoms has been exaggerated by a factor of 3, and shaded arrows have been added to help indicate the atomic shifts. Owing to the large unit cell, only a slice through the (110) plane has been shown. Similar to ScV$_6$Sn$_6$, the primary distortion arises along the Lu1--Sn1--Sn1--Lu1 chains with very  little distortion occurring within the kagome network. The superspace approach results in a qualitatively similar result, utilizing the superspace group $P\bar{3}$1$m$(1/3 1/3 $g$)000 and a large crenel-type occupation and sinusoidal displacement modulation for the Lu1 atom and the neighboring Sn1 atom along the $c$-axis.

The bottom of Fig.~\ref{fig:scatter}(f) is provided to help visualize the displacements. $\Delta z_\text{DW}$ represents the shift in the $z$-coordinate of the atomic position relative to the parent (high-temperature) phase. Positive shifts indicate atoms moving ``up'' on Fig.~\ref{fig:scatter}(f), and negative values indicate atoms moving ``down.'' The displacements for three different chains are shown (0,0,$z$), (2/3,1/3,$z$), and (1/3,2/3,$z$). As one chain shifts up, the other chains shift down (or remain neutral) to compensate for the shifting strain fields within the material. 

\subsection{Magnetic Properties of \nboss (\textit{Ln}: Gd--Tm)}
\label{sec:ResultsDiscussion_Magnetic}

\begin{figure*}
\includegraphics[width=\linewidth]{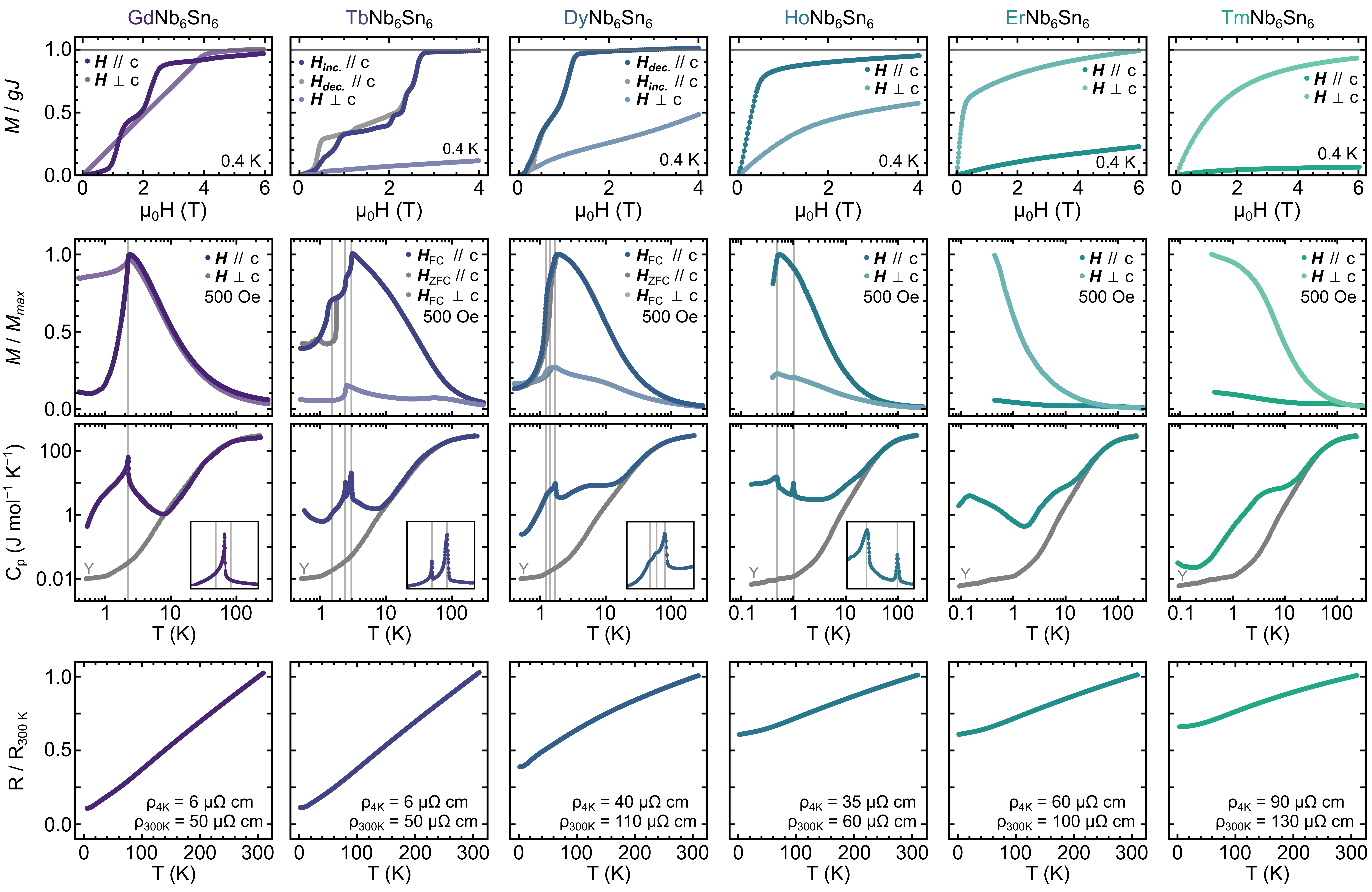}
\caption{As discussed in the text, here we aim to provide a broad, qualitative comparison of the \hfgesn-type magnetic \nboss~ compounds (\textit{Ln}: Gd--Tm). Most materials saturate within 10\% of $gJ$ by 6\,T, and most of the compounds clearly order antiferromagnetically by (0.4~K). Since many of the compounds appear to exhibit multiple transitions, heat capacity measurements are shown to provide additional insight. For compounds that did not order well above 0.4~K (\textit{Ln}:Ho--Tm), additional heat capacity measurements were performed to 0.1~K. Rare-earth nuclear Schottky anomalies are noted for several materials, and nonmagnetic YNb$_6$Sn$_6$ is shown as a qualitative lattice comparison. Finally, we performed electronic resistivity measurements to aid in diagnosing any structural phase transitions, as the strong rare-earth magnetism may mask subtle changes in the magnetization (e.g. DW-transition in \luoss).}
\label{fig:magnetism}
\end{figure*}

Besides the recent search for DW-like distortions in the nonmagnetic kagome metals, one of the hallmark strengths of the \oss~family is the extraordinary chemical flexibility and choice of magnetic $A$ and $M$ sublattices. Like the \textit{Ln}V$_6$Sn$_6$ series, the Nb-based kagome sublattice in the \nboss~ is nonmagnetic, leaving the rare-earth $A$ sublattice to dictate the magnetism. Here we aim to provide a broad, high-level overview of the magnetic properties of the \nboss~system to help the community identify potentially interesting materials for subsequent studies. As alluded to before, we have observed some deleterious disorder for large rare-earth elements, particularly in the cases of PrNb$_6$Sn$_6$ and CeNb$_6$Sn$_6$. As a conservative approach, we will only investigate the magnetic properties of the \nboss~(\textit{Ln}:Gd--Tm), which are well-described by the \hfgesn~structure.

Figure~\ref{fig:magnetism} shows a grid of the primary physical property measurements (magnetization, heat capacity, electronic transport) for single crystals of the \nboss~where (\textit{Ln}:Gd--Tm). For rapid comparison across the series, we have chosen to normalize several of the units ($M/gJ$, $M/M_\text{max}$, and $R/R_{\text{300\,K}}$), where $gJ$ is the expected free-ion saturation magnetization, $M_\text{max}$ is the maximum magnetization (peak), and $R_{\text{300\,K}}$ is the resistivity at 300\,K. The top row presents the isothermal magnetization at the $^3$He base temperature ($T=0.4$\,K). All compounds effectively saturate to $\pm$5-10\% of the expected $gJ$ by a 4-6\,T applied field. For each composition we have shown measurements with both \textbf{\textit{H}}\,$\parallel$\,\textbf{\textit{c}} and \textbf{\textit{H}}\,$\perp$\,\textbf{\textit{c}}, and have further indicated the increasing and decreasing field directions for samples what exhibit field hysteresis. 

The second row presents field-cooled cooling (FC-C) temperature-dependent magnetization measurements for the series, and additionally highlights zero-field-cooled cooling (ZFC-C) measurements where a difference betweem FC-C and ZFC-C was noted. Between the step-like metamagnetic isothermal magnetization curves and the temperature-dependent magnetization, it is clear that the majority of the series are low-temperature antiferromagnets with strong anisotropy. Most compounds clearly order by 0.4~K in the exception of ErNb$_6$Sn$_6$ and TmNb$_6$Sn$_6$. TbNb$_6$Sn$_6$ stands apart as a particularly complicated system, even in the context of other Tb-containing \oss~compounds.\cite{zhang2022electronic,lee2022anisotropic,rosenberg2022uniaxial,pokharel2022highly,Xu2022_ChernMagnetYbMn6Sn6,ElIdrissi1991_NeutronMagStructTbMn6Sn6+HoMn6Sn6,yang2024crystal}

To help visualize the magnetization data, faint gray bars have been drawn through the main features in the temperature-dependent magnetization. Immediately below we have displayed the zero-field heat capacity on the same $\log(T)$ scale to provide rapid comparison of the magnetization data and any corresponding heat capacity anomalies. Each heat capacity plot includes a trace of the nonmagnetic, featureless YNb$_6$Sn$_6$ as a lattice comparison. Any relevant phase transitions are included as a linearly scaled inset. To augment the magnetic measurements (which end at 0.4\,K) the heat capacity of HoNb$_6$Sn$_6$, ErNb$_6$Sn$_6$, and TmNb$_6$Sn$_6$ were performed using a dilution refrigerator down to approximately 0.1\,K to search for any lower temperature transitions. We suspect that ErNb$_6$Sn$_6$ may order near 0.15~K, though the convolution with a rare-earth nuclear Schottky peak (which occurs in several of the compounds) impedes a clear determination.

Due to the strong magnetic signatures of the rare-earth compounds, any signatures of potential DW-like behavior in the Gd--Tm compounds would be completely obscured in the magnetization data. Heat capacity measurements should capture any higher temperature transitions, though we can also use the electrical resistivity as a good screening tool given the strong electrical response of LuNb$_6$Sn$_6$ at the DW transition. The lowest row of Fig.~\ref{fig:magnetism} shows the electrical resistivity normalized to the resistivity at 300\,K. Quantitative data for 4\,K and 300\,K is written in each plot for completeness, though geometrical factors will play a large role in samples of this size (ca. 1$\times$0.3$\times$0.1\,mm).

We do not observe any clear signature of DW-like instabilities in the electronic resistivity of the magnetic compounds down to 4~K. Note that the resistivity measurements do not go sufficiently low in temperature to probe the magnetic transition, though we note a curious trend in $R/R_\text{300K}$ where large rare-earth compounds exhibit stronger temperature-dependence and higher RRR. The effect reverses as we integrate progressively smaller filler atoms, which may be indicative of a tendency towards the ``rattling'' instability that ultimately yields the superlattice in \luoss. Sub-linear resistivity vs temperature in proximity to structural modulations also appears in ScV$_6$Sn$_6$ and $A$V$_3$Sb$_5$.\cite{Mozaffari2024_SublinearResistivityVKagome}

\subsection{Rattling Interpretation of Density Wave}
\label{sec:ResultsDiscussion_Rattling}

\begin{figure*}
    \includegraphics[width=\linewidth]{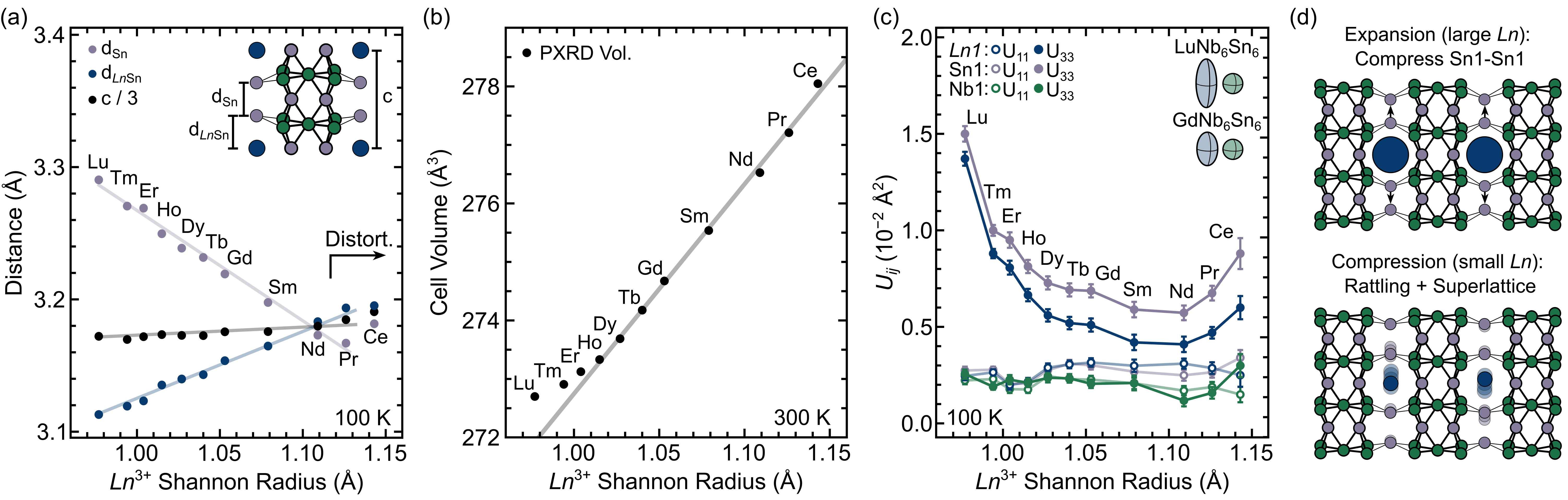}
    \caption{(a) Rare-earth filling serves to stabilize the \oss~structure, but the size of the rare-earth has a dramatic effect on the bonding. Despite large changes in rare-earth size, the \textit{c} axis in the \nboss~ (black) only responds very weakly. Instead, the structure accommodates large \textit{Ln} atoms by displacing Sn1.
    (b) Conversely, when the rare-earth atom shrinks, we note that the \luoss~series diverges from the general linear trend expected for a rare-earth substituted series. This is a consequence of the rigid Nb--Sn scaffolding, which resists further compression after a critical radius is reached. (c) Consistent with both postulates, the ADPs of the \textit{Ln}1 and Sn1 site exhibit a strong dependence on the rare-earth size. Small rare-earth atoms underfill the void (exacerbated by the rigid Nb-Sn network) and have comparatively large displacements. (d) The ``rattling'' of the rare-earth is felt throughout the \textit{Ln}--Sn1--Sn1--\textit{Ln} chain, as the flexible Sn1 atoms adjust to compensate for positioning of \textit{Ln}.}
    \label{fig:Rattling}
\end{figure*}

Until now, ScV$_6$Sn$_6$ was the only member of the \hfgesn~family to exhibit a density wave-like transition. Prior work highlighted that Sc was substantially smaller than all other \textit{Ln}V$_6$Sn$_6$ compounds, leaving extra room along the chains for the Sn1-Sc-Sn1 trimers\cite{Pokharel2023_FrustratedCO+CooperativeDistortionsScV6Sn6} to rattle and driving the Sn1-Sn1 bond modulation.\cite{arachchige2022ScV6Sn6,Liu2024_DrivingMechanismScV6Sn6,Hu2023_ScV6Sn6-Theory-FlatPhonons+UnconventionalCDW,Lee2024_NatureCDWScV6Sn6} The sharp contrast in size between Sc and the next smallest compound (LuV$_6$Sn$_6$) made it difficult to compare the effect of a slightly larger atom without alloying. With the introduction of \luoss~we have a more gradual series of rare-earth sizes, as the size of the next smallest ion (Tm) is only slightly larger than Lu. Yet we are left with the same question; why is \luoss~the only member of this series to display a structural transition? 

Recall that the \oss~family of compounds can be thought of as a long-range, ordered, filled variant of CoSn. Large filler atoms intuitively place an expansive pressure on the surrounding lattice, which drives the motion of the Sn1 atoms and the ultimate evolution of the \hfgesn~structure (see Fig.~\ref{fig:1}(a)). Figure~\ref{fig:Rattling}(a) helps provide a graphical representation of this effect in the \nboss~series extracted from single crystal diffraction results. Here we show three sets of data: 1) the Sn1-Sn1 bond distance (gray), 2) the $Ln$1-Sn1 bond distance (blue), and 3) 1/3 of the $c$-axis lattice parameter (black). 
The $c$ lattice parameter is divided by 3 in order to provide a convenient, scaled proxy for the expansion along $c$. Since PrNb$_6$Sn$_6$ and CeNb$_6$Sn$_6$ still index into the hexagonal $1\times1\times1$ unit cell we have opted to include the results, though we caution that their details are unknown and should be viewed cautiously.

The first anomalous observation is the slow expansion of $c$, expanding by only 0.02\,\AA~across the series even though the Shannon radius is 0.13\,\AA~larger.\cite{Shannon1976_Radii} If we considered the atomic positions largely static, this would be an irreconcilable difference, even for a toy model. However, despite the feeble trend in ($c/3$), both the Sn1-Sn1 bond distance ($d_\text{Sn}$) and the $Ln$1-Sn1 bond distance ($d_{Ln\text{Sn}}$) show dramatic changes. Notably, the Sn1-Sn1 bond is compressed by 0.12\,\AA~across the series, indicating that the structure accommodates large atoms by preferentially compressing the Sn1-Sn1 bond, which is observed in the $A$V$_6$Sn$_6$ series as well.\cite{Meier2023tiny}

Clearly the expansive effect of large rare-earth atoms on the $c$ axis is mitigated by the flexible Sn1--Sn1 bonds, but we should still expect a general expansion of the overall lattice. In many rare-earth containing series the cell volume still trends nearly linearly with the Shannon radius of the rare-earth element.\cite{Pani2017_FormationVolumeRareEarthIntermetallic} However, in contrast to the usual lanthanide trend, Fig.~\ref{fig:Rattling}(b) reveals a markedly nonlinear plot of cell volume vs Shannon radius. Pawley fits to powder X-ray diffraction data have been used for more accurate lattice parameters. As we integrate smaller rare-earth elements, the volumetric contraction of the lattice progressively slows, which originates from the stunted contraction of the $c$-axis. We suspect that upon switching to small rare-earth species, the structure switches into a different regime, where the rigid Nb--Sn scaffolding (Fig.~\ref{fig:1}(a)) accommodates the size mismatch, supporting the structure and preventing further collapse.

Panels (a) and (b) of Fig.~\ref{fig:Rattling} paint a picture of the rigid Nb-Sn network that supports the lattice and the steric compression of the Sn1--Sn1 bonds through the incorporation of larger filler atoms. Both of these effects should have a strong impact on the thermal motion of the atoms and be strongly reflected in the anisotropic displacement parameters (ADPs). Figure \ref{fig:Rattling}(c) presents the single crystal ADPs as a function of the Shannon radius at 100~K. Two of the curves immediately stand apart -- $U_{33}$ for both \textit{Ln}1 and Sn1. These values show a dramatic enhancement for the smallest rare earths, representing increasingly large displacements along the \textbf{\textit{c}} direction. These large displacements are not seen in the $ab$-plane ($U_{11}$) for \textit{Ln}1 or Sn1. The Nb1, Sn2, and Sn3 (Sn2, Sn3 not shown) show largely isotropic and small ADPs, supporting our interpretation of the Nb--Sn rigid scaffolding. The inset in Fig.~\ref{fig:Rattling}(c) depicts the size of the thermal ellipsoids for the rare-earth Lu1 site (light blue) and Nb1 (light green) to scale. The large dynamic displacements of Lu1 and Sn1 in \luoss~are certainly the origin of the diffuse scattering presented in Fig.~\ref{fig:scatter}. Note that the increased ADPs for PrNb$_6$Sn$_6$ and CeNb$_6$Sn$_6$ are likely a consequence of static site disorder, and not true anisotropic motion. 

Figure~\ref{fig:Rattling}(d) summarizes the how rare earth atoms impact filling of the \nboss~structure using a schematic of the Nb--Sn scaffolding and exaggerated filler atoms. Yet, a lingering question remains -- what is the filling threshold for that leads to the structural modulations observed in the \nboss~and $A$V$_6$Sn$_6$ families? This question was difficult to approach in the ScV$_6$Sn$_6$ system, as the difference in size between Sc and the next smallest element (Lu) is quite dramatic. 
In the \luoss~system, however, TmNb$_6$Sn$_6$ does not exhibit a DW despite the similar size of the Lu and Tm ions. Frustratingly, YbNb$_6$Sn$_6$ has not been realized at this time. Thus, we now have the converse problem observed in ScV$_6$Sn$_6$ -- there are no rare-earth (or rare-earth-like) atoms incrementally smaller than Lu.

Based on the similarities with the  ScV$_6$Sn$_6$ series,\cite{arachchige2022ScV6Sn6,Meier2023tiny} we expect pressure and doping to have profound effects on the DW stability. We predict that the structural transition temperature in \luoss~will be suppressed by applied pressure due to the compression of the Nb-Sn scaffolding.\cite{Zhang2022_ScV6Sn6-pressure} In addition, doping with different sized rare earths should modify the room available in the $Ln$-Sn1 columns. Therefore, we anticipate scandium doping for Lu should enhance the transition temperature in \luoss~and Tm should reduce it. Pursuing these avenues forward will direct our research towards our ultimate goal -- realizing a DW-like structural modulation in the \hfgesn~compounds with a magnetic rare-earth sublattice, allowing us to explore how the atomic shifts modify the exchange coupling between rare earth sites how it potentially impacts the magnetic order.

\section{Conclusions}
\label{sec:Conclusions}

The \oss~family of kagome metals is among one of the most influential and chemically diverse kagome platforms currently available. We began this work by providing a thorough evaluation of the known \oss~materials, providing a succinct way to visualize the different structural polymorphs and thermodynamic stability of the broader family. Simultaneously, we provided the development of an extensive collection of new \nboss~single crystals, expanding our phase selectivity into the $4d$ transition metals. The electronic structure of the \luoss~sports a VHS and Dirac cone within 0.1\,eV of the Fermi level, and a flat band approximately 0.6\,eV above $E_\text{F}$. In addition to the complex antiferromagnetism observed in the magnetic \nboss~compounds, we also observed a DW-like transition in \luoss. Despite the confluence of the DW and the kagome band structure, we establish that the DW is likely driven by a structural origin, similar to recent reports in ScV$_6$Sn$_6$. High-quality X-ray scattering data provides a in-depth look into both the diffuse scattering, and superlattice in \luoss~finding a superlattice with a (1/3,1/3,1/3) wave vector. In addition, we observe diffuse scattering at higher temperatures, particularly on $HK(L=3n+0.5)$ planes. The emergence of the superlattice is consistent, and can be predicted from, our previous ``rattling'' theory developed for ScV$_6$Sn$_6$. Utilizing our series of single crystals, we were able to provide a detailed analysis of structural trends within the family, revealing how the \luoss~family adapts to changes in filling atom size, which ultimately intertwines the phase stability and emergence of the DW-like instability. Our work provides the community with a new family of single crystal kagome metals with complex structural, electronic, and magnetic properties -- expanding on our ability to understand and engineer the next-generation of kagome materials.

\section{Acknowledgment}

This research was supported by the U.S. Department of Energy, Office of Science, Basic Energy Sciences, Materials Sciences and Engineering Division (material development, crystal growth, characterization, X-ray, ARPES, and theory). Early aspects of the work (project conception, phase diagram mapping from literature, DFT) was sponsored by the Laboratory Directed Research and Development Program of Oak Ridge National Laboratory, managed by UT-Battelle, LLC, for the US Department of Energy. This work is based on research conducted at the Center for High-Energy X-ray Sciences (CHEXS), which is supported by the National Science Foundation (BIO, ENG, and MPS Directorates) under award DMR-2342336. W.R.M., S.M., A.T., and D.M. acknowledge support from the Gordon and Betty Moore Foundation’s EPiQS Initiative, Grant GBMF9069 awarded to D.M.. S.D.W and S.J.G.A gratefully acknowledge support via the UC Santa Barbara NSF Quantum Foundry funded via the Q-AMASE-i program under award DMR-1906325. G.P. acknowledges support from the University of West Georgia. X-ray scattering and ARPES  measurements used resources at 4-ID and 21-ID-1 beamlines of the National Synchrotron Light Source II, a US Department of Energy Office of Science User Facility operated for the DOE Office of Science by Brookhaven National Laboratory under contract no. DE-SC0012704. This research used resources of the Compute and Data Environment for Science (CADES) at the Oak Ridge National Laboratory, which is supported by the Office of Science of the U.S. Department of Energy under Contract No. DE-AC05-00OR22725. We thank Seunghwan Do for his discussions and feedback. We thank Jong Keum and the X-ray laboratory of the Oak Ridge National Laboratory Spallation Neutron Source (SNS) for use of their Photonic Science Laue camera and Rigaku XtaLab PRO diffractometer.

\bibliography{Frontiers166.bib}% Produces the bibliography via BibTeX.

\providecommand{\noopsort}[1]{}\providecommand{\singleletter}[1]{#1}%
\begin{thebibliography}{155}%
\makeatletter
\providecommand \@ifxundefined [1]{%
 \@ifx{#1\undefined}
}%
\providecommand \@ifnum [1]{%
 \ifnum #1\expandafter \@firstoftwo
 \else \expandafter \@secondoftwo
 \fi
}%
\providecommand \@ifx [1]{%
 \ifx #1\expandafter \@firstoftwo
 \else \expandafter \@secondoftwo
 \fi
}%
\providecommand \natexlab [1]{#1}%
\providecommand \enquote  [1]{``#1''}%
\providecommand \bibnamefont  [1]{#1}%
\providecommand \bibfnamefont [1]{#1}%
\providecommand \citenamefont [1]{#1}%
\providecommand \href@noop [0]{\@secondoftwo}%
\providecommand \href [0]{\begingroup \@sanitize@url \@href}%
\providecommand \@href[1]{\@@startlink{#1}\@@href}%
\providecommand \@@href[1]{\endgroup#1\@@endlink}%
\providecommand \@sanitize@url [0]{\catcode `\\12\catcode `\$12\catcode `\&12\catcode `\#12\catcode `\^12\catcode `\_12\catcode `\%12\relax}%
\providecommand \@@startlink[1]{}%
\providecommand \@@endlink[0]{}%
\providecommand \url  [0]{\begingroup\@sanitize@url \@url }%
\providecommand \@url [1]{\endgroup\@href {#1}{\urlprefix }}%
\providecommand \urlprefix  [0]{URL }%
\providecommand \Eprint [0]{\href }%
\providecommand \doibase [0]{https://doi.org/}%
\providecommand \selectlanguage [0]{\@gobble}%
\providecommand \bibinfo  [0]{\@secondoftwo}%
\providecommand \bibfield  [0]{\@secondoftwo}%
\providecommand \translation [1]{[#1]}%
\providecommand \BibitemOpen [0]{}%
\providecommand \bibitemStop [0]{}%
\providecommand \bibitemNoStop [0]{.\EOS\space}%
\providecommand \EOS [0]{\spacefactor3000\relax}%
\providecommand \BibitemShut  [1]{\csname bibitem#1\endcsname}%
\let\auto@bib@innerbib\@empty
%</preamble>
\bibitem [{\citenamefont {Han}\ \emph {et~al.}(2012)\citenamefont {Han}, \citenamefont {Helton}, \citenamefont {Chu}, \citenamefont {Nocera}, \citenamefont {Rodriguez-Rivera}, \citenamefont {Broholm},\ and\ \citenamefont {Lee}}]{han2012fractionalized}%
  \BibitemOpen
  \bibfield  {author} {\bibinfo {author} {\bibfnamefont {T.-H.}\ \bibnamefont {Han}}, \bibinfo {author} {\bibfnamefont {J.~S.}\ \bibnamefont {Helton}}, \bibinfo {author} {\bibfnamefont {S.}~\bibnamefont {Chu}}, \bibinfo {author} {\bibfnamefont {D.~G.}\ \bibnamefont {Nocera}}, \bibinfo {author} {\bibfnamefont {J.~A.}\ \bibnamefont {Rodriguez-Rivera}}, \bibinfo {author} {\bibfnamefont {C.}~\bibnamefont {Broholm}},\ and\ \bibinfo {author} {\bibfnamefont {Y.~S.}\ \bibnamefont {Lee}},\ }\bibfield  {title} {\bibinfo {title} {{Fractionalized excitations in the spin-liquid state of a kagome-lattice antiferromagnet}},\ }\href {https://doi.org/10.1038/nature11659} {\bibfield  {journal} {\bibinfo  {journal} {Nature}\ }\textbf {\bibinfo {volume} {492}},\ \bibinfo {pages} {406} (\bibinfo {year} {2012})}\BibitemShut {NoStop}%
\bibitem [{\citenamefont {Depenbrock}\ \emph {et~al.}(2012)\citenamefont {Depenbrock}, \citenamefont {McCulloch},\ and\ \citenamefont {Schollw{\"o}ck}}]{depenbrock2012nature}%
  \BibitemOpen
  \bibfield  {author} {\bibinfo {author} {\bibfnamefont {S.}~\bibnamefont {Depenbrock}}, \bibinfo {author} {\bibfnamefont {I.~P.}\ \bibnamefont {McCulloch}},\ and\ \bibinfo {author} {\bibfnamefont {U.}~\bibnamefont {Schollw{\"o}ck}},\ }\bibfield  {title} {\bibinfo {title} {{Nature of the spin-liquid ground state of the S= 1/2 Heisenberg model on the kagome lattice}},\ }\href {https://doi.org/10.1103/PhysRevLett.109.067201} {\bibfield  {journal} {\bibinfo  {journal} {Physical review letters}\ }\textbf {\bibinfo {volume} {109}},\ \bibinfo {pages} {067201} (\bibinfo {year} {2012})}\BibitemShut {NoStop}%
\bibitem [{\citenamefont {Norman}(2016)}]{Norman2016HerbertsmithiteReview}%
  \BibitemOpen
  \bibfield  {author} {\bibinfo {author} {\bibfnamefont {M.~R.}\ \bibnamefont {Norman}},\ }\bibfield  {title} {\bibinfo {title} {{Colloquium : Herbertsmithite and the search for the quantum spin liquid}},\ }\href {https://doi.org/10.1103/revmodphys.88.041002} {\bibfield  {journal} {\bibinfo  {journal} {Reviews of Modern Physics}\ }\textbf {\bibinfo {volume} {88}},\ \bibinfo {pages} {041002} (\bibinfo {year} {2016})}\BibitemShut {NoStop}%
\bibitem [{\citenamefont {Shores}\ \emph {et~al.}(2005)\citenamefont {Shores}, \citenamefont {Nytko}, \citenamefont {Bartlett},\ and\ \citenamefont {Nocera}}]{Shores2005Atacamite}%
  \BibitemOpen
  \bibfield  {author} {\bibinfo {author} {\bibfnamefont {M.~P.}\ \bibnamefont {Shores}}, \bibinfo {author} {\bibfnamefont {E.~A.}\ \bibnamefont {Nytko}}, \bibinfo {author} {\bibfnamefont {B.~M.}\ \bibnamefont {Bartlett}},\ and\ \bibinfo {author} {\bibfnamefont {D.~G.}\ \bibnamefont {Nocera}},\ }\bibfield  {title} {\bibinfo {title} {{A Structurally Perfect S = 1/2 Kagomé Antiferromagnet}},\ }\href {https://doi.org/10.1021/ja053891p} {\bibfield  {journal} {\bibinfo  {journal} {Journal of the American Chemical Society}\ }\textbf {\bibinfo {volume} {127}},\ \bibinfo {pages} {13462} (\bibinfo {year} {2005})}\BibitemShut {NoStop}%
\bibitem [{\citenamefont {Okamoto}\ \emph {et~al.}(2009)\citenamefont {Okamoto}, \citenamefont {Yoshida},\ and\ \citenamefont {Hiroi}}]{Okamoto2009_Vesignieite}%
  \BibitemOpen
  \bibfield  {author} {\bibinfo {author} {\bibfnamefont {Y.}~\bibnamefont {Okamoto}}, \bibinfo {author} {\bibfnamefont {H.}~\bibnamefont {Yoshida}},\ and\ \bibinfo {author} {\bibfnamefont {Z.}~\bibnamefont {Hiroi}},\ }\bibfield  {title} {\bibinfo {title} {{Vesignieite BaCu$_3$V$_2$O$_8$(OH)$_2$ as a Candidate Spin-1/2 Kagome Antiferromagnet}},\ }\href {https://doi.org/10.1143/jpsj.78.033701} {\bibfield  {journal} {\bibinfo  {journal} {Journal of the Physical Society of Japan}\ }\textbf {\bibinfo {volume} {78}},\ \bibinfo {pages} {033701} (\bibinfo {year} {2009})}\BibitemShut {NoStop}%
\bibitem [{\citenamefont {Han}\ \emph {et~al.}(2014)\citenamefont {Han}, \citenamefont {Singleton},\ and\ \citenamefont {Schlueter}}]{Han2014_Barlowite}%
  \BibitemOpen
  \bibfield  {author} {\bibinfo {author} {\bibfnamefont {T.-H.}\ \bibnamefont {Han}}, \bibinfo {author} {\bibfnamefont {J.}~\bibnamefont {Singleton}},\ and\ \bibinfo {author} {\bibfnamefont {J.~A.}\ \bibnamefont {Schlueter}},\ }\bibfield  {title} {\bibinfo {title} {{Barlowite: A Spin-1/2 Antiferromagnet with a Geometrically Perfect Kagome Motif}},\ }\href {https://doi.org/10.1103/physrevlett.113.227203} {\bibfield  {journal} {\bibinfo  {journal} {Physical Review Letters}\ }\textbf {\bibinfo {volume} {113}},\ \bibinfo {pages} {227203} (\bibinfo {year} {2014})}\BibitemShut {NoStop}%
\bibitem [{\citenamefont {Pasco}\ \emph {et~al.}(2018)\citenamefont {Pasco}, \citenamefont {Trump}, \citenamefont {Tran}, \citenamefont {Kelly}, \citenamefont {Hoffmann}, \citenamefont {Heinmaa}, \citenamefont {Stern},\ and\ \citenamefont {McQueen}}]{Pasco2018_Barlowite+Herbertsmithite}%
  \BibitemOpen
  \bibfield  {author} {\bibinfo {author} {\bibfnamefont {C.~M.}\ \bibnamefont {Pasco}}, \bibinfo {author} {\bibfnamefont {B.~A.}\ \bibnamefont {Trump}}, \bibinfo {author} {\bibfnamefont {T.~T.}\ \bibnamefont {Tran}}, \bibinfo {author} {\bibfnamefont {Z.~A.}\ \bibnamefont {Kelly}}, \bibinfo {author} {\bibfnamefont {C.}~\bibnamefont {Hoffmann}}, \bibinfo {author} {\bibfnamefont {I.}~\bibnamefont {Heinmaa}}, \bibinfo {author} {\bibfnamefont {R.}~\bibnamefont {Stern}},\ and\ \bibinfo {author} {\bibfnamefont {T.~M.}\ \bibnamefont {McQueen}},\ }\bibfield  {title} {\bibinfo {title} {{Single-crystal growth of Cu$_{4}$(OH)$_{6}$BrF and universal behavior in quantum spin liquid candidates synthetic barlowite and herbertsmithite}},\ }\href {https://doi.org/10.1103/physrevmaterials.2.044406} {\bibfield  {journal} {\bibinfo  {journal} {Physical Review Materials}\ }\textbf {\bibinfo {volume} {2}},\ \bibinfo {pages} {044406} (\bibinfo {year} {2018})}\BibitemShut {NoStop}%
\bibitem [{\citenamefont {Kiesel}\ and\ \citenamefont {Thomale}(2012)}]{kiesel2012sublattice}%
  \BibitemOpen
  \bibfield  {author} {\bibinfo {author} {\bibfnamefont {M.~L.}\ \bibnamefont {Kiesel}}\ and\ \bibinfo {author} {\bibfnamefont {R.}~\bibnamefont {Thomale}},\ }\bibfield  {title} {\bibinfo {title} {{Sublattice interference in the kagome Hubbard model}},\ }\href {https://doi.org/10.1103/physrevb.86.121105} {\bibfield  {journal} {\bibinfo  {journal} {Physical Review B—Condensed Matter and Materials Physics}\ }\textbf {\bibinfo {volume} {86}},\ \bibinfo {pages} {121105} (\bibinfo {year} {2012})}\BibitemShut {NoStop}%
\bibitem [{\citenamefont {Wang}\ \emph {et~al.}(2013)\citenamefont {Wang}, \citenamefont {Li}, \citenamefont {Xiang},\ and\ \citenamefont {Wang}}]{wang2013competing}%
  \BibitemOpen
  \bibfield  {author} {\bibinfo {author} {\bibfnamefont {W.-S.}\ \bibnamefont {Wang}}, \bibinfo {author} {\bibfnamefont {Z.-Z.}\ \bibnamefont {Li}}, \bibinfo {author} {\bibfnamefont {Y.-Y.}\ \bibnamefont {Xiang}},\ and\ \bibinfo {author} {\bibfnamefont {Q.-H.}\ \bibnamefont {Wang}},\ }\bibfield  {title} {\bibinfo {title} {{Competing electronic orders on kagome lattices at van Hove filling}},\ }\href {https://doi.org/10.1103/physrevb.87.115135} {\bibfield  {journal} {\bibinfo  {journal} {Physical Review B—Condensed Matter and Materials Physics}\ }\textbf {\bibinfo {volume} {87}},\ \bibinfo {pages} {115135} (\bibinfo {year} {2013})}\BibitemShut {NoStop}%
\bibitem [{\citenamefont {Beugeling}\ \emph {et~al.}(2012)\citenamefont {Beugeling}, \citenamefont {Everts},\ and\ \citenamefont {Morais~Smith}}]{Beugeling2012_TopologicalPhases2DLattices}%
  \BibitemOpen
  \bibfield  {author} {\bibinfo {author} {\bibfnamefont {W.}~\bibnamefont {Beugeling}}, \bibinfo {author} {\bibfnamefont {J.~C.}\ \bibnamefont {Everts}},\ and\ \bibinfo {author} {\bibfnamefont {C.}~\bibnamefont {Morais~Smith}},\ }\bibfield  {title} {\bibinfo {title} {{Topological phase transitions driven by next-nearest-neighbor hopping in two-dimensional lattices}},\ }\href {https://doi.org/10.1103/physrevb.86.195129} {\bibfield  {journal} {\bibinfo  {journal} {Physical Review B}\ }\textbf {\bibinfo {volume} {86}},\ \bibinfo {pages} {195129} (\bibinfo {year} {2012})}\BibitemShut {NoStop}%
\bibitem [{\citenamefont {Guo}\ and\ \citenamefont {Franz}(2009)}]{Guo2009_TopoInsuOnKagome}%
  \BibitemOpen
  \bibfield  {author} {\bibinfo {author} {\bibfnamefont {H.-M.}\ \bibnamefont {Guo}}\ and\ \bibinfo {author} {\bibfnamefont {M.}~\bibnamefont {Franz}},\ }\bibfield  {title} {\bibinfo {title} {{Topological insulator on the kagome lattice}},\ }\href {https://doi.org/10.1103/physrevb.80.113102} {\bibfield  {journal} {\bibinfo  {journal} {Physical Review B}\ }\textbf {\bibinfo {volume} {80}},\ \bibinfo {pages} {113102} (\bibinfo {year} {2009})}\BibitemShut {NoStop}%
\bibitem [{\citenamefont {Kiesel}\ \emph {et~al.}(2013{\natexlab{a}})\citenamefont {Kiesel}, \citenamefont {Platt},\ and\ \citenamefont {Thomale}}]{kiesel2013unconventional}%
  \BibitemOpen
  \bibfield  {author} {\bibinfo {author} {\bibfnamefont {M.~L.}\ \bibnamefont {Kiesel}}, \bibinfo {author} {\bibfnamefont {C.}~\bibnamefont {Platt}},\ and\ \bibinfo {author} {\bibfnamefont {R.}~\bibnamefont {Thomale}},\ }\bibfield  {title} {\bibinfo {title} {{Unconventional Fermi surface instabilities in the kagome Hubbard model}},\ }\href {https://doi.org/10.1103/physrevlett.110.126405} {\bibfield  {journal} {\bibinfo  {journal} {Phys. Rev. Lett.}\ }\textbf {\bibinfo {volume} {110}},\ \bibinfo {pages} {126405} (\bibinfo {year} {2013}{\natexlab{a}})}\BibitemShut {NoStop}%
\bibitem [{\citenamefont {Wen}\ \emph {et~al.}(2010)\citenamefont {Wen}, \citenamefont {Rüegg}, \citenamefont {Wang},\ and\ \citenamefont {Fiete}}]{Wen2010_TopoInsulInKagome}%
  \BibitemOpen
  \bibfield  {author} {\bibinfo {author} {\bibfnamefont {J.}~\bibnamefont {Wen}}, \bibinfo {author} {\bibfnamefont {A.}~\bibnamefont {Rüegg}}, \bibinfo {author} {\bibfnamefont {C.-C.~J.}\ \bibnamefont {Wang}},\ and\ \bibinfo {author} {\bibfnamefont {G.~A.}\ \bibnamefont {Fiete}},\ }\bibfield  {title} {\bibinfo {title} {{Interaction-driven topological insulators on the kagome and the decorated honeycomb lattices}},\ }\href {https://doi.org/10.1103/physrevb.82.075125} {\bibfield  {journal} {\bibinfo  {journal} {Physical Review B}\ }\textbf {\bibinfo {volume} {82}},\ \bibinfo {pages} {075125} (\bibinfo {year} {2010})}\BibitemShut {NoStop}%
\bibitem [{\citenamefont {Park}\ \emph {et~al.}(2021)\citenamefont {Park}, \citenamefont {Ye},\ and\ \citenamefont {Balents}}]{Park2021_InstabilitesOfKagomeMetals}%
  \BibitemOpen
  \bibfield  {author} {\bibinfo {author} {\bibfnamefont {T.}~\bibnamefont {Park}}, \bibinfo {author} {\bibfnamefont {M.}~\bibnamefont {Ye}},\ and\ \bibinfo {author} {\bibfnamefont {L.}~\bibnamefont {Balents}},\ }\bibfield  {title} {\bibinfo {title} {{Electronic instabilities of kagome metals: Saddle points and Landau theory}},\ }\href {https://doi.org/10.1103/physrevb.104.035142} {\bibfield  {journal} {\bibinfo  {journal} {Physical Review B}\ }\textbf {\bibinfo {volume} {104}},\ \bibinfo {pages} {035142} (\bibinfo {year} {2021})}\BibitemShut {NoStop}%
\bibitem [{\citenamefont {Christensen}\ \emph {et~al.}(2021)\citenamefont {Christensen}, \citenamefont {Birol}, \citenamefont {Andersen},\ and\ \citenamefont {Fernandes}}]{Christensen2021_135CDWTheory}%
  \BibitemOpen
  \bibfield  {author} {\bibinfo {author} {\bibfnamefont {M.~H.}\ \bibnamefont {Christensen}}, \bibinfo {author} {\bibfnamefont {T.}~\bibnamefont {Birol}}, \bibinfo {author} {\bibfnamefont {B.~M.}\ \bibnamefont {Andersen}},\ and\ \bibinfo {author} {\bibfnamefont {R.~M.}\ \bibnamefont {Fernandes}},\ }\bibfield  {title} {\bibinfo {title} {{Theory of the charge density wave in AV$_3$Sb$_5$ kagome metals}},\ }\href {https://doi.org/10.1103/physrevb.104.214513} {\bibfield  {journal} {\bibinfo  {journal} {Physical Review B}\ }\textbf {\bibinfo {volume} {104}},\ \bibinfo {pages} {214513} (\bibinfo {year} {2021})}\BibitemShut {NoStop}%
\bibitem [{\citenamefont {Denner}\ \emph {et~al.}(2021)\citenamefont {Denner}, \citenamefont {Thomale},\ and\ \citenamefont {Neupert}}]{Denner2021_ChargeOrderAV3Sb5}%
  \BibitemOpen
  \bibfield  {author} {\bibinfo {author} {\bibfnamefont {M.~M.}\ \bibnamefont {Denner}}, \bibinfo {author} {\bibfnamefont {R.}~\bibnamefont {Thomale}},\ and\ \bibinfo {author} {\bibfnamefont {T.}~\bibnamefont {Neupert}},\ }\bibfield  {title} {\bibinfo {title} {{Analysis of Charge Order in the Kagome Metal AV$_3$Sb$_5$ ( A=K,Rb,Cs )}},\ }\href {https://doi.org/10.1103/physrevlett.127.217601} {\bibfield  {journal} {\bibinfo  {journal} {Physical Review Letters}\ }\textbf {\bibinfo {volume} {127}},\ \bibinfo {pages} {217601} (\bibinfo {year} {2021})}\BibitemShut {NoStop}%
\bibitem [{\citenamefont {Ferrari}\ \emph {et~al.}(2022)\citenamefont {Ferrari}, \citenamefont {Becca},\ and\ \citenamefont {Valentí}}]{Ferrari2022_CDW-KagomeHubbard}%
  \BibitemOpen
  \bibfield  {author} {\bibinfo {author} {\bibfnamefont {F.}~\bibnamefont {Ferrari}}, \bibinfo {author} {\bibfnamefont {F.}~\bibnamefont {Becca}},\ and\ \bibinfo {author} {\bibfnamefont {R.}~\bibnamefont {Valentí}},\ }\bibfield  {title} {\bibinfo {title} {{Charge density waves in kagome-lattice extended Hubbard models at the van Hove filling}},\ }\href {https://doi.org/10.1103/physrevb.106.l081107} {\bibfield  {journal} {\bibinfo  {journal} {Physical Review B}\ }\textbf {\bibinfo {volume} {106}},\ \bibinfo {pages} {l081107} (\bibinfo {year} {2022})}\BibitemShut {NoStop}%
\bibitem [{\citenamefont {Feng}\ \emph {et~al.}(2021{\natexlab{a}})\citenamefont {Feng}, \citenamefont {Jiang}, \citenamefont {Wang},\ and\ \citenamefont {Hu}}]{Feng2021_FluxPhaseKagomeAV3Sb5}%
  \BibitemOpen
  \bibfield  {author} {\bibinfo {author} {\bibfnamefont {X.}~\bibnamefont {Feng}}, \bibinfo {author} {\bibfnamefont {K.}~\bibnamefont {Jiang}}, \bibinfo {author} {\bibfnamefont {Z.}~\bibnamefont {Wang}},\ and\ \bibinfo {author} {\bibfnamefont {J.}~\bibnamefont {Hu}},\ }\bibfield  {title} {\bibinfo {title} {{Chiral flux phase in the Kagome superconductor AV$_3$Sb$_5$}},\ }\href {https://doi.org/10.1016/j.scib.2021.04.043} {\bibfield  {journal} {\bibinfo  {journal} {Science Bulletin}\ }\textbf {\bibinfo {volume} {66}},\ \bibinfo {pages} {1384} (\bibinfo {year} {2021}{\natexlab{a}})}\BibitemShut {NoStop}%
\bibitem [{\citenamefont {Feng}\ \emph {et~al.}(2021{\natexlab{b}})\citenamefont {Feng}, \citenamefont {Zhang}, \citenamefont {Jiang},\ and\ \citenamefont {Hu}}]{Feng2021_LowEFluxPhaseInKagome}%
  \BibitemOpen
  \bibfield  {author} {\bibinfo {author} {\bibfnamefont {X.}~\bibnamefont {Feng}}, \bibinfo {author} {\bibfnamefont {Y.}~\bibnamefont {Zhang}}, \bibinfo {author} {\bibfnamefont {K.}~\bibnamefont {Jiang}},\ and\ \bibinfo {author} {\bibfnamefont {J.}~\bibnamefont {Hu}},\ }\bibfield  {title} {\bibinfo {title} {{Low-energy effective theory and symmetry classification of flux phases on the kagome lattice}},\ }\href {https://doi.org/10.1103/physrevb.104.165136} {\bibfield  {journal} {\bibinfo  {journal} {Physical Review B}\ }\textbf {\bibinfo {volume} {104}},\ \bibinfo {pages} {165136} (\bibinfo {year} {2021}{\natexlab{b}})}\BibitemShut {NoStop}%
\bibitem [{\citenamefont {Kiesel}\ \emph {et~al.}(2013{\natexlab{b}})\citenamefont {Kiesel}, \citenamefont {Platt},\ and\ \citenamefont {Thomale}}]{Kiesel2013_FermiSurfaceInstabilites-KagomeHubbard}%
  \BibitemOpen
  \bibfield  {author} {\bibinfo {author} {\bibfnamefont {M.~L.}\ \bibnamefont {Kiesel}}, \bibinfo {author} {\bibfnamefont {C.}~\bibnamefont {Platt}},\ and\ \bibinfo {author} {\bibfnamefont {R.}~\bibnamefont {Thomale}},\ }\bibfield  {title} {\bibinfo {title} {{Unconventional Fermi Surface Instabilities in the Kagome Hubbard Model}},\ }\href {https://doi.org/10.1103/physrevlett.110.126405} {\bibfield  {journal} {\bibinfo  {journal} {Physical Review Letters}\ }\textbf {\bibinfo {volume} {110}},\ \bibinfo {pages} {126405} (\bibinfo {year} {2013}{\natexlab{b}})}\BibitemShut {NoStop}%
\bibitem [{\citenamefont {Ortiz}\ \emph {et~al.}(2019)\citenamefont {Ortiz}, \citenamefont {Gomes}, \citenamefont {Morey}, \citenamefont {Winiarski}, \citenamefont {Bordelon}, \citenamefont {Mangum}, \citenamefont {Oswald}, \citenamefont {Rodriguez-Rivera}, \citenamefont {Neilson}, \citenamefont {Wilson}, \citenamefont {Ertekin}, \citenamefont {McQueen},\ and\ \citenamefont {Toberer}}]{ortiz2019new}%
  \BibitemOpen
  \bibfield  {author} {\bibinfo {author} {\bibfnamefont {B.~R.}\ \bibnamefont {Ortiz}}, \bibinfo {author} {\bibfnamefont {L.~C.}\ \bibnamefont {Gomes}}, \bibinfo {author} {\bibfnamefont {J.~R.}\ \bibnamefont {Morey}}, \bibinfo {author} {\bibfnamefont {M.}~\bibnamefont {Winiarski}}, \bibinfo {author} {\bibfnamefont {M.}~\bibnamefont {Bordelon}}, \bibinfo {author} {\bibfnamefont {J.~S.}\ \bibnamefont {Mangum}}, \bibinfo {author} {\bibfnamefont {I.~W.~H.}\ \bibnamefont {Oswald}}, \bibinfo {author} {\bibfnamefont {J.~A.}\ \bibnamefont {Rodriguez-Rivera}}, \bibinfo {author} {\bibfnamefont {J.~R.}\ \bibnamefont {Neilson}}, \bibinfo {author} {\bibfnamefont {S.~D.}\ \bibnamefont {Wilson}}, \bibinfo {author} {\bibfnamefont {E.}~\bibnamefont {Ertekin}}, \bibinfo {author} {\bibfnamefont {T.~M.}\ \bibnamefont {McQueen}},\ and\ \bibinfo {author} {\bibfnamefont {E.~S.}\ \bibnamefont {Toberer}},\ }\bibfield  {title} {\bibinfo {title} {{New kagome prototype materials: discovery of KV$_3$Sb$_5$, RbV$_3$Sb$_5$, and
  CsV$_3$Sb$_5$}},\ }\href {https://doi.org/10.1103/PhysRevMaterials.3.094407} {\bibfield  {journal} {\bibinfo  {journal} {Phys. Rev. Materials}\ }\textbf {\bibinfo {volume} {3}},\ \bibinfo {pages} {094407} (\bibinfo {year} {2019})}\BibitemShut {NoStop}%
\bibitem [{\citenamefont {Ortiz}\ \emph {et~al.}(2020{\natexlab{a}})\citenamefont {Ortiz}, \citenamefont {Teicher}, \citenamefont {Hu}, \citenamefont {Zuo}, \citenamefont {Sarte}, \citenamefont {Schueller}, \citenamefont {Abeykoon}, \citenamefont {Krogstad}, \citenamefont {Rosenkranz}, \citenamefont {Osborn}, \citenamefont {Seshadri}, \citenamefont {Balents}, \citenamefont {He},\ and\ \citenamefont {Wilson}}]{ortizCsV3Sb5}%
  \BibitemOpen
  \bibfield  {author} {\bibinfo {author} {\bibfnamefont {B.~R.}\ \bibnamefont {Ortiz}}, \bibinfo {author} {\bibfnamefont {S.~M.}\ \bibnamefont {Teicher}}, \bibinfo {author} {\bibfnamefont {Y.}~\bibnamefont {Hu}}, \bibinfo {author} {\bibfnamefont {J.~L.}\ \bibnamefont {Zuo}}, \bibinfo {author} {\bibfnamefont {P.~M.}\ \bibnamefont {Sarte}}, \bibinfo {author} {\bibfnamefont {E.~C.}\ \bibnamefont {Schueller}}, \bibinfo {author} {\bibfnamefont {A.~M.}\ \bibnamefont {Abeykoon}}, \bibinfo {author} {\bibfnamefont {M.~J.}\ \bibnamefont {Krogstad}}, \bibinfo {author} {\bibfnamefont {S.}~\bibnamefont {Rosenkranz}}, \bibinfo {author} {\bibfnamefont {R.}~\bibnamefont {Osborn}}, \bibinfo {author} {\bibfnamefont {R.}~\bibnamefont {Seshadri}}, \bibinfo {author} {\bibfnamefont {L.}~\bibnamefont {Balents}}, \bibinfo {author} {\bibfnamefont {J.}~\bibnamefont {He}},\ and\ \bibinfo {author} {\bibfnamefont {S.~D.}\ \bibnamefont {Wilson}},\ }\bibfield  {title} {\bibinfo {title} {{CsV$_3$Sb$_5$: a $\mathbb{Z}_2$ topological kagome
  metal with a superconducting ground state}},\ }\href {https://doi.org/10.1103/PhysRevLett.125.247002} {\bibfield  {journal} {\bibinfo  {journal} {Phys. Rev. Lett.}\ }\textbf {\bibinfo {volume} {125}},\ \bibinfo {pages} {247002} (\bibinfo {year} {2020}{\natexlab{a}})}\BibitemShut {NoStop}%
\bibitem [{\citenamefont {Ortiz}\ \emph {et~al.}(2020{\natexlab{b}})\citenamefont {Ortiz}, \citenamefont {Kenney}, \citenamefont {Sarte}, \citenamefont {Teicher}, \citenamefont {Seshadri}, \citenamefont {Graf},\ and\ \citenamefont {Wilson}}]{ortiz2020KV3Sb5}%
  \BibitemOpen
  \bibfield  {author} {\bibinfo {author} {\bibfnamefont {B.~R.}\ \bibnamefont {Ortiz}}, \bibinfo {author} {\bibfnamefont {E.}~\bibnamefont {Kenney}}, \bibinfo {author} {\bibfnamefont {P.~M.}\ \bibnamefont {Sarte}}, \bibinfo {author} {\bibfnamefont {S.~M.}\ \bibnamefont {Teicher}}, \bibinfo {author} {\bibfnamefont {R.}~\bibnamefont {Seshadri}}, \bibinfo {author} {\bibfnamefont {M.~J.}\ \bibnamefont {Graf}},\ and\ \bibinfo {author} {\bibfnamefont {S.~D.}\ \bibnamefont {Wilson}},\ }\bibfield  {title} {\bibinfo {title} {{Superconductivity in the $\mathbb{Z}_2$ kagome metal KV$_3$Sb$_5$}},\ }\href {https://doi.org/10.1103/PhysRevMaterials.5.034801} {\bibfield  {journal} {\bibinfo  {journal} {Phys. Rev. Mater.}\ }\textbf {\bibinfo {volume} {5}},\ \bibinfo {pages} {034801} (\bibinfo {year} {2020}{\natexlab{b}})}\BibitemShut {NoStop}%
\bibitem [{\citenamefont {Yin}\ \emph {et~al.}(2021)\citenamefont {Yin}, \citenamefont {Tu}, \citenamefont {Gong}, \citenamefont {Fu}, \citenamefont {Yan},\ and\ \citenamefont {Lei}}]{RbV3Sb5SC}%
  \BibitemOpen
  \bibfield  {author} {\bibinfo {author} {\bibfnamefont {Q.}~\bibnamefont {Yin}}, \bibinfo {author} {\bibfnamefont {Z.}~\bibnamefont {Tu}}, \bibinfo {author} {\bibfnamefont {C.}~\bibnamefont {Gong}}, \bibinfo {author} {\bibfnamefont {Y.}~\bibnamefont {Fu}}, \bibinfo {author} {\bibfnamefont {S.}~\bibnamefont {Yan}},\ and\ \bibinfo {author} {\bibfnamefont {H.}~\bibnamefont {Lei}},\ }\bibfield  {title} {\bibinfo {title} {{Superconductivity and normal-state properties of kagome metal RbV$_3$Sb$_5$ single crystals}},\ }\href {https://doi.org/10.1088/0256-307X/38/3/037403} {\bibfield  {journal} {\bibinfo  {journal} {Chin. Phys. Lett.}\ }\textbf {\bibinfo {volume} {38}},\ \bibinfo {pages} {037403} (\bibinfo {year} {2021})}\BibitemShut {NoStop}%
\bibitem [{\citenamefont {Werhahn}\ \emph {et~al.}(2022)\citenamefont {Werhahn}, \citenamefont {Ortiz}, \citenamefont {Hay}, \citenamefont {Wilson}, \citenamefont {Seshadri},\ and\ \citenamefont {Johrendt}}]{werhahn2022kagome}%
  \BibitemOpen
  \bibfield  {author} {\bibinfo {author} {\bibfnamefont {D.}~\bibnamefont {Werhahn}}, \bibinfo {author} {\bibfnamefont {B.~R.}\ \bibnamefont {Ortiz}}, \bibinfo {author} {\bibfnamefont {A.~K.}\ \bibnamefont {Hay}}, \bibinfo {author} {\bibfnamefont {S.~D.}\ \bibnamefont {Wilson}}, \bibinfo {author} {\bibfnamefont {R.}~\bibnamefont {Seshadri}},\ and\ \bibinfo {author} {\bibfnamefont {D.}~\bibnamefont {Johrendt}},\ }\bibfield  {title} {\bibinfo {title} {{The kagom{\'e} metals RbTi$_3$Bi$_5$ and CsTi$_3$Bi$_5$}},\ }\href {https://doi.org/10.1515/znb-2022-0125} {\bibfield  {journal} {\bibinfo  {journal} {Z. Naturforsch. B}\ }\textbf {\bibinfo {volume} {77}},\ \bibinfo {pages} {757} (\bibinfo {year} {2022})}\BibitemShut {NoStop}%
\bibitem [{\citenamefont {Liu}\ \emph {et~al.}(2024{\natexlab{a}})\citenamefont {Liu}, \citenamefont {Liu}, \citenamefont {Bao}, \citenamefont {Yang}, \citenamefont {Ji}, \citenamefont {Wu}, \citenamefont {Shen}, \citenamefont {Luo}, \citenamefont {Yang}, \citenamefont {Liu}, \citenamefont {Xu}, \citenamefont {Yang}, \citenamefont {Chai}, \citenamefont {Lu}, \citenamefont {Liu}, \citenamefont {Wang}, \citenamefont {Jiang}, \citenamefont {Tao}, \citenamefont {Ren}, \citenamefont {Xu}, \citenamefont {Cao}, \citenamefont {Xu}, \citenamefont {Zhou}, \citenamefont {Cheng},\ and\ \citenamefont {Cao}}]{liu2024superconductivity}%
  \BibitemOpen
  \bibfield  {author} {\bibinfo {author} {\bibfnamefont {Y.}~\bibnamefont {Liu}}, \bibinfo {author} {\bibfnamefont {Z.-Y.}\ \bibnamefont {Liu}}, \bibinfo {author} {\bibfnamefont {J.-K.}\ \bibnamefont {Bao}}, \bibinfo {author} {\bibfnamefont {P.-T.}\ \bibnamefont {Yang}}, \bibinfo {author} {\bibfnamefont {L.-W.}\ \bibnamefont {Ji}}, \bibinfo {author} {\bibfnamefont {S.-Q.}\ \bibnamefont {Wu}}, \bibinfo {author} {\bibfnamefont {Q.-X.}\ \bibnamefont {Shen}}, \bibinfo {author} {\bibfnamefont {J.}~\bibnamefont {Luo}}, \bibinfo {author} {\bibfnamefont {J.}~\bibnamefont {Yang}}, \bibinfo {author} {\bibfnamefont {J.-Y.}\ \bibnamefont {Liu}}, \bibinfo {author} {\bibfnamefont {C.-C.}\ \bibnamefont {Xu}}, \bibinfo {author} {\bibfnamefont {W.-Z.}\ \bibnamefont {Yang}}, \bibinfo {author} {\bibfnamefont {W.-L.}\ \bibnamefont {Chai}}, \bibinfo {author} {\bibfnamefont {J.-Y.}\ \bibnamefont {Lu}}, \bibinfo {author} {\bibfnamefont {C.-C.}\ \bibnamefont {Liu}}, \bibinfo {author} {\bibfnamefont {B.-S.}\ \bibnamefont {Wang}},
  \bibinfo {author} {\bibfnamefont {H.}~\bibnamefont {Jiang}}, \bibinfo {author} {\bibfnamefont {Q.}~\bibnamefont {Tao}}, \bibinfo {author} {\bibfnamefont {Z.}~\bibnamefont {Ren}}, \bibinfo {author} {\bibfnamefont {X.-F.}\ \bibnamefont {Xu}}, \bibinfo {author} {\bibfnamefont {C.}~\bibnamefont {Cao}}, \bibinfo {author} {\bibfnamefont {Z.-A.}\ \bibnamefont {Xu}}, \bibinfo {author} {\bibfnamefont {R.}~\bibnamefont {Zhou}}, \bibinfo {author} {\bibfnamefont {J.-G.}\ \bibnamefont {Cheng}},\ and\ \bibinfo {author} {\bibfnamefont {G.-H.}\ \bibnamefont {Cao}},\ }\bibfield  {title} {\bibinfo {title} {Superconductivity under pressure in a chromium-based kagome metal},\ }\href {https://doi.org/10.1038/s41586-024-07761-x} {\bibfield  {journal} {\bibinfo  {journal} {Nature}\ }\textbf {\bibinfo {volume} {632}},\ \bibinfo {pages} {1032} (\bibinfo {year} {2024}{\natexlab{a}})}\BibitemShut {NoStop}%
\bibitem [{\citenamefont {Teng}\ \emph {et~al.}(2022)\citenamefont {Teng}, \citenamefont {Chen}, \citenamefont {Ye}, \citenamefont {Rosenberg}, \citenamefont {Liu}, \citenamefont {Yin}, \citenamefont {Jiang}, \citenamefont {Oh}, \citenamefont {Hasan}, \citenamefont {Neubauer}, \citenamefont {Gao}, \citenamefont {Xie}, \citenamefont {Hashimoto}, \citenamefont {Lu}, \citenamefont {Jozwiak}, \citenamefont {Bostwick}, \citenamefont {Rotenberg}, \citenamefont {Birgeneau}, \citenamefont {Chu}, \citenamefont {Yi},\ and\ \citenamefont {Dai}}]{teng2022discovery}%
  \BibitemOpen
  \bibfield  {author} {\bibinfo {author} {\bibfnamefont {X.}~\bibnamefont {Teng}}, \bibinfo {author} {\bibfnamefont {L.}~\bibnamefont {Chen}}, \bibinfo {author} {\bibfnamefont {F.}~\bibnamefont {Ye}}, \bibinfo {author} {\bibfnamefont {E.}~\bibnamefont {Rosenberg}}, \bibinfo {author} {\bibfnamefont {Z.}~\bibnamefont {Liu}}, \bibinfo {author} {\bibfnamefont {J.-X.}\ \bibnamefont {Yin}}, \bibinfo {author} {\bibfnamefont {Y.-X.}\ \bibnamefont {Jiang}}, \bibinfo {author} {\bibfnamefont {J.~S.}\ \bibnamefont {Oh}}, \bibinfo {author} {\bibfnamefont {M.~Z.}\ \bibnamefont {Hasan}}, \bibinfo {author} {\bibfnamefont {K.~J.}\ \bibnamefont {Neubauer}}, \bibinfo {author} {\bibfnamefont {B.}~\bibnamefont {Gao}}, \bibinfo {author} {\bibfnamefont {Y.}~\bibnamefont {Xie}}, \bibinfo {author} {\bibfnamefont {M.}~\bibnamefont {Hashimoto}}, \bibinfo {author} {\bibfnamefont {D.}~\bibnamefont {Lu}}, \bibinfo {author} {\bibfnamefont {C.}~\bibnamefont {Jozwiak}}, \bibinfo {author} {\bibfnamefont {A.}~\bibnamefont {Bostwick}}, \bibinfo
  {author} {\bibfnamefont {E.}~\bibnamefont {Rotenberg}}, \bibinfo {author} {\bibfnamefont {R.~J.}\ \bibnamefont {Birgeneau}}, \bibinfo {author} {\bibfnamefont {J.-H.}\ \bibnamefont {Chu}}, \bibinfo {author} {\bibfnamefont {M.}~\bibnamefont {Yi}},\ and\ \bibinfo {author} {\bibfnamefont {P.}~\bibnamefont {Dai}},\ }\bibfield  {title} {\bibinfo {title} {Discovery of charge density wave in a kagome lattice antiferromagnet},\ }\href {https://doi.org/10.1038/s41586-022-05034-z} {\bibfield  {journal} {\bibinfo  {journal} {Nature}\ }\textbf {\bibinfo {volume} {609}},\ \bibinfo {pages} {490} (\bibinfo {year} {2022})}\BibitemShut {NoStop}%
\bibitem [{\citenamefont {Arachchige}\ \emph {et~al.}(2022)\citenamefont {Arachchige}, \citenamefont {Meier}, \citenamefont {Marshall}, \citenamefont {Matsuoka}, \citenamefont {Xue}, \citenamefont {McGuire}, \citenamefont {Hermann}, \citenamefont {Cao},\ and\ \citenamefont {Mandrus}}]{arachchige2022ScV6Sn6}%
  \BibitemOpen
  \bibfield  {author} {\bibinfo {author} {\bibfnamefont {H.~W.~S.}\ \bibnamefont {Arachchige}}, \bibinfo {author} {\bibfnamefont {W.~R.}\ \bibnamefont {Meier}}, \bibinfo {author} {\bibfnamefont {M.}~\bibnamefont {Marshall}}, \bibinfo {author} {\bibfnamefont {T.}~\bibnamefont {Matsuoka}}, \bibinfo {author} {\bibfnamefont {R.}~\bibnamefont {Xue}}, \bibinfo {author} {\bibfnamefont {M.~A.}\ \bibnamefont {McGuire}}, \bibinfo {author} {\bibfnamefont {R.~P.}\ \bibnamefont {Hermann}}, \bibinfo {author} {\bibfnamefont {H.}~\bibnamefont {Cao}},\ and\ \bibinfo {author} {\bibfnamefont {D.}~\bibnamefont {Mandrus}},\ }\bibfield  {title} {\bibinfo {title} {{Charge Density Wave in Kagome Lattice Intermetallic ScV$_6$Sn$_6$}},\ }\href {https://doi.org/10.1103/PhysRevLett.129.216402} {\bibfield  {journal} {\bibinfo  {journal} {Physical Review Letters}\ }\textbf {\bibinfo {volume} {129}},\ \bibinfo {pages} {216402} (\bibinfo {year} {2022})}\BibitemShut {NoStop}%
\bibitem [{\citenamefont {Hu}\ \emph {et~al.}(2024)\citenamefont {Hu}, \citenamefont {Ma}, \citenamefont {Li}, \citenamefont {Jiang}, \citenamefont {Gawryluk}, \citenamefont {Hu}, \citenamefont {Teyssier}, \citenamefont {Multian}, \citenamefont {Yin}, \citenamefont {Xu}, \citenamefont {Shin}, \citenamefont {Plokhikh}, \citenamefont {Han}, \citenamefont {Plumb}, \citenamefont {Liu}, \citenamefont {Yin}, \citenamefont {Guguchia}, \citenamefont {Zhao}, \citenamefont {Schnyder}, \citenamefont {Wu}, \citenamefont {Pomjakushina}, \citenamefont {Hasan}, \citenamefont {Wang},\ and\ \citenamefont {Shi}}]{Hu2024_PhononPromotedCDW-ScV6Sn6}%
  \BibitemOpen
  \bibfield  {author} {\bibinfo {author} {\bibfnamefont {Y.}~\bibnamefont {Hu}}, \bibinfo {author} {\bibfnamefont {J.}~\bibnamefont {Ma}}, \bibinfo {author} {\bibfnamefont {Y.}~\bibnamefont {Li}}, \bibinfo {author} {\bibfnamefont {Y.}~\bibnamefont {Jiang}}, \bibinfo {author} {\bibfnamefont {D.~J.}\ \bibnamefont {Gawryluk}}, \bibinfo {author} {\bibfnamefont {T.}~\bibnamefont {Hu}}, \bibinfo {author} {\bibfnamefont {J.}~\bibnamefont {Teyssier}}, \bibinfo {author} {\bibfnamefont {V.}~\bibnamefont {Multian}}, \bibinfo {author} {\bibfnamefont {Z.}~\bibnamefont {Yin}}, \bibinfo {author} {\bibfnamefont {S.}~\bibnamefont {Xu}}, \bibinfo {author} {\bibfnamefont {S.}~\bibnamefont {Shin}}, \bibinfo {author} {\bibfnamefont {I.}~\bibnamefont {Plokhikh}}, \bibinfo {author} {\bibfnamefont {X.}~\bibnamefont {Han}}, \bibinfo {author} {\bibfnamefont {N.~C.}\ \bibnamefont {Plumb}}, \bibinfo {author} {\bibfnamefont {Y.}~\bibnamefont {Liu}}, \bibinfo {author} {\bibfnamefont {J.-X.}\ \bibnamefont {Yin}}, \bibinfo {author}
  {\bibfnamefont {Z.}~\bibnamefont {Guguchia}}, \bibinfo {author} {\bibfnamefont {Y.}~\bibnamefont {Zhao}}, \bibinfo {author} {\bibfnamefont {A.~P.}\ \bibnamefont {Schnyder}}, \bibinfo {author} {\bibfnamefont {X.}~\bibnamefont {Wu}}, \bibinfo {author} {\bibfnamefont {E.}~\bibnamefont {Pomjakushina}}, \bibinfo {author} {\bibfnamefont {M.~Z.}\ \bibnamefont {Hasan}}, \bibinfo {author} {\bibfnamefont {N.}~\bibnamefont {Wang}},\ and\ \bibinfo {author} {\bibfnamefont {M.}~\bibnamefont {Shi}},\ }\bibfield  {title} {\bibinfo {title} {{Phonon promoted charge density wave in topological kagome metal ScV$_6$Sn$_6$}},\ }\href {https://doi.org/10.1038/s41467-024-45859-y} {\bibfield  {journal} {\bibinfo  {journal} {Nature Communications}\ }\textbf {\bibinfo {volume} {15}},\ \bibinfo {pages} {1658} (\bibinfo {year} {2024})}\BibitemShut {NoStop}%
\bibitem [{\citenamefont {Meier}\ \emph {et~al.}(2023)\citenamefont {Meier}, \citenamefont {Madhogaria}, \citenamefont {Mozaffari}, \citenamefont {Marshall}, \citenamefont {Graf}, \citenamefont {McGuire}, \citenamefont {Arachchige}, \citenamefont {Allen}, \citenamefont {Driver}, \citenamefont {Cao} \emph {et~al.}}]{Meier2023tiny}%
  \BibitemOpen
  \bibfield  {author} {\bibinfo {author} {\bibfnamefont {W.~R.}\ \bibnamefont {Meier}}, \bibinfo {author} {\bibfnamefont {R.~P.}\ \bibnamefont {Madhogaria}}, \bibinfo {author} {\bibfnamefont {S.}~\bibnamefont {Mozaffari}}, \bibinfo {author} {\bibfnamefont {M.}~\bibnamefont {Marshall}}, \bibinfo {author} {\bibfnamefont {D.~E.}\ \bibnamefont {Graf}}, \bibinfo {author} {\bibfnamefont {M.~A.}\ \bibnamefont {McGuire}}, \bibinfo {author} {\bibfnamefont {H.~W.~S.}\ \bibnamefont {Arachchige}}, \bibinfo {author} {\bibfnamefont {C.~L.}\ \bibnamefont {Allen}}, \bibinfo {author} {\bibfnamefont {J.}~\bibnamefont {Driver}}, \bibinfo {author} {\bibfnamefont {H.}~\bibnamefont {Cao}}, \emph {et~al.},\ }\bibfield  {title} {\bibinfo {title} {{Tiny Sc allows the chains to rattle: impact of Lu and Y doping on the charge-density wave in ScV$_6$Sn$_6$}},\ }\href {https://doi.org/10.1021/jacs.3c06394} {\bibfield  {journal} {\bibinfo  {journal} {Journal of the American Chemical Society}\ }\textbf {\bibinfo {volume} {145}},\ \bibinfo
  {pages} {20943} (\bibinfo {year} {2023})}\BibitemShut {NoStop}%
\bibitem [{\citenamefont {Pokharel}\ \emph {et~al.}(2023)\citenamefont {Pokharel}, \citenamefont {Ortiz}, \citenamefont {Kautzsch}, \citenamefont {Gomez~Alvarado}, \citenamefont {Mallayya}, \citenamefont {Wu}, \citenamefont {Kim}, \citenamefont {Ruff}, \citenamefont {Sarker},\ and\ \citenamefont {Wilson}}]{Pokharel2023_FrustratedCO+CooperativeDistortionsScV6Sn6}%
  \BibitemOpen
  \bibfield  {author} {\bibinfo {author} {\bibfnamefont {G.}~\bibnamefont {Pokharel}}, \bibinfo {author} {\bibfnamefont {B.~R.}\ \bibnamefont {Ortiz}}, \bibinfo {author} {\bibfnamefont {L.}~\bibnamefont {Kautzsch}}, \bibinfo {author} {\bibfnamefont {S.~J.}\ \bibnamefont {Gomez~Alvarado}}, \bibinfo {author} {\bibfnamefont {K.}~\bibnamefont {Mallayya}}, \bibinfo {author} {\bibfnamefont {G.}~\bibnamefont {Wu}}, \bibinfo {author} {\bibfnamefont {E.-A.}\ \bibnamefont {Kim}}, \bibinfo {author} {\bibfnamefont {J.~P.~C.}\ \bibnamefont {Ruff}}, \bibinfo {author} {\bibfnamefont {S.}~\bibnamefont {Sarker}},\ and\ \bibinfo {author} {\bibfnamefont {S.~D.}\ \bibnamefont {Wilson}},\ }\bibfield  {title} {\bibinfo {title} {{Frustrated charge order and cooperative distortions in ScV$_6$Sn$_6$}},\ }\href {https://doi.org/10.1103/physrevmaterials.7.104201} {\bibfield  {journal} {\bibinfo  {journal} {Physical Review Materials}\ }\textbf {\bibinfo {volume} {7}},\ \bibinfo {pages} {104201} (\bibinfo {year} {2023})}\BibitemShut
  {NoStop}%
\bibitem [{\citenamefont {Cao}\ \emph {et~al.}(2023)\citenamefont {Cao}, \citenamefont {Xu}, \citenamefont {Fukui}, \citenamefont {Manjo}, \citenamefont {Dong}, \citenamefont {Shi}, \citenamefont {Liu}, \citenamefont {Cao},\ and\ \citenamefont {Song}}]{Cao2023_CompetingChargeOrderScV6Sn6}%
  \BibitemOpen
  \bibfield  {author} {\bibinfo {author} {\bibfnamefont {S.}~\bibnamefont {Cao}}, \bibinfo {author} {\bibfnamefont {C.}~\bibnamefont {Xu}}, \bibinfo {author} {\bibfnamefont {H.}~\bibnamefont {Fukui}}, \bibinfo {author} {\bibfnamefont {T.}~\bibnamefont {Manjo}}, \bibinfo {author} {\bibfnamefont {Y.}~\bibnamefont {Dong}}, \bibinfo {author} {\bibfnamefont {M.}~\bibnamefont {Shi}}, \bibinfo {author} {\bibfnamefont {Y.}~\bibnamefont {Liu}}, \bibinfo {author} {\bibfnamefont {C.}~\bibnamefont {Cao}},\ and\ \bibinfo {author} {\bibfnamefont {Y.}~\bibnamefont {Song}},\ }\bibfield  {title} {\bibinfo {title} {{Competing charge-density wave instabilities in the kagome metal ScV$_6$Sn$_6$}},\ }\href {https://doi.org/10.1038/s41467-023-43454-1} {\bibfield  {journal} {\bibinfo  {journal} {Nature Communications}\ }\textbf {\bibinfo {volume} {14}},\ \bibinfo {pages} {7671} (\bibinfo {year} {2023})}\BibitemShut {NoStop}%
\bibitem [{\citenamefont {Lee}\ \emph {et~al.}(2024)\citenamefont {Lee}, \citenamefont {Won}, \citenamefont {Kim}, \citenamefont {Yoo}, \citenamefont {Park}, \citenamefont {Denlinger}, \citenamefont {Jozwiak}, \citenamefont {Bostwick}, \citenamefont {Rotenberg}, \citenamefont {Comin}, \citenamefont {Kang},\ and\ \citenamefont {Park}}]{Lee2024_NatureCDWScV6Sn6}%
  \BibitemOpen
  \bibfield  {author} {\bibinfo {author} {\bibfnamefont {S.}~\bibnamefont {Lee}}, \bibinfo {author} {\bibfnamefont {C.}~\bibnamefont {Won}}, \bibinfo {author} {\bibfnamefont {J.}~\bibnamefont {Kim}}, \bibinfo {author} {\bibfnamefont {J.}~\bibnamefont {Yoo}}, \bibinfo {author} {\bibfnamefont {S.}~\bibnamefont {Park}}, \bibinfo {author} {\bibfnamefont {J.}~\bibnamefont {Denlinger}}, \bibinfo {author} {\bibfnamefont {C.}~\bibnamefont {Jozwiak}}, \bibinfo {author} {\bibfnamefont {A.}~\bibnamefont {Bostwick}}, \bibinfo {author} {\bibfnamefont {E.}~\bibnamefont {Rotenberg}}, \bibinfo {author} {\bibfnamefont {R.}~\bibnamefont {Comin}}, \bibinfo {author} {\bibfnamefont {M.}~\bibnamefont {Kang}},\ and\ \bibinfo {author} {\bibfnamefont {J.-H.}\ \bibnamefont {Park}},\ }\bibfield  {title} {\bibinfo {title} {Nature of charge density wave in kagome metal scv6sn6},\ }\href {https://doi.org/10.1038/s41535-024-00620-y} {\bibfield  {journal} {\bibinfo  {journal} {npj Quantum Materials}\ }\textbf {\bibinfo {volume} {9}},\
  \bibinfo {pages} {15} (\bibinfo {year} {2024})}\BibitemShut {NoStop}%
\bibitem [{\citenamefont {Korshunov}\ \emph {et~al.}(2023)\citenamefont {Korshunov}, \citenamefont {Hu}, \citenamefont {Subires}, \citenamefont {Jiang}, \citenamefont {Călugăru}, \citenamefont {Feng}, \citenamefont {Rajapitamahuni}, \citenamefont {Yi}, \citenamefont {Roychowdhury}, \citenamefont {Vergniory}, \citenamefont {Strempfer}, \citenamefont {Shekhar}, \citenamefont {Vescovo}, \citenamefont {Chernyshov}, \citenamefont {Said}, \citenamefont {Bosak}, \citenamefont {Felser}, \citenamefont {Bernevig},\ and\ \citenamefont {Blanco-Canosa}}]{Korshunov2023_SofteningPhononScV6Sn6}%
  \BibitemOpen
  \bibfield  {author} {\bibinfo {author} {\bibfnamefont {A.}~\bibnamefont {Korshunov}}, \bibinfo {author} {\bibfnamefont {H.}~\bibnamefont {Hu}}, \bibinfo {author} {\bibfnamefont {D.}~\bibnamefont {Subires}}, \bibinfo {author} {\bibfnamefont {Y.}~\bibnamefont {Jiang}}, \bibinfo {author} {\bibfnamefont {D.}~\bibnamefont {Călugăru}}, \bibinfo {author} {\bibfnamefont {X.}~\bibnamefont {Feng}}, \bibinfo {author} {\bibfnamefont {A.}~\bibnamefont {Rajapitamahuni}}, \bibinfo {author} {\bibfnamefont {C.}~\bibnamefont {Yi}}, \bibinfo {author} {\bibfnamefont {S.}~\bibnamefont {Roychowdhury}}, \bibinfo {author} {\bibfnamefont {M.~G.}\ \bibnamefont {Vergniory}}, \bibinfo {author} {\bibfnamefont {J.}~\bibnamefont {Strempfer}}, \bibinfo {author} {\bibfnamefont {C.}~\bibnamefont {Shekhar}}, \bibinfo {author} {\bibfnamefont {E.}~\bibnamefont {Vescovo}}, \bibinfo {author} {\bibfnamefont {D.}~\bibnamefont {Chernyshov}}, \bibinfo {author} {\bibfnamefont {A.~H.}\ \bibnamefont {Said}}, \bibinfo {author} {\bibfnamefont
  {A.}~\bibnamefont {Bosak}}, \bibinfo {author} {\bibfnamefont {C.}~\bibnamefont {Felser}}, \bibinfo {author} {\bibfnamefont {B.~A.}\ \bibnamefont {Bernevig}},\ and\ \bibinfo {author} {\bibfnamefont {S.}~\bibnamefont {Blanco-Canosa}},\ }\bibfield  {title} {\bibinfo {title} {{Softening of a flat phonon mode in the kagome ScV$_6$Sn$_6$}},\ }\href {https://doi.org/10.1038/s41467-023-42186-6} {\bibfield  {journal} {\bibinfo  {journal} {Nature Communications}\ }\textbf {\bibinfo {volume} {14}},\ \bibinfo {pages} {6646} (\bibinfo {year} {2023})}\BibitemShut {NoStop}%
\bibitem [{\citenamefont {Hu}\ \emph {et~al.}(2023)\citenamefont {Hu}, \citenamefont {Jiang}, \citenamefont {Călugăru}, \citenamefont {Feng}, \citenamefont {Subires}, \citenamefont {Vergniory}, \citenamefont {Felser}, \citenamefont {Blanco-Canosa},\ and\ \citenamefont {Bernevig}}]{Hu2023_ScV6Sn6-Theory-FlatPhonons+UnconventionalCDW}%
  \BibitemOpen
  \bibfield  {author} {\bibinfo {author} {\bibfnamefont {H.}~\bibnamefont {Hu}}, \bibinfo {author} {\bibfnamefont {Y.}~\bibnamefont {Jiang}}, \bibinfo {author} {\bibfnamefont {D.}~\bibnamefont {Călugăru}}, \bibinfo {author} {\bibfnamefont {X.}~\bibnamefont {Feng}}, \bibinfo {author} {\bibfnamefont {D.}~\bibnamefont {Subires}}, \bibinfo {author} {\bibfnamefont {M.~G.}\ \bibnamefont {Vergniory}}, \bibinfo {author} {\bibfnamefont {C.}~\bibnamefont {Felser}}, \bibinfo {author} {\bibfnamefont {S.}~\bibnamefont {Blanco-Canosa}},\ and\ \bibinfo {author} {\bibfnamefont {B.~A.}\ \bibnamefont {Bernevig}},\ }\href {https://doi.org/10.48550/ARXIV.2305.15469} {\bibinfo {title} {Kagome materials i: Sg 191, scv$_6$sn$_6$. flat phonon soft modes and unconventional cdw formation: Microscopic and effective theory}} (\bibinfo {year} {2023})\BibitemShut {NoStop}%
\bibitem [{\citenamefont {Liu}\ \emph {et~al.}(2024{\natexlab{b}})\citenamefont {Liu}, \citenamefont {Wang}, \citenamefont {Yao}, \citenamefont {Jia}, \citenamefont {Zhang},\ and\ \citenamefont {Cho}}]{Liu2024_DrivingMechanismScV6Sn6}%
  \BibitemOpen
  \bibfield  {author} {\bibinfo {author} {\bibfnamefont {S.}~\bibnamefont {Liu}}, \bibinfo {author} {\bibfnamefont {C.}~\bibnamefont {Wang}}, \bibinfo {author} {\bibfnamefont {S.}~\bibnamefont {Yao}}, \bibinfo {author} {\bibfnamefont {Y.}~\bibnamefont {Jia}}, \bibinfo {author} {\bibfnamefont {Z.}~\bibnamefont {Zhang}},\ and\ \bibinfo {author} {\bibfnamefont {J.-H.}\ \bibnamefont {Cho}},\ }\bibfield  {title} {\bibinfo {title} {Driving mechanism and dynamic fluctuations of charge density waves in the kagome metal scv6sn6},\ }\href {https://doi.org/10.1103/physrevb.109.l121103} {\bibfield  {journal} {\bibinfo  {journal} {Physical Review B}\ }\textbf {\bibinfo {volume} {109}},\ \bibinfo {pages} {l121103} (\bibinfo {year} {2024}{\natexlab{b}})}\BibitemShut {NoStop}%
\bibitem [{\citenamefont {Yu}\ \emph {et~al.}(2024)\citenamefont {Yu}, \citenamefont {Lai}, \citenamefont {Liu}, \citenamefont {Liu}, \citenamefont {Chen},\ and\ \citenamefont {Sun}}]{Yu2024_MagAndCorrelationsScV6Sn6}%
  \BibitemOpen
  \bibfield  {author} {\bibinfo {author} {\bibfnamefont {T.}~\bibnamefont {Yu}}, \bibinfo {author} {\bibfnamefont {J.}~\bibnamefont {Lai}}, \bibinfo {author} {\bibfnamefont {X.}~\bibnamefont {Liu}}, \bibinfo {author} {\bibfnamefont {P.}~\bibnamefont {Liu}}, \bibinfo {author} {\bibfnamefont {X.-Q.}\ \bibnamefont {Chen}},\ and\ \bibinfo {author} {\bibfnamefont {Y.}~\bibnamefont {Sun}},\ }\bibfield  {title} {\bibinfo {title} {{Magnetism and weak electronic correlations in the kagome metal ScV$_6$Sn$_6$}},\ }\href {https://doi.org/10.1103/physrevb.109.195145} {\bibfield  {journal} {\bibinfo  {journal} {Physical Review B}\ }\textbf {\bibinfo {volume} {109}},\ \bibinfo {pages} {195145} (\bibinfo {year} {2024})}\BibitemShut {NoStop}%
\bibitem [{\citenamefont {Wang}(2023)}]{Wang2023_EnhancedSpinPolarizationViaDimerizationFeGe}%
  \BibitemOpen
  \bibfield  {author} {\bibinfo {author} {\bibfnamefont {Y.}~\bibnamefont {Wang}},\ }\bibfield  {title} {\bibinfo {title} {Enhanced spin-polarization via partial ge-dimerization as the driving force of the charge density wave in fege},\ }\href {https://doi.org/10.1103/physrevmaterials.7.104006} {\bibfield  {journal} {\bibinfo  {journal} {Physical Review Materials}\ }\textbf {\bibinfo {volume} {7}},\ \bibinfo {pages} {104006} (\bibinfo {year} {2023})}\BibitemShut {NoStop}%
\bibitem [{\citenamefont {Wen}\ \emph {et~al.}(2024)\citenamefont {Wen}, \citenamefont {Zhang}, \citenamefont {Li}, \citenamefont {Gui}, \citenamefont {Li}, \citenamefont {Li}, \citenamefont {Wu}, \citenamefont {Wang}, \citenamefont {Yang}, \citenamefont {Wang}, \citenamefont {Cheng}, \citenamefont {Wang}, \citenamefont {Ying},\ and\ \citenamefont {Chen}}]{Wen2024_UnconventionalCDW-FeGe}%
  \BibitemOpen
  \bibfield  {author} {\bibinfo {author} {\bibfnamefont {X.}~\bibnamefont {Wen}}, \bibinfo {author} {\bibfnamefont {Y.}~\bibnamefont {Zhang}}, \bibinfo {author} {\bibfnamefont {C.}~\bibnamefont {Li}}, \bibinfo {author} {\bibfnamefont {Z.}~\bibnamefont {Gui}}, \bibinfo {author} {\bibfnamefont {Y.}~\bibnamefont {Li}}, \bibinfo {author} {\bibfnamefont {Y.}~\bibnamefont {Li}}, \bibinfo {author} {\bibfnamefont {X.}~\bibnamefont {Wu}}, \bibinfo {author} {\bibfnamefont {A.}~\bibnamefont {Wang}}, \bibinfo {author} {\bibfnamefont {P.}~\bibnamefont {Yang}}, \bibinfo {author} {\bibfnamefont {B.}~\bibnamefont {Wang}}, \bibinfo {author} {\bibfnamefont {J.}~\bibnamefont {Cheng}}, \bibinfo {author} {\bibfnamefont {Y.}~\bibnamefont {Wang}}, \bibinfo {author} {\bibfnamefont {J.}~\bibnamefont {Ying}},\ and\ \bibinfo {author} {\bibfnamefont {X.}~\bibnamefont {Chen}},\ }\bibfield  {title} {\bibinfo {title} {Unconventional charge density wave in a kagome lattice antiferromagnet fege},\ }\href
  {https://doi.org/10.1103/physrevresearch.6.033222} {\bibfield  {journal} {\bibinfo  {journal} {Physical Review Research}\ }\textbf {\bibinfo {volume} {6}},\ \bibinfo {pages} {033222} (\bibinfo {year} {2024})}\BibitemShut {NoStop}%
\bibitem [{\citenamefont {Chen}\ \emph {et~al.}(2024{\natexlab{a}})\citenamefont {Chen}, \citenamefont {Wu}, \citenamefont {Zhou}, \citenamefont {Zhang}, \citenamefont {Yin}, \citenamefont {Li}, \citenamefont {Li}, \citenamefont {Gong}, \citenamefont {He}, \citenamefont {Chai}, \citenamefont {Zhou}, \citenamefont {Wang}, \citenamefont {Wang}, \citenamefont {Yan},\ and\ \citenamefont {Feng}}]{Chen2024_LongRangeGeDimerizationFeGe}%
  \BibitemOpen
  \bibfield  {author} {\bibinfo {author} {\bibfnamefont {Z.}~\bibnamefont {Chen}}, \bibinfo {author} {\bibfnamefont {X.}~\bibnamefont {Wu}}, \bibinfo {author} {\bibfnamefont {S.}~\bibnamefont {Zhou}}, \bibinfo {author} {\bibfnamefont {J.}~\bibnamefont {Zhang}}, \bibinfo {author} {\bibfnamefont {R.}~\bibnamefont {Yin}}, \bibinfo {author} {\bibfnamefont {Y.}~\bibnamefont {Li}}, \bibinfo {author} {\bibfnamefont {M.}~\bibnamefont {Li}}, \bibinfo {author} {\bibfnamefont {J.}~\bibnamefont {Gong}}, \bibinfo {author} {\bibfnamefont {M.}~\bibnamefont {He}}, \bibinfo {author} {\bibfnamefont {Y.}~\bibnamefont {Chai}}, \bibinfo {author} {\bibfnamefont {X.}~\bibnamefont {Zhou}}, \bibinfo {author} {\bibfnamefont {Y.}~\bibnamefont {Wang}}, \bibinfo {author} {\bibfnamefont {A.}~\bibnamefont {Wang}}, \bibinfo {author} {\bibfnamefont {Y.-J.}\ \bibnamefont {Yan}},\ and\ \bibinfo {author} {\bibfnamefont {D.-L.}\ \bibnamefont {Feng}},\ }\bibfield  {title} {\bibinfo {title} {Discovery of a long-ranged charge order with 1/4
  ge1-dimerization in an antiferromagnetic kagome metal},\ }\href {https://doi.org/10.1038/s41467-024-50661-x} {\bibfield  {journal} {\bibinfo  {journal} {Nature Communications}\ }\textbf {\bibinfo {volume} {15}},\ \bibinfo {pages} {6262} (\bibinfo {year} {2024}{\natexlab{a}})}\BibitemShut {NoStop}%
\bibitem [{\citenamefont {Canfield}\ \emph {et~al.}(2016)\citenamefont {Canfield}, \citenamefont {Kong}, \citenamefont {Kaluarachchi},\ and\ \citenamefont {Jo}}]{canfield2016use}%
  \BibitemOpen
  \bibfield  {author} {\bibinfo {author} {\bibfnamefont {P.~C.}\ \bibnamefont {Canfield}}, \bibinfo {author} {\bibfnamefont {T.}~\bibnamefont {Kong}}, \bibinfo {author} {\bibfnamefont {U.~S.}\ \bibnamefont {Kaluarachchi}},\ and\ \bibinfo {author} {\bibfnamefont {N.~H.}\ \bibnamefont {Jo}},\ }\bibfield  {title} {\bibinfo {title} {{Use of frit-disc crucibles for routine and exploratory solution growth of single crystalline samples}},\ }\href {https://doi.org/10.1080/14786435.2015.1122248} {\bibfield  {journal} {\bibinfo  {journal} {Philosophical magazine}\ }\textbf {\bibinfo {volume} {96}},\ \bibinfo {pages} {84} (\bibinfo {year} {2016})}\BibitemShut {NoStop}%
\bibitem [{\citenamefont {CrysAlisPRO}(2014)}]{crysalispro2014agilent}%
  \BibitemOpen
  \bibfield  {author} {\bibinfo {author} {\bibfnamefont {O.~D.}\ \bibnamefont {CrysAlisPRO}},\ }\bibfield  {title} {\bibinfo {title} {Agilent technologies uk ltd},\ }\href@noop {} {\bibfield  {journal} {\bibinfo  {journal} {Yarnton, England}\ }\textbf {\bibinfo {volume} {1}} (\bibinfo {year} {2014})}\BibitemShut {NoStop}%
\bibitem [{\citenamefont {Petříček}\ \emph {et~al.}(2023)\citenamefont {Petříček}, \citenamefont {Palatinus}, \citenamefont {Plášil},\ and\ \citenamefont {Dušek}}]{Petricek2023_JANA2020}%
  \BibitemOpen
  \bibfield  {author} {\bibinfo {author} {\bibfnamefont {V.}~\bibnamefont {Petříček}}, \bibinfo {author} {\bibfnamefont {L.}~\bibnamefont {Palatinus}}, \bibinfo {author} {\bibfnamefont {J.}~\bibnamefont {Plášil}},\ and\ \bibinfo {author} {\bibfnamefont {M.}~\bibnamefont {Dušek}},\ }\bibfield  {title} {\bibinfo {title} {Jana2020 – a new version of the crystallographic computing system jana},\ }\href {https://doi.org/10.1515/zkri-2023-0005} {\bibfield  {journal} {\bibinfo  {journal} {Zeitschrift für Kristallographie - Crystalline Materials}\ }\textbf {\bibinfo {volume} {238}},\ \bibinfo {pages} {271} (\bibinfo {year} {2023})}\BibitemShut {NoStop}%
\bibitem [{\citenamefont {Sheldrick}(2008)}]{sheldrick2008short}%
  \BibitemOpen
  \bibfield  {author} {\bibinfo {author} {\bibfnamefont {G.~M.}\ \bibnamefont {Sheldrick}},\ }\bibfield  {title} {\bibinfo {title} {{A short history of SHELX}},\ }\href {https://doi.org/10.1107/S0108767307043930} {\bibfield  {journal} {\bibinfo  {journal} {Acta Crystallogr. A}\ }\textbf {\bibinfo {volume} {64}},\ \bibinfo {pages} {112} (\bibinfo {year} {2008})}\BibitemShut {NoStop}%
\bibitem [{Nex()}]{NexPy}%
  \BibitemOpen
  \href@noop {} {\bibinfo {title} {Nexpy and nxrefine}},\ \bibinfo {howpublished} {https://github.com/nexpy/nxrefine},\ \bibinfo {note} {accessed: 2010-09-30}\BibitemShut {NoStop}%
\bibitem [{Gom()}]{GomezNexPy}%
  \BibitemOpen
  \href {https://github.com/stevenjgomez/nxs_analysis_tools} {\bibinfo {title} {Nxs analysis tools}},\ \bibinfo {note} {accessed: 2010-09-30}\BibitemShut {NoStop}%
\bibitem [{\citenamefont {Coelho}(2018)}]{coelho2018topas}%
  \BibitemOpen
  \bibfield  {author} {\bibinfo {author} {\bibfnamefont {A.~A.}\ \bibnamefont {Coelho}},\ }\bibfield  {title} {\bibinfo {title} {Topas and topas-academic: an optimization program integrating computer algebra and crystallographic objects written in c++},\ }\href@noop {} {\bibfield  {journal} {\bibinfo  {journal} {Journal of Applied Crystallography}\ }\textbf {\bibinfo {volume} {51}},\ \bibinfo {pages} {210} (\bibinfo {year} {2018})}\BibitemShut {NoStop}%
\bibitem [{\citenamefont {Kohn}\ and\ \citenamefont {Sham}(1965)}]{kohn1965self}%
  \BibitemOpen
  \bibfield  {author} {\bibinfo {author} {\bibfnamefont {W.}~\bibnamefont {Kohn}}\ and\ \bibinfo {author} {\bibfnamefont {L.~J.}\ \bibnamefont {Sham}},\ }\bibfield  {title} {\bibinfo {title} {Self-consistent equations including exchange and correlation effects},\ }\href@noop {} {\bibfield  {journal} {\bibinfo  {journal} {Physical review}\ }\textbf {\bibinfo {volume} {140}},\ \bibinfo {pages} {A1133} (\bibinfo {year} {1965})}\BibitemShut {NoStop}%
\bibitem [{\citenamefont {Andersen}(1975)}]{andersen1975linear}%
  \BibitemOpen
  \bibfield  {author} {\bibinfo {author} {\bibfnamefont {O.~K.}\ \bibnamefont {Andersen}},\ }\bibfield  {title} {\bibinfo {title} {Linear methods in band theory},\ }\href@noop {} {\bibfield  {journal} {\bibinfo  {journal} {Physical Review B}\ }\textbf {\bibinfo {volume} {12}},\ \bibinfo {pages} {3060} (\bibinfo {year} {1975})}\BibitemShut {NoStop}%
\bibitem [{\citenamefont {Singh}\ and\ \citenamefont {Nordstrom}(2006)}]{singh2006planewaves}%
  \BibitemOpen
  \bibfield  {author} {\bibinfo {author} {\bibfnamefont {D.~J.}\ \bibnamefont {Singh}}\ and\ \bibinfo {author} {\bibfnamefont {L.}~\bibnamefont {Nordstrom}},\ }\href@noop {} {\emph {\bibinfo {title} {Planewaves, Pseudopotentials, and the LAPW method}}}\ (\bibinfo  {publisher} {Springer Science \& Business Media},\ \bibinfo {year} {2006})\BibitemShut {NoStop}%
\bibitem [{\citenamefont {Sj{\"o}stedt}\ \emph {et~al.}(2000)\citenamefont {Sj{\"o}stedt}, \citenamefont {Nordstr{\"o}m},\ and\ \citenamefont {Singh}}]{sjostedt2000alternative}%
  \BibitemOpen
  \bibfield  {author} {\bibinfo {author} {\bibfnamefont {E.}~\bibnamefont {Sj{\"o}stedt}}, \bibinfo {author} {\bibfnamefont {L.}~\bibnamefont {Nordstr{\"o}m}},\ and\ \bibinfo {author} {\bibfnamefont {D.}~\bibnamefont {Singh}},\ }\bibfield  {title} {\bibinfo {title} {An alternative way of linearizing the augmented plane-wave method},\ }\href@noop {} {\bibfield  {journal} {\bibinfo  {journal} {Solid state communications}\ }\textbf {\bibinfo {volume} {114}},\ \bibinfo {pages} {15} (\bibinfo {year} {2000})}\BibitemShut {NoStop}%
\bibitem [{\citenamefont {Blaha}\ \emph {et~al.}(2020)\citenamefont {Blaha}, \citenamefont {Schwarz}, \citenamefont {Tran}, \citenamefont {Laskowski}, \citenamefont {Madsen},\ and\ \citenamefont {Marks}}]{blaha2020wien2k}%
  \BibitemOpen
  \bibfield  {author} {\bibinfo {author} {\bibfnamefont {P.}~\bibnamefont {Blaha}}, \bibinfo {author} {\bibfnamefont {K.}~\bibnamefont {Schwarz}}, \bibinfo {author} {\bibfnamefont {F.}~\bibnamefont {Tran}}, \bibinfo {author} {\bibfnamefont {R.}~\bibnamefont {Laskowski}}, \bibinfo {author} {\bibfnamefont {G.~K.}\ \bibnamefont {Madsen}},\ and\ \bibinfo {author} {\bibfnamefont {L.~D.}\ \bibnamefont {Marks}},\ }\bibfield  {title} {\bibinfo {title} {Wien2k: An apw+ lo program for calculating the properties of solids},\ }\href@noop {} {\bibfield  {journal} {\bibinfo  {journal} {The Journal of chemical physics}\ }\textbf {\bibinfo {volume} {152}},\ \bibinfo {pages} {074101} (\bibinfo {year} {2020})}\BibitemShut {NoStop}%
\bibitem [{\citenamefont {Kohn}\ \emph {et~al.}(1996)\citenamefont {Kohn}, \citenamefont {Becke},\ and\ \citenamefont {Parr}}]{kohn1996density}%
  \BibitemOpen
  \bibfield  {author} {\bibinfo {author} {\bibfnamefont {W.}~\bibnamefont {Kohn}}, \bibinfo {author} {\bibfnamefont {A.~D.}\ \bibnamefont {Becke}},\ and\ \bibinfo {author} {\bibfnamefont {R.~G.}\ \bibnamefont {Parr}},\ }\bibfield  {title} {\bibinfo {title} {Density functional theory of electronic structure},\ }\href@noop {} {\bibfield  {journal} {\bibinfo  {journal} {The journal of physical chemistry}\ }\textbf {\bibinfo {volume} {100}},\ \bibinfo {pages} {12974} (\bibinfo {year} {1996})}\BibitemShut {NoStop}%
\bibitem [{\citenamefont {Perdew}\ \emph {et~al.}(1996)\citenamefont {Perdew}, \citenamefont {Burke},\ and\ \citenamefont {Ernzerhof}}]{perdew1996generalized}%
  \BibitemOpen
  \bibfield  {author} {\bibinfo {author} {\bibfnamefont {J.~P.}\ \bibnamefont {Perdew}}, \bibinfo {author} {\bibfnamefont {K.}~\bibnamefont {Burke}},\ and\ \bibinfo {author} {\bibfnamefont {M.}~\bibnamefont {Ernzerhof}},\ }\bibfield  {title} {\bibinfo {title} {Generalized gradient approximation made simple},\ }\href@noop {} {\bibfield  {journal} {\bibinfo  {journal} {Physical review letters}\ }\textbf {\bibinfo {volume} {77}},\ \bibinfo {pages} {3865} (\bibinfo {year} {1996})}\BibitemShut {NoStop}%
\bibitem [{\citenamefont {Koelling}\ and\ \citenamefont {Harmon}(1977)}]{koelling1977technique}%
  \BibitemOpen
  \bibfield  {author} {\bibinfo {author} {\bibfnamefont {D.}~\bibnamefont {Koelling}}\ and\ \bibinfo {author} {\bibfnamefont {B.}~\bibnamefont {Harmon}},\ }\bibfield  {title} {\bibinfo {title} {A technique for relativistic spin-polarised calculations},\ }\href@noop {} {\bibfield  {journal} {\bibinfo  {journal} {Journal of Physics C: Solid State Physics}\ }\textbf {\bibinfo {volume} {10}},\ \bibinfo {pages} {3107} (\bibinfo {year} {1977})}\BibitemShut {NoStop}%
\bibitem [{\citenamefont {Kawamura}(2019)}]{kawamura2019fermisurfer}%
  \BibitemOpen
  \bibfield  {author} {\bibinfo {author} {\bibfnamefont {M.}~\bibnamefont {Kawamura}},\ }\bibfield  {title} {\bibinfo {title} {Fermisurfer: Fermi-surface viewer providing multiple representation schemes},\ }\href@noop {} {\bibfield  {journal} {\bibinfo  {journal} {Computer Physics Communications}\ }\textbf {\bibinfo {volume} {239}},\ \bibinfo {pages} {197} (\bibinfo {year} {2019})}\BibitemShut {NoStop}%
\bibitem [{\citenamefont {Kokalj}(1999)}]{kokalj1999xcrysden}%
  \BibitemOpen
  \bibfield  {author} {\bibinfo {author} {\bibfnamefont {A.}~\bibnamefont {Kokalj}},\ }\bibfield  {title} {\bibinfo {title} {Xcrysden—a new program for displaying crystalline structures and electron densities},\ }\href@noop {} {\bibfield  {journal} {\bibinfo  {journal} {Journal of Molecular Graphics and Modelling}\ }\textbf {\bibinfo {volume} {17}},\ \bibinfo {pages} {176} (\bibinfo {year} {1999})}\BibitemShut {NoStop}%
\bibitem [{\citenamefont {Karl A.~Gschneidner}\ and\ \citenamefont {Eyring}(1999)}]{gschneidner2005handbook}%
  \BibitemOpen
  \bibfield  {author} {\bibinfo {author} {\bibfnamefont {J.}~\bibnamefont {Karl A.~Gschneidner}}\ and\ \bibinfo {author} {\bibfnamefont {L.}~\bibnamefont {Eyring}},\ }\href@noop {} {\emph {\bibinfo {title} {Handbook on the physics and chemistry of rare earths}}},\ Vol.~\bibinfo {volume} {27}\ (\bibinfo  {publisher} {Elsevier},\ \bibinfo {year} {1999})\BibitemShut {NoStop}%
\bibitem [{\citenamefont {Olenich}\ \emph {et~al.}(1981)\citenamefont {Olenich}, \citenamefont {Aksel'rud},\ and\ \citenamefont {Yarmolyuk}}]{olenich1981crystal}%
  \BibitemOpen
  \bibfield  {author} {\bibinfo {author} {\bibfnamefont {R.}~\bibnamefont {Olenich}}, \bibinfo {author} {\bibfnamefont {L.}~\bibnamefont {Aksel'rud}},\ and\ \bibinfo {author} {\bibfnamefont {Y.~R.}\ \bibnamefont {Yarmolyuk}},\ }\bibfield  {title} {\bibinfo {title} {{Crystal structure of ternary germanides RFe$_6$Ge$_6$ (R--Sc, Ti, Zr, Hf, Nd) and RCo$_6$Ge$_6$ (R= Ti, Zr, Hf)}},\ }\href@noop {} {\bibfield  {journal} {\bibinfo  {journal} {Dopovidi Akademii Nauk Ukrains' koj RSR. Seriya A, Fiziko-Tekhnichni ta Matematichni Nauki}\ ,\ \bibinfo {pages} {84}} (\bibinfo {year} {1981})}\BibitemShut {NoStop}%
\bibitem [{\citenamefont {Mazet}\ \emph {et~al.}(1999{\natexlab{a}})\citenamefont {Mazet}, \citenamefont {Welter}, \citenamefont {Venturini}, \citenamefont {Ressouche},\ and\ \citenamefont {Malaman}}]{mazet1999neutron}%
  \BibitemOpen
  \bibfield  {author} {\bibinfo {author} {\bibfnamefont {T.}~\bibnamefont {Mazet}}, \bibinfo {author} {\bibfnamefont {R.}~\bibnamefont {Welter}}, \bibinfo {author} {\bibfnamefont {G.}~\bibnamefont {Venturini}}, \bibinfo {author} {\bibfnamefont {E.}~\bibnamefont {Ressouche}},\ and\ \bibinfo {author} {\bibfnamefont {B.}~\bibnamefont {Malaman}},\ }\bibfield  {title} {\bibinfo {title} {{Neutron diffraction study of the ZrMn$_6$Ge$_6$, LuMn$_6$Ge$_6$ and ScMn$_6$Ge$_6$ compounds}},\ }\href {https://doi.org/10.1016/S0038-1098(99)00093-9} {\bibfield  {journal} {\bibinfo  {journal} {Solid state communications}\ }\textbf {\bibinfo {volume} {110}},\ \bibinfo {pages} {407} (\bibinfo {year} {1999}{\natexlab{a}})}\BibitemShut {NoStop}%
\bibitem [{\citenamefont {Venturini}\ \emph {et~al.}(1992)\citenamefont {Venturini}, \citenamefont {Welter},\ and\ \citenamefont {Malaman}}]{venturini1992crystallographic}%
  \BibitemOpen
  \bibfield  {author} {\bibinfo {author} {\bibfnamefont {G.}~\bibnamefont {Venturini}}, \bibinfo {author} {\bibfnamefont {R.}~\bibnamefont {Welter}},\ and\ \bibinfo {author} {\bibfnamefont {B.}~\bibnamefont {Malaman}},\ }\bibfield  {title} {\bibinfo {title} {{Crystallographic data and magnetic properties of RT$_6$Ge$_6$ compounds (R: Sc, Y, Nd, Sm, Gd \& Lu; T: Mn, Fe)}},\ }\href {https://doi.org/10.1016/0925-8388(92)90558-q} {\bibfield  {journal} {\bibinfo  {journal} {Journal of alloys and compounds}\ }\textbf {\bibinfo {volume} {185}},\ \bibinfo {pages} {99} (\bibinfo {year} {1992})}\BibitemShut {NoStop}%
\bibitem [{\citenamefont {Skolozdra}\ \emph {et~al.}(1991)\citenamefont {Skolozdra}, \citenamefont {Kotur}, \citenamefont {Andrusyak},\ and\ \citenamefont {Gorelenko}}]{skolozdra1991magnetic}%
  \BibitemOpen
  \bibfield  {author} {\bibinfo {author} {\bibfnamefont {R.}~\bibnamefont {Skolozdra}}, \bibinfo {author} {\bibfnamefont {B.~Y.}\ \bibnamefont {Kotur}}, \bibinfo {author} {\bibfnamefont {R.}~\bibnamefont {Andrusyak}},\ and\ \bibinfo {author} {\bibfnamefont {Y.~K.}\ \bibnamefont {Gorelenko}},\ }\bibfield  {title} {\bibinfo {title} {{MAGNETIC AND ELECTRICAL PROPERTIES OF TERNARY GERMANIDES OF SCANDIUM AND 3D TRANSITION METALS}},\ }\href@noop {} {\bibfield  {journal} {\bibinfo  {journal} {Inorganic materials}\ }\textbf {\bibinfo {volume} {27}},\ \bibinfo {pages} {1371} (\bibinfo {year} {1991})}\BibitemShut {NoStop}%
\bibitem [{\citenamefont {Buchholz}\ and\ \citenamefont {Schuster}(1981)}]{buchholz1981intermetallische}%
  \BibitemOpen
  \bibfield  {author} {\bibinfo {author} {\bibfnamefont {W.}~\bibnamefont {Buchholz}}\ and\ \bibinfo {author} {\bibfnamefont {H.-U.}\ \bibnamefont {Schuster}},\ }\bibfield  {title} {\bibinfo {title} {{Intermetallische Phasen mit B35-{\"U}berstruktur und Verwandtschaftsbeziehung zu LiFe$_6$Ge$_6$}},\ }\href {https://doi.org/10.1002/zaac.19814821105} {\bibfield  {journal} {\bibinfo  {journal} {Zeitschrift f{\"u}r anorganische und allgemeine Chemie}\ }\textbf {\bibinfo {volume} {482}},\ \bibinfo {pages} {40} (\bibinfo {year} {1981})}\BibitemShut {NoStop}%
\bibitem [{\citenamefont {Mazet}\ \emph {et~al.}(2008)\citenamefont {Mazet}, \citenamefont {Ihou-Mouko},\ and\ \citenamefont {Malaman}}]{mazet2008first}%
  \BibitemOpen
  \bibfield  {author} {\bibinfo {author} {\bibfnamefont {T.}~\bibnamefont {Mazet}}, \bibinfo {author} {\bibfnamefont {H.}~\bibnamefont {Ihou-Mouko}},\ and\ \bibinfo {author} {\bibfnamefont {B.}~\bibnamefont {Malaman}},\ }\bibfield  {title} {\bibinfo {title} {{First-order ferromagnetic to helimagnetic transition in MgMn$_6$Ge$_6$}},\ }\href {https://doi.org/10.1063/1.2840126} {\bibfield  {journal} {\bibinfo  {journal} {Journal of Applied Physics}\ }\textbf {\bibinfo {volume} {103}},\ \bibinfo {pages} {043903} (\bibinfo {year} {2008})}\BibitemShut {NoStop}%
\bibitem [{\citenamefont {Romaka}\ \emph {et~al.}(2024)\citenamefont {Romaka}, \citenamefont {Romaka}, \citenamefont {Konyk}, \citenamefont {Corredor}, \citenamefont {Srowik}, \citenamefont {Kuzhel}, \citenamefont {Stadnyk},\ and\ \citenamefont {Yatskiv}}]{romaka2024structure}%
  \BibitemOpen
  \bibfield  {author} {\bibinfo {author} {\bibfnamefont {V.}~\bibnamefont {Romaka}}, \bibinfo {author} {\bibfnamefont {L.}~\bibnamefont {Romaka}}, \bibinfo {author} {\bibfnamefont {M.}~\bibnamefont {Konyk}}, \bibinfo {author} {\bibfnamefont {L.}~\bibnamefont {Corredor}}, \bibinfo {author} {\bibfnamefont {K.}~\bibnamefont {Srowik}}, \bibinfo {author} {\bibfnamefont {B.}~\bibnamefont {Kuzhel}}, \bibinfo {author} {\bibfnamefont {Y.}~\bibnamefont {Stadnyk}},\ and\ \bibinfo {author} {\bibfnamefont {Y.}~\bibnamefont {Yatskiv}},\ }\bibfield  {title} {\bibinfo {title} {{Structure, bonding, and properties of RCr$_6$Ge$_6$ intermetallics (R= Gd-Lu)}},\ }\href {https://doi.org/10.1016/j.jssc.2024.124874} {\bibfield  {journal} {\bibinfo  {journal} {Journal of Solid State Chemistry}\ ,\ \bibinfo {pages} {124874}} (\bibinfo {year} {2024})}\BibitemShut {NoStop}%
\bibitem [{\citenamefont {Brabers}\ \emph {et~al.}(1994)\citenamefont {Brabers}, \citenamefont {Buschow},\ and\ \citenamefont {De~Boer}}]{brabers1994magnetic}%
  \BibitemOpen
  \bibfield  {author} {\bibinfo {author} {\bibfnamefont {J.}~\bibnamefont {Brabers}}, \bibinfo {author} {\bibfnamefont {K.}~\bibnamefont {Buschow}},\ and\ \bibinfo {author} {\bibfnamefont {F.}~\bibnamefont {De~Boer}},\ }\bibfield  {title} {\bibinfo {title} {{Magnetic properties of RCr$_6$Ge$_6$ compounds}},\ }\href {https://doi.org/10.1016/0925-8388(94)90769-2} {\bibfield  {journal} {\bibinfo  {journal} {Journal of alloys and compounds}\ }\textbf {\bibinfo {volume} {205}},\ \bibinfo {pages} {77} (\bibinfo {year} {1994})}\BibitemShut {NoStop}%
\bibitem [{\citenamefont {Yang}\ \emph {et~al.}(2024)\citenamefont {Yang}, \citenamefont {Zeng}, \citenamefont {He}, \citenamefont {Xu}, \citenamefont {Du},\ and\ \citenamefont {Qu}}]{yang2024crystal}%
  \BibitemOpen
  \bibfield  {author} {\bibinfo {author} {\bibfnamefont {X.}~\bibnamefont {Yang}}, \bibinfo {author} {\bibfnamefont {Q.}~\bibnamefont {Zeng}}, \bibinfo {author} {\bibfnamefont {M.}~\bibnamefont {He}}, \bibinfo {author} {\bibfnamefont {X.}~\bibnamefont {Xu}}, \bibinfo {author} {\bibfnamefont {H.}~\bibnamefont {Du}},\ and\ \bibinfo {author} {\bibfnamefont {Z.}~\bibnamefont {Qu}},\ }\bibfield  {title} {\bibinfo {title} {{Crystal Growth, Magnetic and Electrical Transport Properties of the Kagome Magnet RCr$_6$Ge$_6$ (R= Gd-Tm)}},\ }\href {https://doi.org/10.1088/1674-1056/ad3dcf} {\bibfield  {journal} {\bibinfo  {journal} {Chinese Physics B}\ }\textbf {\bibinfo {volume} {33}},\ \bibinfo {pages} {077501} (\bibinfo {year} {2024})}\BibitemShut {NoStop}%
\bibitem [{\citenamefont {Romaka}\ \emph {et~al.}(2022)\citenamefont {Romaka}, \citenamefont {Stadnyk}, \citenamefont {Romaka},\ and\ \citenamefont {Konyk}}]{romaka2022interaction}%
  \BibitemOpen
  \bibfield  {author} {\bibinfo {author} {\bibfnamefont {L.}~\bibnamefont {Romaka}}, \bibinfo {author} {\bibfnamefont {Y.}~\bibnamefont {Stadnyk}}, \bibinfo {author} {\bibfnamefont {V.}~\bibnamefont {Romaka}},\ and\ \bibinfo {author} {\bibfnamefont {M.}~\bibnamefont {Konyk}},\ }\bibfield  {title} {\bibinfo {title} {{Interaction between the components in Tm-Cr-Ge system at 1070 K}},\ }\href {https://doi.org/10.15330/pcss.23.4.633-639} {\bibfield  {journal} {\bibinfo  {journal} {Physics and Chemistry of Solid State}\ }\textbf {\bibinfo {volume} {23}},\ \bibinfo {pages} {633} (\bibinfo {year} {2022})}\BibitemShut {NoStop}%
\bibitem [{\citenamefont {Schobinger-Papamantellos}\ \emph {et~al.}(1997{\natexlab{a}})\citenamefont {Schobinger-Papamantellos}, \citenamefont {Rodr{\'\i}guez-Carvajal},\ and\ \citenamefont {Buschow}}]{schobinger1997ferrimagnetism}%
  \BibitemOpen
  \bibfield  {author} {\bibinfo {author} {\bibfnamefont {P.}~\bibnamefont {Schobinger-Papamantellos}}, \bibinfo {author} {\bibfnamefont {J.}~\bibnamefont {Rodr{\'\i}guez-Carvajal}},\ and\ \bibinfo {author} {\bibfnamefont {K.}~\bibnamefont {Buschow}},\ }\bibfield  {title} {\bibinfo {title} {{Ferrimagnetism and disorder in the RCr$_6$Ge$_6$ compounds (R: Dy, Ho, Er, Y): A neutron study}},\ }\href {https://doi.org/10.1016/s0925-8388(96)03109-x} {\bibfield  {journal} {\bibinfo  {journal} {Journal of alloys and compounds}\ }\textbf {\bibinfo {volume} {256}},\ \bibinfo {pages} {92} (\bibinfo {year} {1997}{\natexlab{a}})}\BibitemShut {NoStop}%
\bibitem [{\citenamefont {Konyk}\ \emph {et~al.}(2019)\citenamefont {Konyk}, \citenamefont {Romaka}, \citenamefont {Stadnyk}, \citenamefont {Romaka}, \citenamefont {Serkiz},\ and\ \citenamefont {Horyn}}]{konyk2019er}%
  \BibitemOpen
  \bibfield  {author} {\bibinfo {author} {\bibfnamefont {M.}~\bibnamefont {Konyk}}, \bibinfo {author} {\bibfnamefont {L.}~\bibnamefont {Romaka}}, \bibinfo {author} {\bibfnamefont {Y.}~\bibnamefont {Stadnyk}}, \bibinfo {author} {\bibfnamefont {V.}~\bibnamefont {Romaka}}, \bibinfo {author} {\bibfnamefont {R.}~\bibnamefont {Serkiz}},\ and\ \bibinfo {author} {\bibfnamefont {A.}~\bibnamefont {Horyn}},\ }\bibfield  {title} {\bibinfo {title} {{Er-Cr-Ge Ternary System}},\ }\href {https://doi.org/10.15330/pcss.20.4.376-383} {\bibfield  {journal} {\bibinfo  {journal} {Physics and Chemistry of Solid State}\ }\textbf {\bibinfo {volume} {20}},\ \bibinfo {pages} {376} (\bibinfo {year} {2019})}\BibitemShut {NoStop}%
\bibitem [{\citenamefont {Schobinger-Papamantellos}\ \emph {et~al.}(1997{\natexlab{b}})\citenamefont {Schobinger-Papamantellos}, \citenamefont {Rodr{\'\i}guez-Carvajal},\ and\ \citenamefont {Buschow}}]{schobinger1997atomic}%
  \BibitemOpen
  \bibfield  {author} {\bibinfo {author} {\bibfnamefont {P.}~\bibnamefont {Schobinger-Papamantellos}}, \bibinfo {author} {\bibfnamefont {J.}~\bibnamefont {Rodr{\'\i}guez-Carvajal}},\ and\ \bibinfo {author} {\bibfnamefont {K.}~\bibnamefont {Buschow}},\ }\bibfield  {title} {\bibinfo {title} {{Atomic disorder and canted ferrimagnetism in the TbCr$_6$Ge$_6$ compound. A neutron study}},\ }\href {https://doi.org/10.1016/s0925-8388(96)02872-1} {\bibfield  {journal} {\bibinfo  {journal} {Journal of alloys and compounds}\ }\textbf {\bibinfo {volume} {255}},\ \bibinfo {pages} {67} (\bibinfo {year} {1997}{\natexlab{b}})}\BibitemShut {NoStop}%
\bibitem [{\citenamefont {Konyk}\ \emph {et~al.}(2021)\citenamefont {Konyk}, \citenamefont {Romaka}, \citenamefont {Stadnyk}, \citenamefont {Romaka},\ and\ \citenamefont {Pashkevych}}]{konyk2021phase}%
  \BibitemOpen
  \bibfield  {author} {\bibinfo {author} {\bibfnamefont {M.}~\bibnamefont {Konyk}}, \bibinfo {author} {\bibfnamefont {L.}~\bibnamefont {Romaka}}, \bibinfo {author} {\bibfnamefont {Y.}~\bibnamefont {Stadnyk}}, \bibinfo {author} {\bibfnamefont {V.}~\bibnamefont {Romaka}},\ and\ \bibinfo {author} {\bibfnamefont {V.}~\bibnamefont {Pashkevych}},\ }\bibfield  {title} {\bibinfo {title} {{Phase equilibria in the Gd--Cr--Ge system at 1070 K}},\ }\bibfield  {journal} {\bibinfo  {journal} {Physics and Chemistry of Solid State (PCSS)}\ }\href {https://doi.org/10.15330/pcss.22.2.248-254} {10.15330/pcss.22.2.248-254} (\bibinfo {year} {2021})\BibitemShut {NoStop}%
\bibitem [{\citenamefont {Ihou-Mouko}\ \emph {et~al.}(2006)\citenamefont {Ihou-Mouko}, \citenamefont {Mazet}, \citenamefont {Isnard},\ and\ \citenamefont {Malaman}}]{ihou2006magnetic}%
  \BibitemOpen
  \bibfield  {author} {\bibinfo {author} {\bibfnamefont {H.}~\bibnamefont {Ihou-Mouko}}, \bibinfo {author} {\bibfnamefont {T.}~\bibnamefont {Mazet}}, \bibinfo {author} {\bibfnamefont {O.}~\bibnamefont {Isnard}},\ and\ \bibinfo {author} {\bibfnamefont {B.}~\bibnamefont {Malaman}},\ }\bibfield  {title} {\bibinfo {title} {{Magnetic properties and electronic structure of the new HfFe$_6$Ge$_6$-type HfMn$_6$Ge$_6$ compound}},\ }\href {https://doi.org/10.1016/j.jallcom.2006.02.005} {\bibfield  {journal} {\bibinfo  {journal} {Journal of alloys and compounds}\ }\textbf {\bibinfo {volume} {426}},\ \bibinfo {pages} {26} (\bibinfo {year} {2006})}\BibitemShut {NoStop}%
\bibitem [{\citenamefont {Mazet}\ and\ \citenamefont {Malaman}(2001)}]{mazet2001macroscopic}%
  \BibitemOpen
  \bibfield  {author} {\bibinfo {author} {\bibfnamefont {T.}~\bibnamefont {Mazet}}\ and\ \bibinfo {author} {\bibfnamefont {B.}~\bibnamefont {Malaman}},\ }\bibfield  {title} {\bibinfo {title} {{Macroscopic magnetic properties of the HfFe$_6$Ge$_6$-type RFe$_6$X$_6$ (X= Ge or Sn) compounds involving a non-magnetic R metal}},\ }\href {https://doi.org/10.1016/s0925-8388(01)01378-0} {\bibfield  {journal} {\bibinfo  {journal} {Journal of alloys and compounds}\ }\textbf {\bibinfo {volume} {325}},\ \bibinfo {pages} {67} (\bibinfo {year} {2001})}\BibitemShut {NoStop}%
\bibitem [{\citenamefont {Chabot}\ and\ \citenamefont {Parth{\'e}}(1983)}]{chabot1983lufe6ge6}%
  \BibitemOpen
  \bibfield  {author} {\bibinfo {author} {\bibfnamefont {B.}~\bibnamefont {Chabot}}\ and\ \bibinfo {author} {\bibfnamefont {E.}~\bibnamefont {Parth{\'e}}},\ }\bibfield  {title} {\bibinfo {title} {{LuFe$_6$Ge$_6$ with the HfFe$_6$Ge$_6$-type structure}},\ }\href {https://doi.org/10.1016/0022-5088(83)90215-1} {\bibfield  {journal} {\bibinfo  {journal} {Journal of the Less Common Metals}\ }\textbf {\bibinfo {volume} {93}},\ \bibinfo {pages} {L9} (\bibinfo {year} {1983})}\BibitemShut {NoStop}%
\bibitem [{\citenamefont {Raghavan}(2002)}]{raghavan2002fe}%
  \BibitemOpen
  \bibfield  {author} {\bibinfo {author} {\bibfnamefont {V.}~\bibnamefont {Raghavan}},\ }\bibfield  {title} {\bibinfo {title} {{Fe-Ge-Tm (iron-germanium-thulium)}},\ }\href {https://doi.org/10.1361/105497102770332333} {\bibfield  {journal} {\bibinfo  {journal} {Journal of Phase Equilibria}\ }\textbf {\bibinfo {volume} {23}},\ \bibinfo {pages} {98} (\bibinfo {year} {2002})}\BibitemShut {NoStop}%
\bibitem [{\citenamefont {Oleksyn}\ \emph {et~al.}(1997)\citenamefont {Oleksyn}, \citenamefont {Schobinger-Papamantellos}, \citenamefont {Rodr{\'\i}guez-Carvajal}, \citenamefont {Br{\"u}ck},\ and\ \citenamefont {Buschow}}]{oleksyn1997crystal}%
  \BibitemOpen
  \bibfield  {author} {\bibinfo {author} {\bibfnamefont {O.}~\bibnamefont {Oleksyn}}, \bibinfo {author} {\bibfnamefont {P.}~\bibnamefont {Schobinger-Papamantellos}}, \bibinfo {author} {\bibfnamefont {J.}~\bibnamefont {Rodr{\'\i}guez-Carvajal}}, \bibinfo {author} {\bibfnamefont {E.}~\bibnamefont {Br{\"u}ck}},\ and\ \bibinfo {author} {\bibfnamefont {K.}~\bibnamefont {Buschow}},\ }\bibfield  {title} {\bibinfo {title} {{Crystal structure and magnetic ordering in ErFe$_6$Ge$_6$ studied by X-ray, neutron diffraction and magnetic measurements}},\ }\href {https://doi.org/10.1016/s0925-8388(96)03125-8} {\bibfield  {journal} {\bibinfo  {journal} {Journal of alloys and compounds}\ }\textbf {\bibinfo {volume} {257}},\ \bibinfo {pages} {36} (\bibinfo {year} {1997})}\BibitemShut {NoStop}%
\bibitem [{\citenamefont {Bodak}\ \emph {et~al.}(1992)\citenamefont {Bodak}, \citenamefont {Oleksii}, \citenamefont {Fedyna},\ and\ \citenamefont {Pecharskii}}]{bodak1992er}%
  \BibitemOpen
  \bibfield  {author} {\bibinfo {author} {\bibfnamefont {O.}~\bibnamefont {Bodak}}, \bibinfo {author} {\bibfnamefont {O.~Y.}\ \bibnamefont {Oleksii}}, \bibinfo {author} {\bibfnamefont {M.}~\bibnamefont {Fedyna}},\ and\ \bibinfo {author} {\bibfnamefont {V.}~\bibnamefont {Pecharskii}},\ }\bibfield  {title} {\bibinfo {title} {{Er (Tm)-Fe-Ge System}},\ }\href@noop {} {\bibfield  {journal} {\bibinfo  {journal} {Inorganic materials}\ }\textbf {\bibinfo {volume} {28}},\ \bibinfo {pages} {371} (\bibinfo {year} {1992})}\BibitemShut {NoStop}%
\bibitem [{\citenamefont {El~Idrissi}\ \emph {et~al.}(1991{\natexlab{a}})\citenamefont {El~Idrissi}, \citenamefont {Venturini},\ and\ \citenamefont {Malaman}}]{el1991crystal}%
  \BibitemOpen
  \bibfield  {author} {\bibinfo {author} {\bibfnamefont {B.~C.}\ \bibnamefont {El~Idrissi}}, \bibinfo {author} {\bibfnamefont {G.}~\bibnamefont {Venturini}},\ and\ \bibinfo {author} {\bibfnamefont {B.}~\bibnamefont {Malaman}},\ }\bibfield  {title} {\bibinfo {title} {{Crystal structures of RFe$_6$Sn$_6$ (R= Sc, Y, Gd-Tm, Lu) rare-earth iron stannides}},\ }\href {https://doi.org/10.1016/0025-5408(91)90149-g} {\bibfield  {journal} {\bibinfo  {journal} {Materials Research Bulletin}\ }\textbf {\bibinfo {volume} {26}},\ \bibinfo {pages} {1331} (\bibinfo {year} {1991}{\natexlab{a}})}\BibitemShut {NoStop}%
\bibitem [{\citenamefont {Cadogan}\ and\ \citenamefont {Ryan}(2009)}]{cadogan2009study}%
  \BibitemOpen
  \bibfield  {author} {\bibinfo {author} {\bibfnamefont {J.}~\bibnamefont {Cadogan}}\ and\ \bibinfo {author} {\bibfnamefont {D.}~\bibnamefont {Ryan}},\ }\bibfield  {title} {\bibinfo {title} {{A study on the magnetic behaviour of polymorphic YbFe$_6$Ge$_6$}},\ }\href {https://doi.org/10.1088/0953-8984/22/1/016009} {\bibfield  {journal} {\bibinfo  {journal} {Journal of Physics: Condensed Matter}\ }\textbf {\bibinfo {volume} {22}},\ \bibinfo {pages} {016009} (\bibinfo {year} {2009})}\BibitemShut {NoStop}%
\bibitem [{\citenamefont {Konyk}\ \emph {et~al.}(2020)\citenamefont {Konyk}, \citenamefont {Romaka}, \citenamefont {Kuzhel}, \citenamefont {Stadnyk},\ and\ \citenamefont {Romaka}}]{eproprcr6ce6}%
  \BibitemOpen
  \bibfield  {author} {\bibinfo {author} {\bibfnamefont {M.}~\bibnamefont {Konyk}}, \bibinfo {author} {\bibfnamefont {L.}~\bibnamefont {Romaka}}, \bibinfo {author} {\bibfnamefont {B.}~\bibnamefont {Kuzhel}}, \bibinfo {author} {\bibfnamefont {Y.}~\bibnamefont {Stadnyk}},\ and\ \bibinfo {author} {\bibfnamefont {V.}~\bibnamefont {Romaka}},\ }\bibfield  {title} {\bibinfo {title} {{Electrical transport properties of RCr$_6$Ge$_6$ (R= Y, Gd, Tb, Dy, Lu) compounds}},\ }\bibfield  {journal} {\bibinfo  {journal} {Bulletin of Lviv University}\ }\textbf {\bibinfo {volume} {1}},\ \href {https://doi.org/10.30970/vch.6101.107} {10.30970/vch.6101.107} (\bibinfo {year} {2020})\BibitemShut {NoStop}%
\bibitem [{\citenamefont {Koretskaya}\ and\ \citenamefont {Skolozdra}(1986)}]{koretskaya1986new}%
  \BibitemOpen
  \bibfield  {author} {\bibinfo {author} {\bibfnamefont {O.}~\bibnamefont {Koretskaya}}\ and\ \bibinfo {author} {\bibfnamefont {R.}~\bibnamefont {Skolozdra}},\ }\bibfield  {title} {\bibinfo {title} {{New triple stannides with structure of the YCo$_6$Ge$_6$ type}},\ }\href@noop {} {\bibfield  {journal} {\bibinfo  {journal} {Inorganic Materials}\ }\textbf {\bibinfo {volume} {22}},\ \bibinfo {pages} {606} (\bibinfo {year} {1986})}\BibitemShut {NoStop}%
\bibitem [{\citenamefont {Mruz}\ \emph {et~al.}(1984)\citenamefont {Mruz}, \citenamefont {Starodub},\ and\ \citenamefont {Bodak}}]{mruz1984new}%
  \BibitemOpen
  \bibfield  {author} {\bibinfo {author} {\bibfnamefont {O.~Y.}\ \bibnamefont {Mruz}}, \bibinfo {author} {\bibfnamefont {P.}~\bibnamefont {Starodub}},\ and\ \bibinfo {author} {\bibfnamefont {O.}~\bibnamefont {Bodak}},\ }\bibfield  {title} {\bibinfo {title} {{New representatives of the YCo$_6$Ge$_6$ structure type}},\ }\href@noop {} {\bibfield  {journal} {\bibinfo  {journal} {Dopov. Akad. Nauk Ukr. RSR, Ser. B}\ ,\ \bibinfo {pages} {45}} (\bibinfo {year} {1984})}\BibitemShut {NoStop}%
\bibitem [{\citenamefont {El~Idrissi}\ \emph {et~al.}(1994)\citenamefont {El~Idrissi}, \citenamefont {Venturini}, \citenamefont {Malaman},\ and\ \citenamefont {Ressouche}}]{el1994magnetic}%
  \BibitemOpen
  \bibfield  {author} {\bibinfo {author} {\bibfnamefont {B.~C.}\ \bibnamefont {El~Idrissi}}, \bibinfo {author} {\bibfnamefont {G.}~\bibnamefont {Venturini}}, \bibinfo {author} {\bibfnamefont {B.}~\bibnamefont {Malaman}},\ and\ \bibinfo {author} {\bibfnamefont {E.}~\bibnamefont {Ressouche}},\ }\bibfield  {title} {\bibinfo {title} {{Magnetic properties of NdMn$_6$Ge$_6$ and SmMn$_6$Ge$_6$ compounds from susceptibility measurements and neutron diffraction study}},\ }\href {https://doi.org/10.1016/0925-8388(94)90839-7} {\bibfield  {journal} {\bibinfo  {journal} {Journal of Alloys and Compounds}\ }\textbf {\bibinfo {volume} {215}},\ \bibinfo {pages} {187} (\bibinfo {year} {1994})}\BibitemShut {NoStop}%
\bibitem [{\citenamefont {Hu}\ \emph {et~al.}(2019)\citenamefont {Hu}, \citenamefont {Dong}, \citenamefont {Shen}, \citenamefont {Liu}, \citenamefont {Peng}, \citenamefont {Cai},\ and\ \citenamefont {Jin}}]{hu2019measurement}%
  \BibitemOpen
  \bibfield  {author} {\bibinfo {author} {\bibfnamefont {K.}~\bibnamefont {Hu}}, \bibinfo {author} {\bibfnamefont {S.}~\bibnamefont {Dong}}, \bibinfo {author} {\bibfnamefont {C.}~\bibnamefont {Shen}}, \bibinfo {author} {\bibfnamefont {H.}~\bibnamefont {Liu}}, \bibinfo {author} {\bibfnamefont {H.}~\bibnamefont {Peng}}, \bibinfo {author} {\bibfnamefont {G.}~\bibnamefont {Cai}},\ and\ \bibinfo {author} {\bibfnamefont {Z.}~\bibnamefont {Jin}},\ }\bibfield  {title} {\bibinfo {title} {{Measurement of phase equilibria in Ti-Co-Ge ternary system}},\ }\href {https://doi.org/10.1016/j.jallcom.2019.04.223} {\bibfield  {journal} {\bibinfo  {journal} {Journal of Alloys and Compounds}\ }\textbf {\bibinfo {volume} {793}},\ \bibinfo {pages} {653} (\bibinfo {year} {2019})}\BibitemShut {NoStop}%
\bibitem [{\citenamefont {Kun}\ \emph {et~al.}(2023)\citenamefont {Kun}, \citenamefont {Chen}, \citenamefont {LIU},\ and\ \citenamefont {ZHANG}}]{kun2023phase}%
  \BibitemOpen
  \bibfield  {author} {\bibinfo {author} {\bibfnamefont {H.}~\bibnamefont {Kun}}, \bibinfo {author} {\bibfnamefont {S.}~\bibnamefont {Chen}}, \bibinfo {author} {\bibfnamefont {H.-s.}\ \bibnamefont {LIU}},\ and\ \bibinfo {author} {\bibfnamefont {H.-b.}\ \bibnamefont {ZHANG}},\ }\bibfield  {title} {\bibinfo {title} {{Phase equilibria of Zr-Co-Ge ternary system at 1023, 1173 and 1373 K}},\ }\href {https://doi.org/10.1016/s1003-6326(23)66198-9} {\bibfield  {journal} {\bibinfo  {journal} {Transactions of Nonferrous Metals Society of China}\ }\textbf {\bibinfo {volume} {33}},\ \bibinfo {pages} {1492} (\bibinfo {year} {2023})}\BibitemShut {NoStop}%
\bibitem [{\citenamefont {Buchholz}\ and\ \citenamefont {Schuster}(1978)}]{buchholz1978verbindungen}%
  \BibitemOpen
  \bibfield  {author} {\bibinfo {author} {\bibfnamefont {W.}~\bibnamefont {Buchholz}}\ and\ \bibinfo {author} {\bibfnamefont {H.-U.}\ \bibnamefont {Schuster}},\ }\bibfield  {title} {\bibinfo {title} {{Die Verbindungen MgFe$_6$Ge$_6$ und LiCo$_6$Ge$_6$: The Compounds MgFe$_6$Ge$_6$ and LiCo$_6$Ge$_6$}},\ }\href {https://doi.org/10.1515/znb-1978-0812} {\bibfield  {journal} {\bibinfo  {journal} {Zeitschrift f{\"u}r Naturforschung B}\ }\textbf {\bibinfo {volume} {33}},\ \bibinfo {pages} {877} (\bibinfo {year} {1978})}\BibitemShut {NoStop}%
\bibitem [{\citenamefont {Szytu{\l}a}\ \emph {et~al.}(2004)\citenamefont {Szytu{\l}a}, \citenamefont {Wawrzy{\'n}ska},\ and\ \citenamefont {Zygmunt}}]{szytula2004crystal}%
  \BibitemOpen
  \bibfield  {author} {\bibinfo {author} {\bibfnamefont {A.}~\bibnamefont {Szytu{\l}a}}, \bibinfo {author} {\bibfnamefont {E.}~\bibnamefont {Wawrzy{\'n}ska}},\ and\ \bibinfo {author} {\bibfnamefont {A.}~\bibnamefont {Zygmunt}},\ }\bibfield  {title} {\bibinfo {title} {{Crystal structure and magnetic properties of GdCo$_6$X$_6$ (X= Ge, Sn) and TbCo$_6$Ge$_6$}},\ }\href {https://doi.org/10.1016/s0925-8388(03)00752-7} {\bibfield  {journal} {\bibinfo  {journal} {Journal of alloys and compounds}\ }\textbf {\bibinfo {volume} {366}},\ \bibinfo {pages} {L16} (\bibinfo {year} {2004})}\BibitemShut {NoStop}%
\bibitem [{\citenamefont {Weiland}\ \emph {et~al.}(2020)\citenamefont {Weiland}, \citenamefont {Eddy}, \citenamefont {McCandless}, \citenamefont {Hodovanets}, \citenamefont {Paglione},\ and\ \citenamefont {Chan}}]{weiland2020refine}%
  \BibitemOpen
  \bibfield  {author} {\bibinfo {author} {\bibfnamefont {A.}~\bibnamefont {Weiland}}, \bibinfo {author} {\bibfnamefont {L.~J.}\ \bibnamefont {Eddy}}, \bibinfo {author} {\bibfnamefont {G.~T.}\ \bibnamefont {McCandless}}, \bibinfo {author} {\bibfnamefont {H.}~\bibnamefont {Hodovanets}}, \bibinfo {author} {\bibfnamefont {J.}~\bibnamefont {Paglione}},\ and\ \bibinfo {author} {\bibfnamefont {J.~Y.}\ \bibnamefont {Chan}},\ }\bibfield  {title} {\bibinfo {title} {{Refine intervention: characterizing disordered Yb$_{0.5}$Co$_3$Ge$_3$}},\ }\href {https://doi.org/10.1021/acs.cgd.0c00865} {\bibfield  {journal} {\bibinfo  {journal} {Crystal Growth \& Design}\ }\textbf {\bibinfo {volume} {20}},\ \bibinfo {pages} {6715} (\bibinfo {year} {2020})}\BibitemShut {NoStop}%
\bibitem [{\citenamefont {Dzyanyj}\ \emph {et~al.}(1995)\citenamefont {Dzyanyj}, \citenamefont {Bodak}, \citenamefont {Aksel'rud},\ and\ \citenamefont {Pavlyuk}}]{dzyanyj1995crystal}%
  \BibitemOpen
  \bibfield  {author} {\bibinfo {author} {\bibfnamefont {R.}~\bibnamefont {Dzyanyj}}, \bibinfo {author} {\bibfnamefont {O.}~\bibnamefont {Bodak}}, \bibinfo {author} {\bibfnamefont {L.}~\bibnamefont {Aksel'rud}},\ and\ \bibinfo {author} {\bibfnamefont {V.}~\bibnamefont {Pavlyuk}},\ }\bibfield  {title} {\bibinfo {title} {{Crystal structure of YbM$_6$Ge$_6$ (M= Fe, Co, Mn) compounds}},\ }\href@noop {} {\bibfield  {journal} {\bibinfo  {journal} {Neorganicheskie Materialy}\ }\textbf {\bibinfo {volume} {31}},\ \bibinfo {pages} {987} (\bibinfo {year} {1995})}\BibitemShut {NoStop}%
\bibitem [{\citenamefont {He}\ \emph {et~al.}(2024)\citenamefont {He}, \citenamefont {Xu}, \citenamefont {Li}, \citenamefont {Zeng}, \citenamefont {Liu}, \citenamefont {Zhao}, \citenamefont {Zhou}, \citenamefont {Zhou},\ and\ \citenamefont {Qu}}]{he2024quantum}%
  \BibitemOpen
  \bibfield  {author} {\bibinfo {author} {\bibfnamefont {M.}~\bibnamefont {He}}, \bibinfo {author} {\bibfnamefont {X.}~\bibnamefont {Xu}}, \bibinfo {author} {\bibfnamefont {D.}~\bibnamefont {Li}}, \bibinfo {author} {\bibfnamefont {Q.}~\bibnamefont {Zeng}}, \bibinfo {author} {\bibfnamefont {Y.}~\bibnamefont {Liu}}, \bibinfo {author} {\bibfnamefont {H.}~\bibnamefont {Zhao}}, \bibinfo {author} {\bibfnamefont {S.}~\bibnamefont {Zhou}}, \bibinfo {author} {\bibfnamefont {J.}~\bibnamefont {Zhou}},\ and\ \bibinfo {author} {\bibfnamefont {Z.}~\bibnamefont {Qu}},\ }\bibfield  {title} {\bibinfo {title} {{Quantum oscillations in the kagome metals (Ti, Zr, Hf)V$_6$Sn$_6$ at Van Hove filling}},\ }\href {https://doi.org/10.1103/PhysRevB.109.155117} {\bibfield  {journal} {\bibinfo  {journal} {Physical Review B}\ }\textbf {\bibinfo {volume} {109}},\ \bibinfo {pages} {155117} (\bibinfo {year} {2024})}\BibitemShut {NoStop}%
\bibitem [{\citenamefont {Romaka}\ \emph {et~al.}(2011)\citenamefont {Romaka}, \citenamefont {Stadnyk}, \citenamefont {Romaka}, \citenamefont {Demchenko}, \citenamefont {Stadnyshyn},\ and\ \citenamefont {Konyk}}]{romaka2011peculiarities}%
  \BibitemOpen
  \bibfield  {author} {\bibinfo {author} {\bibfnamefont {L.}~\bibnamefont {Romaka}}, \bibinfo {author} {\bibfnamefont {Y.}~\bibnamefont {Stadnyk}}, \bibinfo {author} {\bibfnamefont {V.}~\bibnamefont {Romaka}}, \bibinfo {author} {\bibfnamefont {P.}~\bibnamefont {Demchenko}}, \bibinfo {author} {\bibfnamefont {M.}~\bibnamefont {Stadnyshyn}},\ and\ \bibinfo {author} {\bibfnamefont {M.}~\bibnamefont {Konyk}},\ }\bibfield  {title} {\bibinfo {title} {{Peculiarities of component interaction in $\{$Gd, Er$\}$--V--Sn Ternary systems at 870 K and crystal structure of RV$_6$Sn$_6$ stannides}},\ }\href {https://doi.org/10.1016/j.jallcom.2011.06.095} {\bibfield  {journal} {\bibinfo  {journal} {Journal of alloys and compounds}\ }\textbf {\bibinfo {volume} {509}},\ \bibinfo {pages} {8862} (\bibinfo {year} {2011})}\BibitemShut {NoStop}%
\bibitem [{\citenamefont {Romaka}\ \emph {et~al.}(2019)\citenamefont {Romaka}, \citenamefont {Konyk}, \citenamefont {Stadnyk}, \citenamefont {Romaka},\ and\ \citenamefont {Serkiz}}]{romaka2019lu}%
  \BibitemOpen
  \bibfield  {author} {\bibinfo {author} {\bibfnamefont {L.}~\bibnamefont {Romaka}}, \bibinfo {author} {\bibfnamefont {M.}~\bibnamefont {Konyk}}, \bibinfo {author} {\bibfnamefont {Y.}~\bibnamefont {Stadnyk}}, \bibinfo {author} {\bibfnamefont {V.}~\bibnamefont {Romaka}},\ and\ \bibinfo {author} {\bibfnamefont {R.}~\bibnamefont {Serkiz}},\ }\bibfield  {title} {\bibinfo {title} {{Lu-V-$\{$Ge, Sn$\}$ ternary systems}},\ }\href {https://doi.org/10.15330/pcss.20.1.76} {\bibfield  {journal} {\bibinfo  {journal} {Physics and Chemistry of Solid State}\ }\textbf {\bibinfo {volume} {20}},\ \bibinfo {pages} {69} (\bibinfo {year} {2019})}\BibitemShut {NoStop}%
\bibitem [{\citenamefont {Guo}\ \emph {et~al.}(2023)\citenamefont {Guo}, \citenamefont {Ye}, \citenamefont {Guan},\ and\ \citenamefont {Jia}}]{guo2023triangular}%
  \BibitemOpen
  \bibfield  {author} {\bibinfo {author} {\bibfnamefont {K.}~\bibnamefont {Guo}}, \bibinfo {author} {\bibfnamefont {J.}~\bibnamefont {Ye}}, \bibinfo {author} {\bibfnamefont {S.}~\bibnamefont {Guan}},\ and\ \bibinfo {author} {\bibfnamefont {S.}~\bibnamefont {Jia}},\ }\bibfield  {title} {\bibinfo {title} {{Triangular Kondo lattice in YbV$_6$Sn$_6$ and its quantum critical behavior in a magnetic field}},\ }\href {https://doi.org/10.1103/PhysRevB.107.205151} {\bibfield  {journal} {\bibinfo  {journal} {Physical Review B}\ }\textbf {\bibinfo {volume} {107}},\ \bibinfo {pages} {205151} (\bibinfo {year} {2023})}\BibitemShut {NoStop}%
\bibitem [{\citenamefont {Zhang}\ \emph {et~al.}(2022{\natexlab{a}})\citenamefont {Zhang}, \citenamefont {Liu}, \citenamefont {Cui}, \citenamefont {Guo}, \citenamefont {Wang}, \citenamefont {Shi}, \citenamefont {Zhang}, \citenamefont {Wang}, \citenamefont {Dong}, \citenamefont {Sun} \emph {et~al.}}]{zhang2022electronic}%
  \BibitemOpen
  \bibfield  {author} {\bibinfo {author} {\bibfnamefont {X.}~\bibnamefont {Zhang}}, \bibinfo {author} {\bibfnamefont {Z.}~\bibnamefont {Liu}}, \bibinfo {author} {\bibfnamefont {Q.}~\bibnamefont {Cui}}, \bibinfo {author} {\bibfnamefont {Q.}~\bibnamefont {Guo}}, \bibinfo {author} {\bibfnamefont {N.}~\bibnamefont {Wang}}, \bibinfo {author} {\bibfnamefont {L.}~\bibnamefont {Shi}}, \bibinfo {author} {\bibfnamefont {H.}~\bibnamefont {Zhang}}, \bibinfo {author} {\bibfnamefont {W.}~\bibnamefont {Wang}}, \bibinfo {author} {\bibfnamefont {X.}~\bibnamefont {Dong}}, \bibinfo {author} {\bibfnamefont {J.}~\bibnamefont {Sun}}, \emph {et~al.},\ }\bibfield  {title} {\bibinfo {title} {{Electronic and magnetic properties of intermetallic kagome magnets RV$_6$Sn$_6$ (R= Tb-Tm)}},\ }\href@noop {} {\bibfield  {journal} {\bibinfo  {journal} {Physical Review Materials}\ }\textbf {\bibinfo {volume} {6}},\ \bibinfo {pages} {105001} (\bibinfo {year} {2022}{\natexlab{a}})}\BibitemShut {NoStop}%
\bibitem [{\citenamefont {Huang}\ \emph {et~al.}(2023)\citenamefont {Huang}, \citenamefont {Cui}, \citenamefont {Huang}, \citenamefont {Huo}, \citenamefont {Liu}, \citenamefont {Li}, \citenamefont {Liang}, \citenamefont {Chen}, \citenamefont {Sun}, \citenamefont {Shen} \emph {et~al.}}]{huang2023anisotropic}%
  \BibitemOpen
  \bibfield  {author} {\bibinfo {author} {\bibfnamefont {X.}~\bibnamefont {Huang}}, \bibinfo {author} {\bibfnamefont {Z.}~\bibnamefont {Cui}}, \bibinfo {author} {\bibfnamefont {C.}~\bibnamefont {Huang}}, \bibinfo {author} {\bibfnamefont {M.}~\bibnamefont {Huo}}, \bibinfo {author} {\bibfnamefont {H.}~\bibnamefont {Liu}}, \bibinfo {author} {\bibfnamefont {J.}~\bibnamefont {Li}}, \bibinfo {author} {\bibfnamefont {F.}~\bibnamefont {Liang}}, \bibinfo {author} {\bibfnamefont {L.}~\bibnamefont {Chen}}, \bibinfo {author} {\bibfnamefont {H.}~\bibnamefont {Sun}}, \bibinfo {author} {\bibfnamefont {B.}~\bibnamefont {Shen}}, \emph {et~al.},\ }\bibfield  {title} {\bibinfo {title} {{Anisotropic magnetism and electronic properties of the kagome metal SmV$_6$Sn$_6$}},\ }\href {https://doi.org/10.1103/PhysRevMaterials.7.054403} {\bibfield  {journal} {\bibinfo  {journal} {Physical Review Materials}\ }\textbf {\bibinfo {volume} {7}},\ \bibinfo {pages} {054403} (\bibinfo {year} {2023})}\BibitemShut {NoStop}%
\bibitem [{\citenamefont {Zeng}\ \emph {et~al.}(2024)\citenamefont {Zeng}, \citenamefont {Wang}, \citenamefont {Wang}, \citenamefont {Lin}, \citenamefont {Gong}, \citenamefont {Ma}, \citenamefont {Han}, \citenamefont {Wang}, \citenamefont {Dai},\ and\ \citenamefont {Xia}}]{zeng2024magnetic}%
  \BibitemOpen
  \bibfield  {author} {\bibinfo {author} {\bibfnamefont {X.-Y.}\ \bibnamefont {Zeng}}, \bibinfo {author} {\bibfnamefont {H.}~\bibnamefont {Wang}}, \bibinfo {author} {\bibfnamefont {X.-Y.}\ \bibnamefont {Wang}}, \bibinfo {author} {\bibfnamefont {J.-F.}\ \bibnamefont {Lin}}, \bibinfo {author} {\bibfnamefont {J.}~\bibnamefont {Gong}}, \bibinfo {author} {\bibfnamefont {X.-P.}\ \bibnamefont {Ma}}, \bibinfo {author} {\bibfnamefont {K.}~\bibnamefont {Han}}, \bibinfo {author} {\bibfnamefont {Y.-T.}\ \bibnamefont {Wang}}, \bibinfo {author} {\bibfnamefont {Z.-Y.}\ \bibnamefont {Dai}},\ and\ \bibinfo {author} {\bibfnamefont {T.-L.}\ \bibnamefont {Xia}},\ }\bibfield  {title} {\bibinfo {title} {{Magnetic and magnetotransport properties in the vanadium-based kagome metals DyV$_6$Sn$_6$ and HoV$_6$Sn$_6$}},\ }\href@noop {} {\bibfield  {journal} {\bibinfo  {journal} {Physical Review B}\ }\textbf {\bibinfo {volume} {109}},\ \bibinfo {pages} {104412} (\bibinfo {year} {2024})}\BibitemShut {NoStop}%
\bibitem [{\citenamefont {Pokharel}\ \emph {et~al.}(2022)\citenamefont {Pokharel}, \citenamefont {Ortiz}, \citenamefont {Chamorro}, \citenamefont {Sarte}, \citenamefont {Kautzsch}, \citenamefont {Wu}, \citenamefont {Ruff},\ and\ \citenamefont {Wilson}}]{pokharel2022highly}%
  \BibitemOpen
  \bibfield  {author} {\bibinfo {author} {\bibfnamefont {G.}~\bibnamefont {Pokharel}}, \bibinfo {author} {\bibfnamefont {B.}~\bibnamefont {Ortiz}}, \bibinfo {author} {\bibfnamefont {J.}~\bibnamefont {Chamorro}}, \bibinfo {author} {\bibfnamefont {P.}~\bibnamefont {Sarte}}, \bibinfo {author} {\bibfnamefont {L.}~\bibnamefont {Kautzsch}}, \bibinfo {author} {\bibfnamefont {G.}~\bibnamefont {Wu}}, \bibinfo {author} {\bibfnamefont {J.}~\bibnamefont {Ruff}},\ and\ \bibinfo {author} {\bibfnamefont {S.~D.}\ \bibnamefont {Wilson}},\ }\bibfield  {title} {\bibinfo {title} {{Highly anisotropic magnetism in the vanadium-based kagome metal TbV$_6$Sn$_6$}},\ }\href {https://doi.org/10.1103/PhysRevMaterials.6.104202} {\bibfield  {journal} {\bibinfo  {journal} {Physical Review Materials}\ }\textbf {\bibinfo {volume} {6}},\ \bibinfo {pages} {104202} (\bibinfo {year} {2022})}\BibitemShut {NoStop}%
\bibitem [{\citenamefont {Rosenberg}\ \emph {et~al.}(2022)\citenamefont {Rosenberg}, \citenamefont {DeStefano}, \citenamefont {Guo}, \citenamefont {Oh}, \citenamefont {Hashimoto}, \citenamefont {Lu}, \citenamefont {Birgeneau}, \citenamefont {Lee}, \citenamefont {Ke}, \citenamefont {Yi} \emph {et~al.}}]{rosenberg2022uniaxial}%
  \BibitemOpen
  \bibfield  {author} {\bibinfo {author} {\bibfnamefont {E.}~\bibnamefont {Rosenberg}}, \bibinfo {author} {\bibfnamefont {J.~M.}\ \bibnamefont {DeStefano}}, \bibinfo {author} {\bibfnamefont {Y.}~\bibnamefont {Guo}}, \bibinfo {author} {\bibfnamefont {J.~S.}\ \bibnamefont {Oh}}, \bibinfo {author} {\bibfnamefont {M.}~\bibnamefont {Hashimoto}}, \bibinfo {author} {\bibfnamefont {D.}~\bibnamefont {Lu}}, \bibinfo {author} {\bibfnamefont {R.~J.}\ \bibnamefont {Birgeneau}}, \bibinfo {author} {\bibfnamefont {Y.}~\bibnamefont {Lee}}, \bibinfo {author} {\bibfnamefont {L.}~\bibnamefont {Ke}}, \bibinfo {author} {\bibfnamefont {M.}~\bibnamefont {Yi}}, \emph {et~al.},\ }\bibfield  {title} {\bibinfo {title} {{Uniaxial ferromagnetism in the kagome metal TbV$_6$Sn$_6$}},\ }\href {https://doi.org/10.1103/PhysRevB.106.115139} {\bibfield  {journal} {\bibinfo  {journal} {Physical Review B}\ }\textbf {\bibinfo {volume} {106}},\ \bibinfo {pages} {115139} (\bibinfo {year} {2022})}\BibitemShut {NoStop}%
\bibitem [{\citenamefont {Pokharel}\ \emph {et~al.}(2021)\citenamefont {Pokharel}, \citenamefont {Teicher}, \citenamefont {Ortiz}, \citenamefont {Sarte}, \citenamefont {Wu}, \citenamefont {Peng}, \citenamefont {He}, \citenamefont {Seshadri},\ and\ \citenamefont {Wilson}}]{pokharel2021electronic}%
  \BibitemOpen
  \bibfield  {author} {\bibinfo {author} {\bibfnamefont {G.}~\bibnamefont {Pokharel}}, \bibinfo {author} {\bibfnamefont {S.~M.}\ \bibnamefont {Teicher}}, \bibinfo {author} {\bibfnamefont {B.~R.}\ \bibnamefont {Ortiz}}, \bibinfo {author} {\bibfnamefont {P.~M.}\ \bibnamefont {Sarte}}, \bibinfo {author} {\bibfnamefont {G.}~\bibnamefont {Wu}}, \bibinfo {author} {\bibfnamefont {S.}~\bibnamefont {Peng}}, \bibinfo {author} {\bibfnamefont {J.}~\bibnamefont {He}}, \bibinfo {author} {\bibfnamefont {R.}~\bibnamefont {Seshadri}},\ and\ \bibinfo {author} {\bibfnamefont {S.~D.}\ \bibnamefont {Wilson}},\ }\bibfield  {title} {\bibinfo {title} {{Electronic properties of the topological kagome metals YV$_6$Sn$_6$ and GdV$_6$Sn$_6$}},\ }\href {https://doi.org/10.1103/PhysRevB.104.235139} {\bibfield  {journal} {\bibinfo  {journal} {Physical Review B}\ }\textbf {\bibinfo {volume} {104}},\ \bibinfo {pages} {235139} (\bibinfo {year} {2021})}\BibitemShut {NoStop}%
\bibitem [{\citenamefont {Lee}\ and\ \citenamefont {Mun}(2022)}]{lee2022anisotropic}%
  \BibitemOpen
  \bibfield  {author} {\bibinfo {author} {\bibfnamefont {J.}~\bibnamefont {Lee}}\ and\ \bibinfo {author} {\bibfnamefont {E.}~\bibnamefont {Mun}},\ }\bibfield  {title} {\bibinfo {title} {{Anisotropic magnetic property of single crystals RV$_6$Sn$_6$ (R= Y, Gd-Tm, Lu)}},\ }\href {https://doi.org/10.1103/PhysRevMaterials.6.083401} {\bibfield  {journal} {\bibinfo  {journal} {Physical Review Materials}\ }\textbf {\bibinfo {volume} {6}},\ \bibinfo {pages} {083401} (\bibinfo {year} {2022})}\BibitemShut {NoStop}%
\bibitem [{\citenamefont {Mazet}\ \emph {et~al.}(1999{\natexlab{b}})\citenamefont {Mazet}, \citenamefont {Welter},\ and\ \citenamefont {Malaman}}]{mazet1999study}%
  \BibitemOpen
  \bibfield  {author} {\bibinfo {author} {\bibfnamefont {T.}~\bibnamefont {Mazet}}, \bibinfo {author} {\bibfnamefont {R.}~\bibnamefont {Welter}},\ and\ \bibinfo {author} {\bibfnamefont {B.}~\bibnamefont {Malaman}},\ }\bibfield  {title} {\bibinfo {title} {{A study of the new HfFe$_6$Ge$_6$-type ZrMn$_6$Sn$_6$ and HfMn$_6$Sn$_6$ compounds by magnetization and neutron diffraction measurements}},\ }\href {https://doi.org/10.1016/s0925-8388(98)00923-2} {\bibfield  {journal} {\bibinfo  {journal} {Journal of alloys and compounds}\ }\textbf {\bibinfo {volume} {284}},\ \bibinfo {pages} {54} (\bibinfo {year} {1999}{\natexlab{b}})}\BibitemShut {NoStop}%
\bibitem [{\citenamefont {El~Idrissi}\ \emph {et~al.}(1991{\natexlab{b}})\citenamefont {El~Idrissi}, \citenamefont {Venturini},\ and\ \citenamefont {Malaman}}]{el1991refinement}%
  \BibitemOpen
  \bibfield  {author} {\bibinfo {author} {\bibfnamefont {B.~C.}\ \bibnamefont {El~Idrissi}}, \bibinfo {author} {\bibfnamefont {G.}~\bibnamefont {Venturini}},\ and\ \bibinfo {author} {\bibfnamefont {B.}~\bibnamefont {Malaman}},\ }\bibfield  {title} {\bibinfo {title} {{Refinement of HfFe$_6$Ge$_6$ isostructural ScMn$_6$Sn$_6$ and TbMn$_6$Sn$_6$}},\ }\href {https://doi.org/10.1016/0025-5408(91)90181-k} {\bibfield  {journal} {\bibinfo  {journal} {Materials research bulletin}\ }\textbf {\bibinfo {volume} {26}},\ \bibinfo {pages} {431} (\bibinfo {year} {1991}{\natexlab{b}})}\BibitemShut {NoStop}%
\bibitem [{\citenamefont {Xia}\ and\ \citenamefont {Bobev}(2006)}]{xia2006ybmn6sn6}%
  \BibitemOpen
  \bibfield  {author} {\bibinfo {author} {\bibfnamefont {S.-Q.}\ \bibnamefont {Xia}}\ and\ \bibinfo {author} {\bibfnamefont {S.}~\bibnamefont {Bobev}},\ }\bibfield  {title} {\bibinfo {title} {{YbMn$_6$Sn$_6$}},\ }\href {https://doi.org/10.1107/S1600536805040006} {\bibfield  {journal} {\bibinfo  {journal} {Acta Crystallographica Section E: Structure Reports Online}\ }\textbf {\bibinfo {volume} {62}},\ \bibinfo {pages} {i7} (\bibinfo {year} {2006})}\BibitemShut {NoStop}%
\bibitem [{\citenamefont {Weitzer}\ \emph {et~al.}(1993)\citenamefont {Weitzer}, \citenamefont {Leithe-Jasper}, \citenamefont {Hiebl}, \citenamefont {Rogl}, \citenamefont {Qi},\ and\ \citenamefont {Coey}}]{weitzer1993structural}%
  \BibitemOpen
  \bibfield  {author} {\bibinfo {author} {\bibfnamefont {F.}~\bibnamefont {Weitzer}}, \bibinfo {author} {\bibfnamefont {A.}~\bibnamefont {Leithe-Jasper}}, \bibinfo {author} {\bibfnamefont {K.}~\bibnamefont {Hiebl}}, \bibinfo {author} {\bibfnamefont {P.}~\bibnamefont {Rogl}}, \bibinfo {author} {\bibfnamefont {Q.}~\bibnamefont {Qi}},\ and\ \bibinfo {author} {\bibfnamefont {J.}~\bibnamefont {Coey}},\ }\bibfield  {title} {\bibinfo {title} {{Structural chemistry, magnetism and 119Sn M{\"o}ssbauer spectroscopy of ternary compounds REMn$_6$Sn$_6$ (RE= Pr, Nd, Sm)}},\ }\href {https://doi.org/10.1063/1.353416} {\bibfield  {journal} {\bibinfo  {journal} {Journal of applied physics}\ }\textbf {\bibinfo {volume} {73}},\ \bibinfo {pages} {8447} (\bibinfo {year} {1993})}\BibitemShut {NoStop}%
\bibitem [{\citenamefont {Clatterbuck}\ and\ \citenamefont {Gschneidner~Jr}(1999)}]{clatterbuck1999magnetic}%
  \BibitemOpen
  \bibfield  {author} {\bibinfo {author} {\bibfnamefont {D.}~\bibnamefont {Clatterbuck}}\ and\ \bibinfo {author} {\bibfnamefont {K.}~\bibnamefont {Gschneidner~Jr}},\ }\bibfield  {title} {\bibinfo {title} {{Magnetic properties of RMn$_6$Sn$_6$ (R= Tb, Ho, Er, Tm, Lu) single crystals}},\ }\href {https://doi.org/10.1016/s0304-8853(99)00571-5} {\bibfield  {journal} {\bibinfo  {journal} {Journal of magnetism and magnetic materials}\ }\textbf {\bibinfo {volume} {207}},\ \bibinfo {pages} {78} (\bibinfo {year} {1999})}\BibitemShut {NoStop}%
\bibitem [{\citenamefont {Venturini}\ \emph {et~al.}(1993)\citenamefont {Venturini}, \citenamefont {Welter}, \citenamefont {Malaman},\ and\ \citenamefont {Ressouche}}]{venturini1993magnetic}%
  \BibitemOpen
  \bibfield  {author} {\bibinfo {author} {\bibfnamefont {G.}~\bibnamefont {Venturini}}, \bibinfo {author} {\bibfnamefont {R.}~\bibnamefont {Welter}}, \bibinfo {author} {\bibfnamefont {B.}~\bibnamefont {Malaman}},\ and\ \bibinfo {author} {\bibfnamefont {E.}~\bibnamefont {Ressouche}},\ }\bibfield  {title} {\bibinfo {title} {{Magnetic structure of YMn$_6$Ge$_6$ and room temperature magnetic structure of LuMn$_6$Sn$_6$ obtained from neutron diffraction study}},\ }\href {https://doi.org/10.1016/0925-8388(93)90470-8} {\bibfield  {journal} {\bibinfo  {journal} {Journal of alloys and compounds}\ }\textbf {\bibinfo {volume} {200}},\ \bibinfo {pages} {51} (\bibinfo {year} {1993})}\BibitemShut {NoStop}%
\bibitem [{\citenamefont {Mazet}\ \emph {et~al.}(2006)\citenamefont {Mazet}, \citenamefont {Ihou-Mouko}, \citenamefont {Mar{\^e}ch{\'e}},\ and\ \citenamefont {Malaman}}]{mazet2006magnetic}%
  \BibitemOpen
  \bibfield  {author} {\bibinfo {author} {\bibfnamefont {T.}~\bibnamefont {Mazet}}, \bibinfo {author} {\bibfnamefont {H.}~\bibnamefont {Ihou-Mouko}}, \bibinfo {author} {\bibfnamefont {J.}~\bibnamefont {Mar{\^e}ch{\'e}}},\ and\ \bibinfo {author} {\bibfnamefont {B.}~\bibnamefont {Malaman}},\ }\bibfield  {title} {\bibinfo {title} {{Magnetic properties and 119Sn hyperfine interaction parameters of LiMn$_6$Sn$_6$}},\ }\href {https://doi.org/10.1140/epjb/e2006-00207-9} {\bibfield  {journal} {\bibinfo  {journal} {The European Physical Journal B-Condensed Matter and Complex Systems}\ }\textbf {\bibinfo {volume} {51}},\ \bibinfo {pages} {173} (\bibinfo {year} {2006})}\BibitemShut {NoStop}%
\bibitem [{\citenamefont {Mazet}\ and\ \citenamefont {Malaman}(2000{\natexlab{a}})}]{mazet2000local}%
  \BibitemOpen
  \bibfield  {author} {\bibinfo {author} {\bibfnamefont {T.}~\bibnamefont {Mazet}}\ and\ \bibinfo {author} {\bibfnamefont {B.}~\bibnamefont {Malaman}},\ }\bibfield  {title} {\bibinfo {title} {{Local chemical and magnetic disorder within the HfFe$_6$Ge$_6$-type RFe$_6$Sn$_6$ compounds (R= Sc, Tm, Lu and Zr)}},\ }\href {https://doi.org/10.1016/s0304-8853(00)00415-7} {\bibfield  {journal} {\bibinfo  {journal} {Journal of magnetism and magnetic materials}\ }\textbf {\bibinfo {volume} {219}},\ \bibinfo {pages} {33} (\bibinfo {year} {2000}{\natexlab{a}})}\BibitemShut {NoStop}%
\bibitem [{\citenamefont {Mazet}\ \emph {et~al.}(2002)\citenamefont {Mazet}, \citenamefont {Isnard},\ and\ \citenamefont {Malaman}}]{mazet2002study}%
  \BibitemOpen
  \bibfield  {author} {\bibinfo {author} {\bibfnamefont {T.}~\bibnamefont {Mazet}}, \bibinfo {author} {\bibfnamefont {O.}~\bibnamefont {Isnard}},\ and\ \bibinfo {author} {\bibfnamefont {B.}~\bibnamefont {Malaman}},\ }\bibfield  {title} {\bibinfo {title} {{A study of the new Yb$_{0.6}$Fe$_6$Sn$_6$ compound by neutron diffraction, 57Fe and 119Sn M{\"o}ssbauer spectroscopy experiments}},\ }\href {https://doi.org/10.1016/S0304-8853(01)00946-5} {\bibfield  {journal} {\bibinfo  {journal} {Journal of magnetism and magnetic materials}\ }\textbf {\bibinfo {volume} {241}},\ \bibinfo {pages} {51} (\bibinfo {year} {2002})}\BibitemShut {NoStop}%
\bibitem [{\citenamefont {Mazet}\ and\ \citenamefont {Malaman}(2000{\natexlab{b}})}]{mazet2000evidence}%
  \BibitemOpen
  \bibfield  {author} {\bibinfo {author} {\bibfnamefont {T.}~\bibnamefont {Mazet}}\ and\ \bibinfo {author} {\bibfnamefont {B.}~\bibnamefont {Malaman}},\ }\bibfield  {title} {\bibinfo {title} {{Evidence of spin reorientation in YbFe$_6$Ge$_6$ from neutron diffraction and 57Fe M{\"o}ssbauer experiments}},\ }\href {https://doi.org/10.1088/0953-8984/12/6/325} {\bibfield  {journal} {\bibinfo  {journal} {Journal of Physics: Condensed Matter}\ }\textbf {\bibinfo {volume} {12}},\ \bibinfo {pages} {1085} (\bibinfo {year} {2000}{\natexlab{b}})}\BibitemShut {NoStop}%
\bibitem [{\citenamefont {Schobinger-Papamantellos}\ \emph {et~al.}(1998)\citenamefont {Schobinger-Papamantellos}, \citenamefont {Buschow}, \citenamefont {De~Boer}, \citenamefont {Ritter}, \citenamefont {Isnard},\ and\ \citenamefont {Fauth}}]{schobinger1998fe}%
  \BibitemOpen
  \bibfield  {author} {\bibinfo {author} {\bibfnamefont {P.}~\bibnamefont {Schobinger-Papamantellos}}, \bibinfo {author} {\bibfnamefont {K.}~\bibnamefont {Buschow}}, \bibinfo {author} {\bibfnamefont {F.}~\bibnamefont {De~Boer}}, \bibinfo {author} {\bibfnamefont {C.}~\bibnamefont {Ritter}}, \bibinfo {author} {\bibfnamefont {O.}~\bibnamefont {Isnard}},\ and\ \bibinfo {author} {\bibfnamefont {F.}~\bibnamefont {Fauth}},\ }\bibfield  {title} {\bibinfo {title} {{The Fe ordering in RFe$_6$Ge$_6$ compounds with non-magnetic R (R= Y, Lu, Hf) studied by neutron diffraction and magnetic measurements}},\ }\href {https://doi.org/10.1016/s0925-8388(97)00548-3} {\bibfield  {journal} {\bibinfo  {journal} {Journal of alloys and compounds}\ }\textbf {\bibinfo {volume} {267}},\ \bibinfo {pages} {59} (\bibinfo {year} {1998})}\BibitemShut {NoStop}%
\bibitem [{\citenamefont {Wang}\ \emph {et~al.}(1994)\citenamefont {Wang}, \citenamefont {Wiards}, \citenamefont {Ryan},\ and\ \citenamefont {Cadogan}}]{wang1994structural}%
  \BibitemOpen
  \bibfield  {author} {\bibinfo {author} {\bibfnamefont {Y.}~\bibnamefont {Wang}}, \bibinfo {author} {\bibfnamefont {D.}~\bibnamefont {Wiards}}, \bibinfo {author} {\bibfnamefont {D.}~\bibnamefont {Ryan}},\ and\ \bibinfo {author} {\bibfnamefont {J.}~\bibnamefont {Cadogan}},\ }\bibfield  {title} {\bibinfo {title} {{Structural and magnetic properties of RFe$_6$Ge$_6$ (R= Y, Gd, Tb, Er)}},\ }\href {https://doi.org/10.1109/20.334276} {\bibfield  {journal} {\bibinfo  {journal} {IEEE Transactions on Magnetics}\ }\textbf {\bibinfo {volume} {30}},\ \bibinfo {pages} {4951} (\bibinfo {year} {1994})}\BibitemShut {NoStop}%
\bibitem [{\citenamefont {Raghavan}(2001)}]{raghavan2001fe}%
  \BibitemOpen
  \bibfield  {author} {\bibinfo {author} {\bibfnamefont {V.}~\bibnamefont {Raghavan}},\ }\bibfield  {title} {\bibinfo {title} {{Fe-Sm-Sn (iron-samarium-tin)}},\ }\href {https://doi.org/10.1361/105497101770339139} {\bibfield  {journal} {\bibinfo  {journal} {Journal of phase equilibria}\ }\textbf {\bibinfo {volume} {22}},\ \bibinfo {pages} {169} (\bibinfo {year} {2001})}\BibitemShut {NoStop}%
\bibitem [{\citenamefont {Stepie{\'n}-Damm}\ \emph {et~al.}(2000)\citenamefont {Stepie{\'n}-Damm}, \citenamefont {Ga{\l}decka}, \citenamefont {Bodak},\ and\ \citenamefont {Belan}}]{stȩpien2000ternary}%
  \BibitemOpen
  \bibfield  {author} {\bibinfo {author} {\bibfnamefont {J.}~\bibnamefont {Stepie{\'n}-Damm}}, \bibinfo {author} {\bibfnamefont {E.}~\bibnamefont {Ga{\l}decka}}, \bibinfo {author} {\bibfnamefont {O.}~\bibnamefont {Bodak}},\ and\ \bibinfo {author} {\bibfnamefont {B.}~\bibnamefont {Belan}},\ }\bibfield  {title} {\bibinfo {title} {{The ternary Sm--Fe--Sn system: phase diagram, structural characterisation and magnetic properties of ternary compounds}},\ }\href {https://doi.org/10.1016/s0925-8388(99)00626-x} {\bibfield  {journal} {\bibinfo  {journal} {Journal of alloys and compounds}\ }\textbf {\bibinfo {volume} {298}},\ \bibinfo {pages} {26} (\bibinfo {year} {2000})}\BibitemShut {NoStop}%
\bibitem [{\citenamefont {Skolozdra}\ and\ \citenamefont {Koretskaya}(1984)}]{skolozdra1984crystal}%
  \BibitemOpen
  \bibfield  {author} {\bibinfo {author} {\bibfnamefont {R.}~\bibnamefont {Skolozdra}}\ and\ \bibinfo {author} {\bibfnamefont {O.~E.}\ \bibnamefont {Koretskaya}},\ }\bibfield  {title} {\bibinfo {title} {{Crystal structure and magnetic susceptibility of RCo$_6$Sn$_6$ compounds (R= Y, Tb, Dy, Ho, Er, Tm, Lu)}},\ }\href@noop {} {\bibfield  {journal} {\bibinfo  {journal} {Ukrainskij Fizicheskij Zhurnal}\ }\textbf {\bibinfo {volume} {29}},\ \bibinfo {pages} {877} (\bibinfo {year} {1984})}\BibitemShut {NoStop}%
\bibitem [{\citenamefont {Skolozdra}(1997)}]{Skolozdra1997StannidesRE+TM}%
  \BibitemOpen
  \bibfield  {author} {\bibinfo {author} {\bibfnamefont {R.}~\bibnamefont {Skolozdra}},\ }\bibfield  {title} {\bibinfo {title} {{Chapter 164 Stannides of rare-earth and transition metals}},\ }\href {https://doi.org/10.1016/s0168-1273(97)24009-2} {\bibfield  {journal} {\bibinfo  {journal} {Department of Inorganic Chemistry, IL) an Franko State University, Kyryl and Mefodiy str., 6, 290005, Lvio, Ukraine}\ ,\ \bibinfo {pages} {470}} (\bibinfo {year} {1997})}\BibitemShut {NoStop}%
\bibitem [{\citenamefont {Skolozdra}\ \emph {et~al.}(2000)\citenamefont {Skolozdra}, \citenamefont {Mudryk},\ and\ \citenamefont {Romaka}}]{skolozdra2000ternary}%
  \BibitemOpen
  \bibfield  {author} {\bibinfo {author} {\bibfnamefont {R.}~\bibnamefont {Skolozdra}}, \bibinfo {author} {\bibfnamefont {Y.~S.}\ \bibnamefont {Mudryk}},\ and\ \bibinfo {author} {\bibfnamefont {L.}~\bibnamefont {Romaka}},\ }\bibfield  {title} {\bibinfo {title} {{The ternary Er--Co--Sn system}},\ }\href {https://doi.org/10.1016/s0925-8388(99)00545-9} {\bibfield  {journal} {\bibinfo  {journal} {Journal of alloys and compounds}\ }\textbf {\bibinfo {volume} {296}},\ \bibinfo {pages} {290} (\bibinfo {year} {2000})}\BibitemShut {NoStop}%
\bibitem [{\citenamefont {Zhuang}\ \emph {et~al.}(2008)\citenamefont {Zhuang}, \citenamefont {Zhu}, \citenamefont {Yan}, \citenamefont {Xu},\ and\ \citenamefont {Li}}]{zhuang2008phase}%
  \BibitemOpen
  \bibfield  {author} {\bibinfo {author} {\bibfnamefont {Y.}~\bibnamefont {Zhuang}}, \bibinfo {author} {\bibfnamefont {J.}~\bibnamefont {Zhu}}, \bibinfo {author} {\bibfnamefont {J.}~\bibnamefont {Yan}}, \bibinfo {author} {\bibfnamefont {Y.}~\bibnamefont {Xu}},\ and\ \bibinfo {author} {\bibfnamefont {J.}~\bibnamefont {Li}},\ }\bibfield  {title} {\bibinfo {title} {{Phase relationships in the Dy--Co--Sn system at 773 K}},\ }\href@noop {} {\bibfield  {journal} {\bibinfo  {journal} {Journal of alloys and compounds}\ }\textbf {\bibinfo {volume} {459}},\ \bibinfo {pages} {461} (\bibinfo {year} {2008})}\BibitemShut {NoStop}%
\bibitem [{\citenamefont {Oshchapovsky}\ \emph {et~al.}(2010)\citenamefont {Oshchapovsky}, \citenamefont {Pavlyuk}, \citenamefont {F{\"a}ssler},\ and\ \citenamefont {Hlukhyy}}]{oshchapovsky2010tbnb6sn6}%
  \BibitemOpen
  \bibfield  {author} {\bibinfo {author} {\bibfnamefont {I.}~\bibnamefont {Oshchapovsky}}, \bibinfo {author} {\bibfnamefont {V.}~\bibnamefont {Pavlyuk}}, \bibinfo {author} {\bibfnamefont {T.~F.}\ \bibnamefont {F{\"a}ssler}},\ and\ \bibinfo {author} {\bibfnamefont {V.}~\bibnamefont {Hlukhyy}},\ }\bibfield  {title} {\bibinfo {title} {{TbNb$_6$Sn$_6$: the first ternary compound from the rare earth--niobium--tin system}},\ }\href {https://doi.org/10.1107/S1600536810045964} {\bibfield  {journal} {\bibinfo  {journal} {Acta Crystallographica Section E: Structure Reports Online}\ }\textbf {\bibinfo {volume} {66}},\ \bibinfo {pages} {i82} (\bibinfo {year} {2010})}\BibitemShut {NoStop}%
\bibitem [{\citenamefont {Yue}\ and\ \citenamefont {Lei}(2012)}]{yue2012syntheses}%
  \BibitemOpen
  \bibfield  {author} {\bibinfo {author} {\bibfnamefont {C.-Y.}\ \bibnamefont {Yue}}\ and\ \bibinfo {author} {\bibfnamefont {X.-W.}\ \bibnamefont {Lei}},\ }\bibfield  {title} {\bibinfo {title} {{Syntheses and Structures of Sc$_2$Nb$_{4-x}$Sn$_5$, YNb$_6$Sn$_6$, and ErNb$_6$Sn$_5$: Exploratory Studies in Ternary Rare-Earth Niobium Stannides}},\ }\href@noop {} {\bibfield  {journal} {\bibinfo  {journal} {Inorganic Chemistry}\ }\textbf {\bibinfo {volume} {51}},\ \bibinfo {pages} {2461} (\bibinfo {year} {2012})}\BibitemShut {NoStop}%
\bibitem [{\citenamefont {Savidan}\ \emph {et~al.}(2010)\citenamefont {Savidan}, \citenamefont {Joubert},\ and\ \citenamefont {Toffolon-Masclet}}]{savidan2010experimental}%
  \BibitemOpen
  \bibfield  {author} {\bibinfo {author} {\bibfnamefont {J.-C.}\ \bibnamefont {Savidan}}, \bibinfo {author} {\bibfnamefont {J.-M.}\ \bibnamefont {Joubert}},\ and\ \bibinfo {author} {\bibfnamefont {C.}~\bibnamefont {Toffolon-Masclet}},\ }\bibfield  {title} {\bibinfo {title} {{An experimental study of the Fe--Sn--Zr ternary system at 900° C}},\ }\href {https://doi.org/10.1016/j.intermet.2010.07.007} {\bibfield  {journal} {\bibinfo  {journal} {Intermetallics}\ }\textbf {\bibinfo {volume} {18}},\ \bibinfo {pages} {2224} (\bibinfo {year} {2010})}\BibitemShut {NoStop}%
\bibitem [{\citenamefont {Malaman}\ \emph {et~al.}(1997)\citenamefont {Malaman}, \citenamefont {Venturini}, \citenamefont {Chafik El~Idrissi},\ and\ \citenamefont {Ressouche}}]{malaman1997magnetic}%
  \BibitemOpen
  \bibfield  {author} {\bibinfo {author} {\bibfnamefont {B.}~\bibnamefont {Malaman}}, \bibinfo {author} {\bibfnamefont {G.}~\bibnamefont {Venturini}}, \bibinfo {author} {\bibfnamefont {B.}~\bibnamefont {Chafik El~Idrissi}},\ and\ \bibinfo {author} {\bibfnamefont {E.}~\bibnamefont {Ressouche}},\ }\bibfield  {title} {\bibinfo {title} {Magnetic properties of ndmn$_6$sn$_6$ and smmn$_6$sn$_6$ compounds from susceptibility measurements and neutron diffraction study},\ }\href {https://doi.org/10.1016/s0925-8388(96)02717-x} {\bibfield  {journal} {\bibinfo  {journal} {Journal of Alloys and Compounds}\ }\textbf {\bibinfo {volume} {252}},\ \bibinfo {pages} {41} (\bibinfo {year} {1997})}\BibitemShut {NoStop}%
\bibitem [{\citenamefont {Jovanovic}\ and\ \citenamefont {Schoop}(2022)}]{jovanovic2022simple}%
  \BibitemOpen
  \bibfield  {author} {\bibinfo {author} {\bibfnamefont {M.}~\bibnamefont {Jovanovic}}\ and\ \bibinfo {author} {\bibfnamefont {L.~M.}\ \bibnamefont {Schoop}},\ }\bibfield  {title} {\bibinfo {title} {{Simple chemical rules for predicting band structures of kagome materials}},\ }\href {https://doi.org/10.1021/jacs.2c04183} {\bibfield  {journal} {\bibinfo  {journal} {Journal of the American Chemical Society}\ }\textbf {\bibinfo {volume} {144}},\ \bibinfo {pages} {10978} (\bibinfo {year} {2022})}\BibitemShut {NoStop}%
\bibitem [{\citenamefont {Meier}\ \emph {et~al.}(2020)\citenamefont {Meier}, \citenamefont {Du}, \citenamefont {Okamoto}, \citenamefont {Mohanta}, \citenamefont {May}, \citenamefont {McGuire}, \citenamefont {Bridges}, \citenamefont {Samolyuk},\ and\ \citenamefont {Sales}}]{Meier2020_FlatBandsCoSnType}%
  \BibitemOpen
  \bibfield  {author} {\bibinfo {author} {\bibfnamefont {W.~R.}\ \bibnamefont {Meier}}, \bibinfo {author} {\bibfnamefont {M.-H.}\ \bibnamefont {Du}}, \bibinfo {author} {\bibfnamefont {S.}~\bibnamefont {Okamoto}}, \bibinfo {author} {\bibfnamefont {N.}~\bibnamefont {Mohanta}}, \bibinfo {author} {\bibfnamefont {A.~F.}\ \bibnamefont {May}}, \bibinfo {author} {\bibfnamefont {M.~A.}\ \bibnamefont {McGuire}}, \bibinfo {author} {\bibfnamefont {C.~A.}\ \bibnamefont {Bridges}}, \bibinfo {author} {\bibfnamefont {G.~D.}\ \bibnamefont {Samolyuk}},\ and\ \bibinfo {author} {\bibfnamefont {B.~C.}\ \bibnamefont {Sales}},\ }\bibfield  {title} {\bibinfo {title} {{Flat bands in the CoSn-type compounds}},\ }\href {https://doi.org/10.1103/physrevb.102.075148} {\bibfield  {journal} {\bibinfo  {journal} {Physical Review B}\ }\textbf {\bibinfo {volume} {102}},\ \bibinfo {pages} {075148} (\bibinfo {year} {2020})}\BibitemShut {NoStop}%
\bibitem [{\citenamefont {Venturini}(2006)}]{venturini2006filling}%
  \BibitemOpen
  \bibfield  {author} {\bibinfo {author} {\bibfnamefont {G.}~\bibnamefont {Venturini}},\ }\bibfield  {title} {\bibinfo {title} {{Filling the CoSn host-cell: the HfFe6Ge6-type and the related structures}},\ }\href {https://doi.org/10.1524/zkri.2006.221.5-7.511} {\bibfield  {journal} {\bibinfo  {journal} {Zeitschrift f{\"u}r Kristallographie-Crystalline Materials}\ }\textbf {\bibinfo {volume} {221}},\ \bibinfo {pages} {511} (\bibinfo {year} {2006})}\BibitemShut {NoStop}%
\bibitem [{\citenamefont {Simak}\ \emph {et~al.}(1997)\citenamefont {Simak}, \citenamefont {H{\"a}u{\ss}ermann}, \citenamefont {Abrikosov}, \citenamefont {Eriksson}, \citenamefont {Wills}, \citenamefont {Lidin},\ and\ \citenamefont {Johansson}}]{simak1997stability}%
  \BibitemOpen
  \bibfield  {author} {\bibinfo {author} {\bibfnamefont {S.}~\bibnamefont {Simak}}, \bibinfo {author} {\bibfnamefont {U.}~\bibnamefont {H{\"a}u{\ss}ermann}}, \bibinfo {author} {\bibfnamefont {I.}~\bibnamefont {Abrikosov}}, \bibinfo {author} {\bibfnamefont {O.}~\bibnamefont {Eriksson}}, \bibinfo {author} {\bibfnamefont {J.}~\bibnamefont {Wills}}, \bibinfo {author} {\bibfnamefont {S.}~\bibnamefont {Lidin}},\ and\ \bibinfo {author} {\bibfnamefont {B.}~\bibnamefont {Johansson}},\ }\bibfield  {title} {\bibinfo {title} {{Stability of the anomalous large-void CoSn structure}},\ }\href {https://doi.org/10.1103/physrevlett.79.1333} {\bibfield  {journal} {\bibinfo  {journal} {Physical review letters}\ }\textbf {\bibinfo {volume} {79}},\ \bibinfo {pages} {1333} (\bibinfo {year} {1997})}\BibitemShut {NoStop}%
\bibitem [{\citenamefont {Mazet}\ \emph {et~al.}(1999{\natexlab{c}})\citenamefont {Mazet}, \citenamefont {Welter},\ and\ \citenamefont {Malaman}}]{mazet1999study2}%
  \BibitemOpen
  \bibfield  {author} {\bibinfo {author} {\bibfnamefont {T.}~\bibnamefont {Mazet}}, \bibinfo {author} {\bibfnamefont {R.}~\bibnamefont {Welter}},\ and\ \bibinfo {author} {\bibfnamefont {B.}~\bibnamefont {Malaman}},\ }\bibfield  {title} {\bibinfo {title} {{A study of the new ferromagnetic YbMn$_6$Sn$_6$ compound by magnetization and neutron diffraction measurements}},\ }\href {https://doi.org/10.1016/s0304-8853(99)00452-7} {\bibfield  {journal} {\bibinfo  {journal} {Journal of magnetism and magnetic materials}\ }\textbf {\bibinfo {volume} {204}},\ \bibinfo {pages} {11} (\bibinfo {year} {1999}{\natexlab{c}})}\BibitemShut {NoStop}%
\bibitem [{\citenamefont {Fredrickson}\ \emph {et~al.}(2008)\citenamefont {Fredrickson}, \citenamefont {Lidin}, \citenamefont {Venturini}, \citenamefont {Malaman},\ and\ \citenamefont {Christensen}}]{fredrickson2008origins}%
  \BibitemOpen
  \bibfield  {author} {\bibinfo {author} {\bibfnamefont {D.~C.}\ \bibnamefont {Fredrickson}}, \bibinfo {author} {\bibfnamefont {S.}~\bibnamefont {Lidin}}, \bibinfo {author} {\bibfnamefont {G.}~\bibnamefont {Venturini}}, \bibinfo {author} {\bibfnamefont {B.}~\bibnamefont {Malaman}},\ and\ \bibinfo {author} {\bibfnamefont {J.}~\bibnamefont {Christensen}},\ }\bibfield  {title} {\bibinfo {title} {{Origins of superstructure ordering and incommensurability in stuffed CoSn-Type phases}},\ }\href {https://doi.org/10.1021/ja077380} {\bibfield  {journal} {\bibinfo  {journal} {Journal of the American Chemical Society}\ }\textbf {\bibinfo {volume} {130}},\ \bibinfo {pages} {8195} (\bibinfo {year} {2008})}\BibitemShut {NoStop}%
\bibitem [{\citenamefont {Goto}\ \emph {et~al.}(2004)\citenamefont {Goto}, \citenamefont {Nemoto}, \citenamefont {Yamaguchi}, \citenamefont {Akatsu}, \citenamefont {Yanagisawa}, \citenamefont {Suzuki},\ and\ \citenamefont {Kitazawa}}]{Goto2004_Tunneling+RattlingInClathrate}%
  \BibitemOpen
  \bibfield  {author} {\bibinfo {author} {\bibfnamefont {T.}~\bibnamefont {Goto}}, \bibinfo {author} {\bibfnamefont {Y.}~\bibnamefont {Nemoto}}, \bibinfo {author} {\bibfnamefont {T.}~\bibnamefont {Yamaguchi}}, \bibinfo {author} {\bibfnamefont {M.}~\bibnamefont {Akatsu}}, \bibinfo {author} {\bibfnamefont {T.}~\bibnamefont {Yanagisawa}}, \bibinfo {author} {\bibfnamefont {O.}~\bibnamefont {Suzuki}},\ and\ \bibinfo {author} {\bibfnamefont {H.}~\bibnamefont {Kitazawa}},\ }\bibfield  {title} {\bibinfo {title} {Tunneling and rattling in clathrate crystal},\ }\href {https://doi.org/10.1103/physrevb.70.184126} {\bibfield  {journal} {\bibinfo  {journal} {Physical Review B}\ }\textbf {\bibinfo {volume} {70}},\ \bibinfo {pages} {184126} (\bibinfo {year} {2004})}\BibitemShut {NoStop}%
\bibitem [{\citenamefont {Ciesielski}\ \emph {et~al.}(2023)\citenamefont {Ciesielski}, \citenamefont {Ortiz}, \citenamefont {Gomes}, \citenamefont {Meschke}, \citenamefont {Adamczyk}, \citenamefont {Braden}, \citenamefont {Kaczorowski}, \citenamefont {Ertekin},\ and\ \citenamefont {Toberer}}]{Ciesielski2023_RattlingThermalConductivityClathrate}%
  \BibitemOpen
  \bibfield  {author} {\bibinfo {author} {\bibfnamefont {K.~M.}\ \bibnamefont {Ciesielski}}, \bibinfo {author} {\bibfnamefont {B.~R.}\ \bibnamefont {Ortiz}}, \bibinfo {author} {\bibfnamefont {L.~C.}\ \bibnamefont {Gomes}}, \bibinfo {author} {\bibfnamefont {V.}~\bibnamefont {Meschke}}, \bibinfo {author} {\bibfnamefont {J.}~\bibnamefont {Adamczyk}}, \bibinfo {author} {\bibfnamefont {T.~L.}\ \bibnamefont {Braden}}, \bibinfo {author} {\bibfnamefont {D.}~\bibnamefont {Kaczorowski}}, \bibinfo {author} {\bibfnamefont {E.}~\bibnamefont {Ertekin}},\ and\ \bibinfo {author} {\bibfnamefont {E.~S.}\ \bibnamefont {Toberer}},\ }\bibfield  {title} {\bibinfo {title} {Strong scattering from low-frequency rattling modes results in low thermal conductivity in antimonide clathrate compounds},\ }\href {https://doi.org/10.1021/acs.chemmater.2c03821} {\bibfield  {journal} {\bibinfo  {journal} {Chemistry of Materials}\ }\textbf {\bibinfo {volume} {35}},\ \bibinfo {pages} {2918} (\bibinfo {year} {2023})}\BibitemShut {NoStop}%
\bibitem [{\citenamefont {Dong}\ \emph {et~al.}(2000)\citenamefont {Dong}, \citenamefont {Sankey}, \citenamefont {Ramachandran},\ and\ \citenamefont {McMillan}}]{Dong2000_ChemTrendsRattlingClathrates}%
  \BibitemOpen
  \bibfield  {author} {\bibinfo {author} {\bibfnamefont {J.}~\bibnamefont {Dong}}, \bibinfo {author} {\bibfnamefont {O.~F.}\ \bibnamefont {Sankey}}, \bibinfo {author} {\bibfnamefont {G.~K.}\ \bibnamefont {Ramachandran}},\ and\ \bibinfo {author} {\bibfnamefont {P.~F.}\ \bibnamefont {McMillan}},\ }\bibfield  {title} {\bibinfo {title} {Chemical trends of the rattling phonon modes in alloyed germanium clathrates},\ }\href {https://doi.org/10.1063/1.373447} {\bibfield  {journal} {\bibinfo  {journal} {Journal of Applied Physics}\ }\textbf {\bibinfo {volume} {87}},\ \bibinfo {pages} {7726} (\bibinfo {year} {2000})}\BibitemShut {NoStop}%
\bibitem [{\citenamefont {Sales}\ \emph {et~al.}(1999)\citenamefont {Sales}, \citenamefont {Chakoumakos}, \citenamefont {Mandrus},\ and\ \citenamefont {Sharp}}]{Sales1999_DisplacementsThermalConductivityClathrateLike}%
  \BibitemOpen
  \bibfield  {author} {\bibinfo {author} {\bibfnamefont {B.}~\bibnamefont {Sales}}, \bibinfo {author} {\bibfnamefont {B.}~\bibnamefont {Chakoumakos}}, \bibinfo {author} {\bibfnamefont {D.}~\bibnamefont {Mandrus}},\ and\ \bibinfo {author} {\bibfnamefont {J.}~\bibnamefont {Sharp}},\ }\bibfield  {title} {\bibinfo {title} {Atomic displacement parameters and the lattice thermal conductivity of clathrate-like thermoelectric compounds},\ }\href {https://doi.org/10.1006/jssc.1999.8354} {\bibfield  {journal} {\bibinfo  {journal} {Journal of Solid State Chemistry}\ }\textbf {\bibinfo {volume} {146}},\ \bibinfo {pages} {528} (\bibinfo {year} {1999})}\BibitemShut {NoStop}%
\bibitem [{\citenamefont {Morozkin}\ \emph {et~al.}(2016)\citenamefont {Morozkin}, \citenamefont {Knotko}, \citenamefont {Garshev}, \citenamefont {Yapaskurt}, \citenamefont {Nirmala}, \citenamefont {Quezado},\ and\ \citenamefont {Malik}}]{morozkin2016ni}%
  \BibitemOpen
  \bibfield  {author} {\bibinfo {author} {\bibfnamefont {A.}~\bibnamefont {Morozkin}}, \bibinfo {author} {\bibfnamefont {A.}~\bibnamefont {Knotko}}, \bibinfo {author} {\bibfnamefont {A.}~\bibnamefont {Garshev}}, \bibinfo {author} {\bibfnamefont {V.}~\bibnamefont {Yapaskurt}}, \bibinfo {author} {\bibfnamefont {R.}~\bibnamefont {Nirmala}}, \bibinfo {author} {\bibfnamefont {S.}~\bibnamefont {Quezado}},\ and\ \bibinfo {author} {\bibfnamefont {S.}~\bibnamefont {Malik}},\ }\bibfield  {title} {\bibinfo {title} {The ce-ni-si system as a representative of the rare earth-ni-si family: Isothermal section and new rare-earth nickel silicides},\ }\href@noop {} {\bibfield  {journal} {\bibinfo  {journal} {Journal of Solid State Chemistry}\ }\textbf {\bibinfo {volume} {243}},\ \bibinfo {pages} {290} (\bibinfo {year} {2016})}\BibitemShut {NoStop}%
\bibitem [{\citenamefont {Ortiz}\ \emph {et~al.}(2023{\natexlab{a}})\citenamefont {Ortiz}, \citenamefont {Pokharel}, \citenamefont {Gundayao}, \citenamefont {Li}, \citenamefont {Kaboudvand}, \citenamefont {Kautzsch}, \citenamefont {Sarker}, \citenamefont {Ruff}, \citenamefont {Hogan}, \citenamefont {Alvarado}, \citenamefont {Sarte}, \citenamefont {Wu}, \citenamefont {Braden}, \citenamefont {Seshadri}, \citenamefont {Toberer}, \citenamefont {Zeljkovic},\ and\ \citenamefont {Wilson}}]{ortiz2023ybv}%
  \BibitemOpen
  \bibfield  {author} {\bibinfo {author} {\bibfnamefont {B.~R.}\ \bibnamefont {Ortiz}}, \bibinfo {author} {\bibfnamefont {G.}~\bibnamefont {Pokharel}}, \bibinfo {author} {\bibfnamefont {M.}~\bibnamefont {Gundayao}}, \bibinfo {author} {\bibfnamefont {H.}~\bibnamefont {Li}}, \bibinfo {author} {\bibfnamefont {F.}~\bibnamefont {Kaboudvand}}, \bibinfo {author} {\bibfnamefont {L.}~\bibnamefont {Kautzsch}}, \bibinfo {author} {\bibfnamefont {S.}~\bibnamefont {Sarker}}, \bibinfo {author} {\bibfnamefont {J.~P.~C.}\ \bibnamefont {Ruff}}, \bibinfo {author} {\bibfnamefont {T.}~\bibnamefont {Hogan}}, \bibinfo {author} {\bibfnamefont {S.~J.~G.}\ \bibnamefont {Alvarado}}, \bibinfo {author} {\bibfnamefont {P.~M.}\ \bibnamefont {Sarte}}, \bibinfo {author} {\bibfnamefont {G.}~\bibnamefont {Wu}}, \bibinfo {author} {\bibfnamefont {T.}~\bibnamefont {Braden}}, \bibinfo {author} {\bibfnamefont {R.}~\bibnamefont {Seshadri}}, \bibinfo {author} {\bibfnamefont {E.~S.}\ \bibnamefont {Toberer}}, \bibinfo {author} {\bibfnamefont
  {I.}~\bibnamefont {Zeljkovic}},\ and\ \bibinfo {author} {\bibfnamefont {S.~D.}\ \bibnamefont {Wilson}},\ }\bibfield  {title} {\bibinfo {title} {{YbV$_3$Sb$_4$ and EuV$_3$Sb$_4$ vanadium-based kagome metals with Yb$^{2+}$ and Eu$^{2+}$ zigzag chains}},\ }\href {https://doi.org/10.1103/PhysRevMaterials.7.064201} {\bibfield  {journal} {\bibinfo  {journal} {Phys. Rev. Mater.}\ }\textbf {\bibinfo {volume} {7}},\ \bibinfo {pages} {064201} (\bibinfo {year} {2023}{\natexlab{a}})}\BibitemShut {NoStop}%
\bibitem [{\citenamefont {Ortiz}\ \emph {et~al.}(2023{\natexlab{b}})\citenamefont {Ortiz}, \citenamefont {Miao}, \citenamefont {Parker}, \citenamefont {Yang}, \citenamefont {Samolyuk}, \citenamefont {Clements}, \citenamefont {Rajapitamahuni}, \citenamefont {Yilmaz}, \citenamefont {Vescovo}, \citenamefont {Yan} \emph {et~al.}}]{ortiz2023evolution}%
  \BibitemOpen
  \bibfield  {author} {\bibinfo {author} {\bibfnamefont {B.~R.}\ \bibnamefont {Ortiz}}, \bibinfo {author} {\bibfnamefont {H.}~\bibnamefont {Miao}}, \bibinfo {author} {\bibfnamefont {D.~S.}\ \bibnamefont {Parker}}, \bibinfo {author} {\bibfnamefont {F.}~\bibnamefont {Yang}}, \bibinfo {author} {\bibfnamefont {G.~D.}\ \bibnamefont {Samolyuk}}, \bibinfo {author} {\bibfnamefont {E.~M.}\ \bibnamefont {Clements}}, \bibinfo {author} {\bibfnamefont {A.}~\bibnamefont {Rajapitamahuni}}, \bibinfo {author} {\bibfnamefont {T.}~\bibnamefont {Yilmaz}}, \bibinfo {author} {\bibfnamefont {E.}~\bibnamefont {Vescovo}}, \bibinfo {author} {\bibfnamefont {J.}~\bibnamefont {Yan}}, \emph {et~al.},\ }\bibfield  {title} {\bibinfo {title} {{Evolution of Highly Anisotropic Magnetism in the Titanium-Based Kagome Metals LnTi$_3$Bi$_4$ (Ln: La{\textperiodcentered}{\textperiodcentered}{\textperiodcentered} Gd$^{3+}$, Eu$^{2+}$, Yb$^{2+}$)}},\ }\href {https://doi.org/10.1021/acs.chemmater.3c02289} {\bibfield  {journal} {\bibinfo  {journal}
  {Chemistry of Materials}\ }\textbf {\bibinfo {volume} {35}},\ \bibinfo {pages} {9756} (\bibinfo {year} {2023}{\natexlab{b}})}\BibitemShut {NoStop}%
\bibitem [{\citenamefont {Ortiz}\ \emph {et~al.}(2024)\citenamefont {Ortiz}, \citenamefont {Zhang}, \citenamefont {Górnicka}, \citenamefont {Parker}, \citenamefont {Samolyuk}, \citenamefont {Yang}, \citenamefont {Miao}, \citenamefont {Lu}, \citenamefont {Moore}, \citenamefont {May},\ and\ \citenamefont {McGuire}}]{ortiz2024intricate}%
  \BibitemOpen
  \bibfield  {author} {\bibinfo {author} {\bibfnamefont {B.~R.}\ \bibnamefont {Ortiz}}, \bibinfo {author} {\bibfnamefont {H.}~\bibnamefont {Zhang}}, \bibinfo {author} {\bibfnamefont {K.}~\bibnamefont {Górnicka}}, \bibinfo {author} {\bibfnamefont {D.~S.}\ \bibnamefont {Parker}}, \bibinfo {author} {\bibfnamefont {G.~D.}\ \bibnamefont {Samolyuk}}, \bibinfo {author} {\bibfnamefont {F.}~\bibnamefont {Yang}}, \bibinfo {author} {\bibfnamefont {H.}~\bibnamefont {Miao}}, \bibinfo {author} {\bibfnamefont {Q.}~\bibnamefont {Lu}}, \bibinfo {author} {\bibfnamefont {R.~G.}\ \bibnamefont {Moore}}, \bibinfo {author} {\bibfnamefont {A.~F.}\ \bibnamefont {May}},\ and\ \bibinfo {author} {\bibfnamefont {M.~A.}\ \bibnamefont {McGuire}},\ }\bibfield  {title} {\bibinfo {title} {{Intricate Magnetic Landscape in Antiferromagnetic Kagome Metal TbTi$_3$Bi$_4$ and Interplay with $Ln_{2-x}$Ti$_{6+x}$Bi$_9$ ($Ln$: Tb{\textperiodcentered}{\textperiodcentered}{\textperiodcentered} Lu) Shurikagome Metals}},\ }\href
  {https://doi.org/10.1021/acs.chemmater.4c01449} {\bibfield  {journal} {\bibinfo  {journal} {Chemistry of Materials}\ }\textbf {\bibinfo {volume} {36}},\ \bibinfo {pages} {8002} (\bibinfo {year} {2024})}\BibitemShut {NoStop}%
\bibitem [{\citenamefont {Ovchinnikov}\ and\ \citenamefont {Bobev}(2018)}]{ovchinnikov2018synthesis}%
  \BibitemOpen
  \bibfield  {author} {\bibinfo {author} {\bibfnamefont {A.}~\bibnamefont {Ovchinnikov}}\ and\ \bibinfo {author} {\bibfnamefont {S.}~\bibnamefont {Bobev}},\ }\bibfield  {title} {\bibinfo {title} {{Synthesis, Crystal and Electronic Structure of the Titanium Bismuthides Sr$_5$Ti$_{12}$Bi$_{19+x}$, Ba$_5$Ti$_{12}$Bi$_{19+x}$, and Sr$_{5-\delta}$Eu$_\delta$Ti$_{12}$Bi$_{19+x}$ (x=0.5--1.0; $\delta$=2.4, 4.0)}},\ }\href {https://doi.org/10.1002/ejic.201701426} {\bibfield  {journal} {\bibinfo  {journal} {Eur. J. Inorg. Chem.}\ }\textbf {\bibinfo {volume} {2018}},\ \bibinfo {pages} {1266} (\bibinfo {year} {2018})}\BibitemShut {NoStop}%
\bibitem [{\citenamefont {Ovchinnikov}\ and\ \citenamefont {Bobev}(2019)}]{ovchinnikov2019bismuth}%
  \BibitemOpen
  \bibfield  {author} {\bibinfo {author} {\bibfnamefont {A.}~\bibnamefont {Ovchinnikov}}\ and\ \bibinfo {author} {\bibfnamefont {S.}~\bibnamefont {Bobev}},\ }\bibfield  {title} {\bibinfo {title} {{Bismuth as a reactive solvent in the synthesis of multicomponent transition-metal-bearing bismuthides}},\ }\href {https://doi.org/10.1021/acs.inorgchem.9b02881} {\bibfield  {journal} {\bibinfo  {journal} {Inorg. Chem.}\ }\textbf {\bibinfo {volume} {59}},\ \bibinfo {pages} {3459} (\bibinfo {year} {2019})}\BibitemShut {NoStop}%
\bibitem [{\citenamefont {Motoyama}\ \emph {et~al.}(2018)\citenamefont {Motoyama}, \citenamefont {Sezaki}, \citenamefont {Gouchi}, \citenamefont {Miyoshi}, \citenamefont {Nishigori}, \citenamefont {Mutou}, \citenamefont {Fujiwara},\ and\ \citenamefont {Uwatoko}}]{motoyama2018magnetic}%
  \BibitemOpen
  \bibfield  {author} {\bibinfo {author} {\bibfnamefont {G.}~\bibnamefont {Motoyama}}, \bibinfo {author} {\bibfnamefont {M.}~\bibnamefont {Sezaki}}, \bibinfo {author} {\bibfnamefont {J.}~\bibnamefont {Gouchi}}, \bibinfo {author} {\bibfnamefont {K.}~\bibnamefont {Miyoshi}}, \bibinfo {author} {\bibfnamefont {S.}~\bibnamefont {Nishigori}}, \bibinfo {author} {\bibfnamefont {T.}~\bibnamefont {Mutou}}, \bibinfo {author} {\bibfnamefont {K.}~\bibnamefont {Fujiwara}},\ and\ \bibinfo {author} {\bibfnamefont {Y.}~\bibnamefont {Uwatoko}},\ }\bibfield  {title} {\bibinfo {title} {{Magnetic properties of new antiferromagnetic heavy-fermion compounds, Ce$_3$TiBi$_5$ and CeTi$_3$Bi$_4$}},\ }\href {https://doi.org/10.1016/j.physb.2017.10.005} {\bibfield  {journal} {\bibinfo  {journal} {Physica B Condens.}\ }\textbf {\bibinfo {volume} {536}},\ \bibinfo {pages} {142} (\bibinfo {year} {2018})}\BibitemShut {NoStop}%
\bibitem [{\citenamefont {Chen}\ \emph {et~al.}(2024{\natexlab{b}})\citenamefont {Chen}, \citenamefont {Zhou}, \citenamefont {Zhang}, \citenamefont {Ji}, \citenamefont {Liao}, \citenamefont {Ji}, \citenamefont {Li}, \citenamefont {Guo}, \citenamefont {Shen}, \citenamefont {Yu}, \citenamefont {Yu}, \citenamefont {Weng},\ and\ \citenamefont {Wang}}]{chen2023134}%
  \BibitemOpen
  \bibfield  {author} {\bibinfo {author} {\bibfnamefont {L.}~\bibnamefont {Chen}}, \bibinfo {author} {\bibfnamefont {Y.}~\bibnamefont {Zhou}}, \bibinfo {author} {\bibfnamefont {H.}~\bibnamefont {Zhang}}, \bibinfo {author} {\bibfnamefont {X.}~\bibnamefont {Ji}}, \bibinfo {author} {\bibfnamefont {K.}~\bibnamefont {Liao}}, \bibinfo {author} {\bibfnamefont {Y.}~\bibnamefont {Ji}}, \bibinfo {author} {\bibfnamefont {Y.}~\bibnamefont {Li}}, \bibinfo {author} {\bibfnamefont {Z.}~\bibnamefont {Guo}}, \bibinfo {author} {\bibfnamefont {X.}~\bibnamefont {Shen}}, \bibinfo {author} {\bibfnamefont {R.}~\bibnamefont {Yu}}, \bibinfo {author} {\bibfnamefont {X.}~\bibnamefont {Yu}}, \bibinfo {author} {\bibfnamefont {H.}~\bibnamefont {Weng}},\ and\ \bibinfo {author} {\bibfnamefont {G.}~\bibnamefont {Wang}},\ }\bibfield  {title} {\bibinfo {title} {Tunable magnetism in titanium-based kagome metals by rare-earth engineering and high pressure},\ }\href {https://doi.org/10.1038/s43246-024-00513-4} {\bibfield  {journal} {\bibinfo
  {journal} {Communications Materials}\ }\textbf {\bibinfo {volume} {5}},\ \bibinfo {pages} {73} (\bibinfo {year} {2024}{\natexlab{b}})}\BibitemShut {NoStop}%
\bibitem [{\citenamefont {Guo}\ \emph {et~al.}(2024)\citenamefont {Guo}, \citenamefont {Zhou}, \citenamefont {Ding}, \citenamefont {Qu}, \citenamefont {Liu}, \citenamefont {Du}, \citenamefont {Zhang}, \citenamefont {Li}, \citenamefont {Zhang}, \citenamefont {Zhou} \emph {et~al.}}]{guo2024tunable}%
  \BibitemOpen
  \bibfield  {author} {\bibinfo {author} {\bibfnamefont {J.}~\bibnamefont {Guo}}, \bibinfo {author} {\bibfnamefont {L.}~\bibnamefont {Zhou}}, \bibinfo {author} {\bibfnamefont {J.}~\bibnamefont {Ding}}, \bibinfo {author} {\bibfnamefont {G.}~\bibnamefont {Qu}}, \bibinfo {author} {\bibfnamefont {Z.}~\bibnamefont {Liu}}, \bibinfo {author} {\bibfnamefont {Y.}~\bibnamefont {Du}}, \bibinfo {author} {\bibfnamefont {H.}~\bibnamefont {Zhang}}, \bibinfo {author} {\bibfnamefont {J.}~\bibnamefont {Li}}, \bibinfo {author} {\bibfnamefont {Y.}~\bibnamefont {Zhang}}, \bibinfo {author} {\bibfnamefont {F.}~\bibnamefont {Zhou}}, \emph {et~al.},\ }\bibfield  {title} {\bibinfo {title} {Tunable magnetism and band structure in kagome materials reti$_3$bi$_4$ family with weak interlayer interactions},\ }\href@noop {} {\bibfield  {journal} {\bibinfo  {journal} {Science bulletin}\ }\textbf {\bibinfo {volume} {69}},\ \bibinfo {pages} {2660} (\bibinfo {year} {2024})}\BibitemShut {NoStop}%
\bibitem [{\citenamefont {Feng}\ \emph {et~al.}()\citenamefont {Feng}, \citenamefont {Jiang}, \citenamefont {Hu}, \citenamefont {Călugăru}, \citenamefont {Regnault}, \citenamefont {Vergniory}, \citenamefont {Felser}, \citenamefont {Blanco-Canosa},\ and\ \citenamefont {Bernevig}}]{Feng2024_MT6Z6-PhononInstabilities}%
  \BibitemOpen
  \bibfield  {author} {\bibinfo {author} {\bibfnamefont {X.}~\bibnamefont {Feng}}, \bibinfo {author} {\bibfnamefont {Y.}~\bibnamefont {Jiang}}, \bibinfo {author} {\bibfnamefont {H.}~\bibnamefont {Hu}}, \bibinfo {author} {\bibfnamefont {D.}~\bibnamefont {Călugăru}}, \bibinfo {author} {\bibfnamefont {N.}~\bibnamefont {Regnault}}, \bibinfo {author} {\bibfnamefont {M.~G.}\ \bibnamefont {Vergniory}}, \bibinfo {author} {\bibfnamefont {C.}~\bibnamefont {Felser}}, \bibinfo {author} {\bibfnamefont {S.}~\bibnamefont {Blanco-Canosa}},\ and\ \bibinfo {author} {\bibfnamefont {B.~A.}\ \bibnamefont {Bernevig}},\ }\bibfield  {title} {\bibinfo {title} {Catalogue of phonon instabilities in symmetry group 191 kagome mt$_6$z$_6$ materials},\ }\bibfield  {journal} {\bibinfo  {journal} {arXiv (cond-mat.mtrl-sci) 17 Sep 2024. 10.48550/arXiv.2409.13078. (accessed 2025-01-10).}\ }\href {https://doi.org/10.48550/ARXIV.2409.13078} {10.48550/ARXIV.2409.13078},\ \Eprint {https://arxiv.org/abs/2409.13078} {2409.13078} \BibitemShut
  {NoStop}%
\bibitem [{\citenamefont {Cheng}\ \emph {et~al.}(2024{\natexlab{a}})\citenamefont {Cheng}, \citenamefont {Shao}, \citenamefont {Kim}, \citenamefont {Cochran}, \citenamefont {Yang}, \citenamefont {Yi}, \citenamefont {Jiang}, \citenamefont {Zhang}, \citenamefont {Hossain}, \citenamefont {Roychowdhury}, \citenamefont {Yilmaz}, \citenamefont {Vescovo}, \citenamefont {Fedorov}, \citenamefont {Shekhar}, \citenamefont {Felser}, \citenamefont {Chang},\ and\ \citenamefont {Hasan}}]{Cheng2024_ScV6Sn6-ARPES-BulkSurface}%
  \BibitemOpen
  \bibfield  {author} {\bibinfo {author} {\bibfnamefont {Z.-J.}\ \bibnamefont {Cheng}}, \bibinfo {author} {\bibfnamefont {S.}~\bibnamefont {Shao}}, \bibinfo {author} {\bibfnamefont {B.}~\bibnamefont {Kim}}, \bibinfo {author} {\bibfnamefont {T.~A.}\ \bibnamefont {Cochran}}, \bibinfo {author} {\bibfnamefont {X.~P.}\ \bibnamefont {Yang}}, \bibinfo {author} {\bibfnamefont {C.}~\bibnamefont {Yi}}, \bibinfo {author} {\bibfnamefont {Y.-X.}\ \bibnamefont {Jiang}}, \bibinfo {author} {\bibfnamefont {J.}~\bibnamefont {Zhang}}, \bibinfo {author} {\bibfnamefont {M.~S.}\ \bibnamefont {Hossain}}, \bibinfo {author} {\bibfnamefont {S.}~\bibnamefont {Roychowdhury}}, \bibinfo {author} {\bibfnamefont {T.}~\bibnamefont {Yilmaz}}, \bibinfo {author} {\bibfnamefont {E.}~\bibnamefont {Vescovo}}, \bibinfo {author} {\bibfnamefont {A.}~\bibnamefont {Fedorov}}, \bibinfo {author} {\bibfnamefont {C.}~\bibnamefont {Shekhar}}, \bibinfo {author} {\bibfnamefont {C.}~\bibnamefont {Felser}}, \bibinfo {author} {\bibfnamefont {G.}~\bibnamefont
  {Chang}},\ and\ \bibinfo {author} {\bibfnamefont {M.~Z.}\ \bibnamefont {Hasan}},\ }\bibfield  {title} {\bibinfo {title} {{Untangling charge-order dependent bulk states from surface effects in a topological kagome metal ScV$_6$Sn$_6$}},\ }\href {https://doi.org/10.1103/physrevb.109.075150} {\bibfield  {journal} {\bibinfo  {journal} {Physical Review B}\ }\textbf {\bibinfo {volume} {109}},\ \bibinfo {pages} {075150} (\bibinfo {year} {2024}{\natexlab{a}})}\BibitemShut {NoStop}%
\bibitem [{\citenamefont {Tan}\ and\ \citenamefont {Yan}(2023)}]{Tan2023_AbundantInstabilitesScV6Sn6}%
  \BibitemOpen
  \bibfield  {author} {\bibinfo {author} {\bibfnamefont {H.}~\bibnamefont {Tan}}\ and\ \bibinfo {author} {\bibfnamefont {B.}~\bibnamefont {Yan}},\ }\bibfield  {title} {\bibinfo {title} {{Abundant Lattice Instability in Kagome Metal ScV$_6$Sn$_6$}},\ }\href {https://doi.org/10.1103/physrevlett.130.266402} {\bibfield  {journal} {\bibinfo  {journal} {Physical Review Letters}\ }\textbf {\bibinfo {volume} {130}},\ \bibinfo {pages} {266402} (\bibinfo {year} {2023})}\BibitemShut {NoStop}%
\bibitem [{\citenamefont {Tuniz}\ \emph {et~al.}(2023)\citenamefont {Tuniz}, \citenamefont {Consiglio}, \citenamefont {Puntel}, \citenamefont {Bigi}, \citenamefont {Enzner}, \citenamefont {Pokharel}, \citenamefont {Orgiani}, \citenamefont {Bronsch}, \citenamefont {Parmigiani}, \citenamefont {Polewczyk}, \citenamefont {King}, \citenamefont {Wells}, \citenamefont {Zeljkovic}, \citenamefont {Carrara}, \citenamefont {Rossi}, \citenamefont {Fujii}, \citenamefont {Vobornik}, \citenamefont {Wilson}, \citenamefont {Thomale}, \citenamefont {Wehling}, \citenamefont {Sangiovanni}, \citenamefont {Panaccione}, \citenamefont {Cilento}, \citenamefont {Di~Sante},\ and\ \citenamefont {Mazzola}}]{Tuniz2023_CDW-DynamicsScV6Sn6}%
  \BibitemOpen
  \bibfield  {author} {\bibinfo {author} {\bibfnamefont {M.}~\bibnamefont {Tuniz}}, \bibinfo {author} {\bibfnamefont {A.}~\bibnamefont {Consiglio}}, \bibinfo {author} {\bibfnamefont {D.}~\bibnamefont {Puntel}}, \bibinfo {author} {\bibfnamefont {C.}~\bibnamefont {Bigi}}, \bibinfo {author} {\bibfnamefont {S.}~\bibnamefont {Enzner}}, \bibinfo {author} {\bibfnamefont {G.}~\bibnamefont {Pokharel}}, \bibinfo {author} {\bibfnamefont {P.}~\bibnamefont {Orgiani}}, \bibinfo {author} {\bibfnamefont {W.}~\bibnamefont {Bronsch}}, \bibinfo {author} {\bibfnamefont {F.}~\bibnamefont {Parmigiani}}, \bibinfo {author} {\bibfnamefont {V.}~\bibnamefont {Polewczyk}}, \bibinfo {author} {\bibfnamefont {P.~D.~C.}\ \bibnamefont {King}}, \bibinfo {author} {\bibfnamefont {J.~W.}\ \bibnamefont {Wells}}, \bibinfo {author} {\bibfnamefont {I.}~\bibnamefont {Zeljkovic}}, \bibinfo {author} {\bibfnamefont {P.}~\bibnamefont {Carrara}}, \bibinfo {author} {\bibfnamefont {G.}~\bibnamefont {Rossi}}, \bibinfo {author} {\bibfnamefont
  {J.}~\bibnamefont {Fujii}}, \bibinfo {author} {\bibfnamefont {I.}~\bibnamefont {Vobornik}}, \bibinfo {author} {\bibfnamefont {S.~D.}\ \bibnamefont {Wilson}}, \bibinfo {author} {\bibfnamefont {R.}~\bibnamefont {Thomale}}, \bibinfo {author} {\bibfnamefont {T.}~\bibnamefont {Wehling}}, \bibinfo {author} {\bibfnamefont {G.}~\bibnamefont {Sangiovanni}}, \bibinfo {author} {\bibfnamefont {G.}~\bibnamefont {Panaccione}}, \bibinfo {author} {\bibfnamefont {F.}~\bibnamefont {Cilento}}, \bibinfo {author} {\bibfnamefont {D.}~\bibnamefont {Di~Sante}},\ and\ \bibinfo {author} {\bibfnamefont {F.}~\bibnamefont {Mazzola}},\ }\bibfield  {title} {\bibinfo {title} {{Dynamics and resilience of the unconventional charge density wave in ScV$_6$Sn$_6$ bilayer kagome metal}},\ }\href {https://doi.org/10.1038/s43246-023-00430-y} {\bibfield  {journal} {\bibinfo  {journal} {Communications Materials}\ }\textbf {\bibinfo {volume} {4}},\ \bibinfo {pages} {103} (\bibinfo {year} {2023})}\BibitemShut {NoStop}%
\bibitem [{\citenamefont {Cheng}\ \emph {et~al.}(2024{\natexlab{b}})\citenamefont {Cheng}, \citenamefont {Ren}, \citenamefont {Li}, \citenamefont {Oh}, \citenamefont {Tan}, \citenamefont {Pokharel}, \citenamefont {DeStefano}, \citenamefont {Rosenberg}, \citenamefont {Guo}, \citenamefont {Zhang}, \citenamefont {Yue}, \citenamefont {Lee}, \citenamefont {Gorovikov}, \citenamefont {Zonno}, \citenamefont {Hashimoto}, \citenamefont {Lu}, \citenamefont {Ke}, \citenamefont {Mazzola}, \citenamefont {Kono}, \citenamefont {Birgeneau}, \citenamefont {Chu}, \citenamefont {Wilson}, \citenamefont {Wang}, \citenamefont {Yan}, \citenamefont {Yi},\ and\ \citenamefont {Zeljkovic}}]{Cheng2024_STM+ARPES-ScV6Sn6}%
  \BibitemOpen
  \bibfield  {author} {\bibinfo {author} {\bibfnamefont {S.}~\bibnamefont {Cheng}}, \bibinfo {author} {\bibfnamefont {Z.}~\bibnamefont {Ren}}, \bibinfo {author} {\bibfnamefont {H.}~\bibnamefont {Li}}, \bibinfo {author} {\bibfnamefont {J.~S.}\ \bibnamefont {Oh}}, \bibinfo {author} {\bibfnamefont {H.}~\bibnamefont {Tan}}, \bibinfo {author} {\bibfnamefont {G.}~\bibnamefont {Pokharel}}, \bibinfo {author} {\bibfnamefont {J.~M.}\ \bibnamefont {DeStefano}}, \bibinfo {author} {\bibfnamefont {E.}~\bibnamefont {Rosenberg}}, \bibinfo {author} {\bibfnamefont {Y.}~\bibnamefont {Guo}}, \bibinfo {author} {\bibfnamefont {Y.}~\bibnamefont {Zhang}}, \bibinfo {author} {\bibfnamefont {Z.}~\bibnamefont {Yue}}, \bibinfo {author} {\bibfnamefont {Y.}~\bibnamefont {Lee}}, \bibinfo {author} {\bibfnamefont {S.}~\bibnamefont {Gorovikov}}, \bibinfo {author} {\bibfnamefont {M.}~\bibnamefont {Zonno}}, \bibinfo {author} {\bibfnamefont {M.}~\bibnamefont {Hashimoto}}, \bibinfo {author} {\bibfnamefont {D.}~\bibnamefont {Lu}}, \bibinfo {author}
  {\bibfnamefont {L.}~\bibnamefont {Ke}}, \bibinfo {author} {\bibfnamefont {F.}~\bibnamefont {Mazzola}}, \bibinfo {author} {\bibfnamefont {J.}~\bibnamefont {Kono}}, \bibinfo {author} {\bibfnamefont {R.~J.}\ \bibnamefont {Birgeneau}}, \bibinfo {author} {\bibfnamefont {J.-H.}\ \bibnamefont {Chu}}, \bibinfo {author} {\bibfnamefont {S.~D.}\ \bibnamefont {Wilson}}, \bibinfo {author} {\bibfnamefont {Z.}~\bibnamefont {Wang}}, \bibinfo {author} {\bibfnamefont {B.}~\bibnamefont {Yan}}, \bibinfo {author} {\bibfnamefont {M.}~\bibnamefont {Yi}},\ and\ \bibinfo {author} {\bibfnamefont {I.}~\bibnamefont {Zeljkovic}},\ }\bibfield  {title} {\bibinfo {title} {{Nanoscale visualization and spectral fingerprints of the charge order in ScV$_6$Sn$_6$ distinct from other kagome metals}},\ }\href {https://doi.org/10.1038/s41535-024-00623-9} {\bibfield  {journal} {\bibinfo  {journal} {npj Quantum Materials}\ }\textbf {\bibinfo {volume} {9}},\ \bibinfo {pages} {14} (\bibinfo {year} {2024}{\natexlab{b}})}\BibitemShut {NoStop}%
\bibitem [{\citenamefont {Kang}\ \emph {et~al.}()\citenamefont {Kang}, \citenamefont {Li}, \citenamefont {Meier}, \citenamefont {Villanova}, \citenamefont {Hus}, \citenamefont {Jeon}, \citenamefont {Arachchige}, \citenamefont {Lu}, \citenamefont {Gai}, \citenamefont {Denlinger}, \citenamefont {Moore}, \citenamefont {Yoon},\ and\ \citenamefont {Mandrus}}]{Kang2023_LifshitzScV6Sn6}%
  \BibitemOpen
  \bibfield  {author} {\bibinfo {author} {\bibfnamefont {S.-H.}\ \bibnamefont {Kang}}, \bibinfo {author} {\bibfnamefont {H.}~\bibnamefont {Li}}, \bibinfo {author} {\bibfnamefont {W.~R.}\ \bibnamefont {Meier}}, \bibinfo {author} {\bibfnamefont {J.~W.}\ \bibnamefont {Villanova}}, \bibinfo {author} {\bibfnamefont {S.}~\bibnamefont {Hus}}, \bibinfo {author} {\bibfnamefont {H.}~\bibnamefont {Jeon}}, \bibinfo {author} {\bibfnamefont {H.~W.~S.}\ \bibnamefont {Arachchige}}, \bibinfo {author} {\bibfnamefont {Q.}~\bibnamefont {Lu}}, \bibinfo {author} {\bibfnamefont {Z.}~\bibnamefont {Gai}}, \bibinfo {author} {\bibfnamefont {J.}~\bibnamefont {Denlinger}}, \bibinfo {author} {\bibfnamefont {R.}~\bibnamefont {Moore}}, \bibinfo {author} {\bibfnamefont {M.}~\bibnamefont {Yoon}},\ and\ \bibinfo {author} {\bibfnamefont {D.}~\bibnamefont {Mandrus}},\ }\bibfield  {title} {\bibinfo {title} {Emergence of a new band and the lifshitz transition in kagome metal scv$_6$sn$_6$ with charge density wave},\ }\bibfield  {journal} {\bibinfo
   {journal} {arXiv (cond-mat.str-el) 27 Feb 2023. 10.48550/arxiv.2302.14041. (accessed 2025-01-10).}\ }\href {https://doi.org/10.48550/arxiv.2302.14041} {10.48550/arxiv.2302.14041},\ \Eprint {https://arxiv.org/abs/2302.14041} {2302.14041 [cond-mat.str-el]} \BibitemShut {NoStop}%
\bibitem [{\citenamefont {Alvarado}\ \emph {et~al.}(2024)\citenamefont {Alvarado}, \citenamefont {Pokharel}, \citenamefont {Ortiz}, \citenamefont {Paddison}, \citenamefont {Sarker}, \citenamefont {Ruff},\ and\ \citenamefont {Wilson}}]{Alvarado2024_FrustratedIsingChargeCorrScV6Sn6}%
  \BibitemOpen
  \bibfield  {author} {\bibinfo {author} {\bibfnamefont {S.~J.~G.}\ \bibnamefont {Alvarado}}, \bibinfo {author} {\bibfnamefont {G.}~\bibnamefont {Pokharel}}, \bibinfo {author} {\bibfnamefont {B.~R.}\ \bibnamefont {Ortiz}}, \bibinfo {author} {\bibfnamefont {J.~A.~M.}\ \bibnamefont {Paddison}}, \bibinfo {author} {\bibfnamefont {S.}~\bibnamefont {Sarker}}, \bibinfo {author} {\bibfnamefont {J.~P.~C.}\ \bibnamefont {Ruff}},\ and\ \bibinfo {author} {\bibfnamefont {S.~D.}\ \bibnamefont {Wilson}},\ }\bibfield  {title} {\bibinfo {title} {{Frustrated Ising charge correlations in the kagome metal ScV$_6$Sn$_6$}},\ }\href {https://doi.org/10.1103/physrevb.110.l140304} {\bibfield  {journal} {\bibinfo  {journal} {Physical Review B}\ }\textbf {\bibinfo {volume} {110}},\ \bibinfo {pages} {l140304} (\bibinfo {year} {2024})}\BibitemShut {NoStop}%
\bibitem [{\citenamefont {Xu}\ \emph {et~al.}(2022)\citenamefont {Xu}, \citenamefont {Yin}, \citenamefont {Ma}, \citenamefont {Tien}, \citenamefont {Qiang}, \citenamefont {Reddy}, \citenamefont {Zhou}, \citenamefont {Shen}, \citenamefont {Lu}, \citenamefont {Chang}, \citenamefont {Qu},\ and\ \citenamefont {Jia}}]{Xu2022_ChernMagnetYbMn6Sn6}%
  \BibitemOpen
  \bibfield  {author} {\bibinfo {author} {\bibfnamefont {X.}~\bibnamefont {Xu}}, \bibinfo {author} {\bibfnamefont {J.-X.}\ \bibnamefont {Yin}}, \bibinfo {author} {\bibfnamefont {W.}~\bibnamefont {Ma}}, \bibinfo {author} {\bibfnamefont {H.-J.}\ \bibnamefont {Tien}}, \bibinfo {author} {\bibfnamefont {X.-B.}\ \bibnamefont {Qiang}}, \bibinfo {author} {\bibfnamefont {P.~V.~S.}\ \bibnamefont {Reddy}}, \bibinfo {author} {\bibfnamefont {H.}~\bibnamefont {Zhou}}, \bibinfo {author} {\bibfnamefont {J.}~\bibnamefont {Shen}}, \bibinfo {author} {\bibfnamefont {H.-Z.}\ \bibnamefont {Lu}}, \bibinfo {author} {\bibfnamefont {T.-R.}\ \bibnamefont {Chang}}, \bibinfo {author} {\bibfnamefont {Z.}~\bibnamefont {Qu}},\ and\ \bibinfo {author} {\bibfnamefont {S.}~\bibnamefont {Jia}},\ }\bibfield  {title} {\bibinfo {title} {{Topological charge-entropy scaling in kagome Chern magnet TbMn$_6$Sn$_6$}},\ }\href {https://doi.org/10.1038/s41467-022-28796-6} {\bibfield  {journal} {\bibinfo  {journal} {Nature Communications}\ }\textbf
  {\bibinfo {volume} {13}},\ \bibinfo {pages} {1197} (\bibinfo {year} {2022})}\BibitemShut {NoStop}%
\bibitem [{\citenamefont {El~Idrissi}\ \emph {et~al.}(1991{\natexlab{c}})\citenamefont {El~Idrissi}, \citenamefont {Venturini}, \citenamefont {Malaman},\ and\ \citenamefont {Fruchart}}]{ElIdrissi1991_NeutronMagStructTbMn6Sn6+HoMn6Sn6}%
  \BibitemOpen
  \bibfield  {author} {\bibinfo {author} {\bibfnamefont {B.}~\bibnamefont {El~Idrissi}}, \bibinfo {author} {\bibfnamefont {G.}~\bibnamefont {Venturini}}, \bibinfo {author} {\bibfnamefont {B.}~\bibnamefont {Malaman}},\ and\ \bibinfo {author} {\bibfnamefont {D.}~\bibnamefont {Fruchart}},\ }\bibfield  {title} {\bibinfo {title} {{Magnetic structures of TbMn$_6$Sn$_6$ and HoMn$_6$Sn$_6$ compounds from neutron diffraction study}},\ }\href {https://doi.org/10.1016/0022-5088(91)90359-c} {\bibfield  {journal} {\bibinfo  {journal} {Journal of the Less Common Metals}\ }\textbf {\bibinfo {volume} {175}},\ \bibinfo {pages} {143} (\bibinfo {year} {1991}{\natexlab{c}})}\BibitemShut {NoStop}%
\bibitem [{\citenamefont {Mozaffari}\ \emph {et~al.}(2024)\citenamefont {Mozaffari}, \citenamefont {Meier}, \citenamefont {Madhogaria}, \citenamefont {Peshcherenko}, \citenamefont {Kang}, \citenamefont {Villanova}, \citenamefont {Arachchige}, \citenamefont {Zheng}, \citenamefont {Zhu}, \citenamefont {Chen}, \citenamefont {Jenkins}, \citenamefont {Zhang}, \citenamefont {Chan}, \citenamefont {Li}, \citenamefont {Yoon}, \citenamefont {Zhang},\ and\ \citenamefont {Mandrus}}]{Mozaffari2024_SublinearResistivityVKagome}%
  \BibitemOpen
  \bibfield  {author} {\bibinfo {author} {\bibfnamefont {S.}~\bibnamefont {Mozaffari}}, \bibinfo {author} {\bibfnamefont {W.~R.}\ \bibnamefont {Meier}}, \bibinfo {author} {\bibfnamefont {R.~P.}\ \bibnamefont {Madhogaria}}, \bibinfo {author} {\bibfnamefont {N.}~\bibnamefont {Peshcherenko}}, \bibinfo {author} {\bibfnamefont {S.-H.}\ \bibnamefont {Kang}}, \bibinfo {author} {\bibfnamefont {J.~W.}\ \bibnamefont {Villanova}}, \bibinfo {author} {\bibfnamefont {H.~W.~S.}\ \bibnamefont {Arachchige}}, \bibinfo {author} {\bibfnamefont {G.}~\bibnamefont {Zheng}}, \bibinfo {author} {\bibfnamefont {Y.}~\bibnamefont {Zhu}}, \bibinfo {author} {\bibfnamefont {K.-W.}\ \bibnamefont {Chen}}, \bibinfo {author} {\bibfnamefont {K.}~\bibnamefont {Jenkins}}, \bibinfo {author} {\bibfnamefont {D.}~\bibnamefont {Zhang}}, \bibinfo {author} {\bibfnamefont {A.}~\bibnamefont {Chan}}, \bibinfo {author} {\bibfnamefont {L.}~\bibnamefont {Li}}, \bibinfo {author} {\bibfnamefont {M.}~\bibnamefont {Yoon}}, \bibinfo {author} {\bibfnamefont
  {Y.}~\bibnamefont {Zhang}},\ and\ \bibinfo {author} {\bibfnamefont {D.~G.}\ \bibnamefont {Mandrus}},\ }\bibfield  {title} {\bibinfo {title} {Universal sublinear resistivity in vanadium kagome materials hosting charge density waves},\ }\href {https://doi.org/10.1103/physrevb.110.035135} {\bibfield  {journal} {\bibinfo  {journal} {Physical Review B}\ }\textbf {\bibinfo {volume} {110}},\ \bibinfo {pages} {035135} (\bibinfo {year} {2024})}\BibitemShut {NoStop}%
\bibitem [{\citenamefont {Shannon}(1976)}]{Shannon1976_Radii}%
  \BibitemOpen
  \bibfield  {author} {\bibinfo {author} {\bibfnamefont {R.~D.}\ \bibnamefont {Shannon}},\ }\bibfield  {title} {\bibinfo {title} {Revised effective ionic radii and systematic studies of interatomic distances in halides and chalcogenides},\ }\href {https://doi.org/10.1107/S0567739476001551} {\bibfield  {journal} {\bibinfo  {journal} {Acta Cryst.}\ }\textbf {\bibinfo {volume} {A32}},\ \bibinfo {pages} {751} (\bibinfo {year} {1976})}\BibitemShut {NoStop}%
\bibitem [{\citenamefont {Artini}(2017)}]{Pani2017_FormationVolumeRareEarthIntermetallic}%
  \BibitemOpen
  \bibinfo {editor} {\bibfnamefont {C.}~\bibnamefont {Artini}},\ ed.,\ \href@noop {} {\emph {\bibinfo {title} {Alloys and intermetallic compounds: from modeling to engineering}}}\ (\bibinfo  {publisher} {CRC press},\ \bibinfo {year} {2017})\BibitemShut {NoStop}%
\bibitem [{\citenamefont {Zhang}\ \emph {et~al.}(2022{\natexlab{b}})\citenamefont {Zhang}, \citenamefont {Hou}, \citenamefont {Xia}, \citenamefont {Xu}, \citenamefont {Yang}, \citenamefont {Wang}, \citenamefont {Liu}, \citenamefont {Shen}, \citenamefont {Zhang}, \citenamefont {Dong}, \citenamefont {Uwatoko}, \citenamefont {Sun}, \citenamefont {Wang}, \citenamefont {Guo},\ and\ \citenamefont {Cheng}}]{Zhang2022_ScV6Sn6-pressure}%
  \BibitemOpen
  \bibfield  {author} {\bibinfo {author} {\bibfnamefont {X.}~\bibnamefont {Zhang}}, \bibinfo {author} {\bibfnamefont {J.}~\bibnamefont {Hou}}, \bibinfo {author} {\bibfnamefont {W.}~\bibnamefont {Xia}}, \bibinfo {author} {\bibfnamefont {Z.}~\bibnamefont {Xu}}, \bibinfo {author} {\bibfnamefont {P.}~\bibnamefont {Yang}}, \bibinfo {author} {\bibfnamefont {A.}~\bibnamefont {Wang}}, \bibinfo {author} {\bibfnamefont {Z.}~\bibnamefont {Liu}}, \bibinfo {author} {\bibfnamefont {J.}~\bibnamefont {Shen}}, \bibinfo {author} {\bibfnamefont {H.}~\bibnamefont {Zhang}}, \bibinfo {author} {\bibfnamefont {X.}~\bibnamefont {Dong}}, \bibinfo {author} {\bibfnamefont {Y.}~\bibnamefont {Uwatoko}}, \bibinfo {author} {\bibfnamefont {J.}~\bibnamefont {Sun}}, \bibinfo {author} {\bibfnamefont {B.}~\bibnamefont {Wang}}, \bibinfo {author} {\bibfnamefont {Y.}~\bibnamefont {Guo}},\ and\ \bibinfo {author} {\bibfnamefont {J.}~\bibnamefont {Cheng}},\ }\bibfield  {title} {\bibinfo {title} {Destabilization of the charge density wave and the
  absence of superconductivity in {ScV}$_{6}${Sn}$_{6}$ under high pressures up to 11 {GPa}},\ }\href {https://doi.org/10.3390/ma15207372} {\bibfield  {journal} {\bibinfo  {journal} {Materials}\ }\textbf {\bibinfo {volume} {15}},\ \bibinfo {pages} {7372} (\bibinfo {year} {2022}{\natexlab{b}})}\BibitemShut {NoStop}%
\end{thebibliography}%

\end{document}